\documentclass{article}
\usepackage[utf8]{inputenc}
\usepackage{authblk}
\title{Review: How dynamic prestress governs the shape of living systems, from the subcellular to tissue scale}

\author[1,2,3]{Alexander Erlich}
\author[3,*]{Jocelyn \'Etienne}
\author[4,*]{Jonathan Fouchard}
\author[5]{Tom Wyatt}
\affil[1]{Institut de Recherche sur les Phénomènes Hors Équilibre (IRPHÉ), Aix-Marseille Université  (ORCID Erlich: https://orcid.org/0000-0002-2294-1894)}
\affil[2]{Institut de Biologie du Développement de Marseille (IBDM), Aix-Marseille Université}
\affil[3]{Univ.\ Grenoble Alpes, CNRS, LIPHY, 38000 Grenoble, France (ORCID Étienne: https://orcid.org/0000-0002-1866-5604)}
\affil[4]{Laboratoire de Biologie du Développement, Institut de Biologie Paris Seine (IBPS), Sorbonne Université, CNRS (UMR 7622), INSERM (URL 1156), 7 quai Saint Bernard, 75005 Paris, France
	(ORCID Fouchard: https://orcid.org/0000-0002-9976-462X)}
\affil[5]{Wellcome Trust-Medical Research Council Cambridge Stem Cell Institute, University of Cambridge, Cambridge, UK (ORCID Wyatt: https://orcid.org/0000-0002-2589-2370)}
\affil[*]{Corresponding authors : jocelyn.etienne@univ-grenoble-alpes.fr, jonathan.fouchard@sorbonne-universite.fr}

\usepackage[a4paper,left=2in,right=0.5in]{geometry}
\usepackage{graphicx,subfigure}
\usepackage{amsmath}
\usepackage{amssymb}
\usepackage{dsfont} % for \mathds, like the Identity tensor symbol
\usepackage[authoryear]{natbib}
\usepackage[colorlinks,urlcolor=blue]{hyperref}
\usepackage[table, x11names]{xcolor}

\usepackage{tabularx}
\usepackage{multirow}
\usepackage{tikz}
\usetikzlibrary{decorations.pathmorphing,patterns}
\usepackage{multicol}

\newcommand{\fig}[1]{\protect Fig.\ \ref{#1}}
\newcommand{\eq}[1]{Eq.\ (\ref{#1})}

\begin{document}

\maketitle

%\JF{List of the things we do not review for now: microtubule networks (mitotic spindle) where prestress has been demonstrated}
%\JE{mechano-sensation: I think it's slightly off-topic for people from rheology/mechanics, since it's obvious for them that a deformable element reacts to outside mechanical environment. Mechano-transduction is on the contrary very original but rather as an opening (discussion?). Microtubules: I think that'd be good to mention that in the networks part, I'll maybe make an attempt and you adjust it to make it representative of papers you want to cite? Thermodynamics: that'd be too much, but we have to refer more explicitly to active gel theory.}

%\JE{Checklist for consistency:
%\\- prestress/prestrain/actomyosin/nonlinear vs pre-stress/pre-strain/acto-/non-/micro-... I'd go without - everywhere.
%\\- myosin/Myosin/Myosin II and also e.g.\ Blebbistatin...: since we don't do genetics, I'm not sure it makes sense to use the capitalization convention of biologists. I'd go for plain "myosin" everywhere, and I don't think we should specify "II" -- we say "adhesion", e.g., not saying what molecules are involved, I believe for myosin similarly we use is as "some specific molecular motor", which is all is needed to understand the mechanics. Then  Drosophila with italics.}

\begin{abstract}
Cells and tissues change shape both to {carry out} their function and during pathology. In most cases, these deformations are driven from within the systems themselves. This is permitted by a range of molecular actors, such as active crosslinkers and ion pumps, whose activity is biologically controlled in space and time. The resulting stresses are propagated within complex and dynamical architectures like networks or cell aggregates. From a mechanical point of view, these effects can be seen as the generation of prestress or prestrain, resulting from either a contractile or growth activity. In this review, we present this concept of prestress and the theoretical tools available to conceptualise the statics and dynamics of living systems. We then describe a range of phenomena where prestress controls shape changes in biopolymer networks (especially the actomyosin cytoskeleton and fibrous tissues) and cellularised tissues. Despite the diversity of scale and organisation, we demonstrate that these phenomena stem from a limited number of spatial distributions of prestress, which can be categorised as heterogeneous, anisotropic or differential. We suggest that in addition to growth and contraction, a third type of prestress -- topological prestress -- can result from active processes altering the microstructure of tissue. 
\end{abstract}

\section{Introduction}

A peculiarity of living systems is their ability to constantly rearrange their structure in order to perform biological function. Cells, for example, transition from static to migrating while changing shape, or split in the process of cell division. At the tissue scale, specific shapes are acquired during development and maintained at adult age to accomplish organ function, but can be lost as in the case of cancer.
These rearrangements are in most cases the result of active processes taking place within the cells or tissues themselves, rather than being imposed from the exterior through boundary conditions. 

Across the scales and specificities of systems, one finds a number of ways for these internal stresses to be generated, ranging from protein synthesis or pumping of ions that give rise to osmotic pressure gradients to ATP hydrolysis-fuelled changes of conformation of crosslinkers within biopolymer networks. One common point of these mechanisms is that they are controlled by biological pathways, and that they can be triggered or modulated dynamically, enabling systems to change shape or state.
In this review, we aim to show how these very different force-generation mechanisms can be usefully understood with a common concept of prestress (and prestrain). 
{While this concept has been useful in describing stable tissue shape in adult tissues \citep{taber1995biomechanics}, here we focus on how dynamic changes in prestress can alter shape.}
Often, prestress generation interplays with other less specific properties of living systems like their complex rheological properties, thin-sheet geometry and foam or network architecture. In some cases, it appears challenging to distinguish between phenomena stemming from dynamic prestress and those stemming from these complex material properties. An example developed below is the cell neighbour exchange (also called T1 transition) occurring in epithelia, which can have either active or passive origins. 

In section 2 of the review, we will present the concept of prestress and the theoretical frameworks available to describe prestress in statics and dynamics. Following on, we will briefly present in section 3 the strategies available to experimentalists to identify and measure prestress in living systems. Then, in the final two sections, we will describe strategies of prestress generation in biological systems and the macroscopic effects obtained. On our way, we will show that common categories of spatial regulation of prestress (heterogeneous, anisotropic or differential) are used in systems with different compositions and scales. 

{Two key concepts will be used throughout this review: \emph{growth} and \emph{contraction}. 
By \emph{growth}, we mean any active process that leads to increase the equilibrium size of some of the components of a system. This can be due to the addition of material within the system by some out-of-equilibrium process, e.g.\ the polymerisation of filaments fed by monomers diffusing into it. This can also be more simply due to osmotically-driven attraction of more water into the system.
By \emph{contraction}, we mean any active process that leads to the reduction of the equilibrium size of some of the components of a system.}

In section 4 of the review, we will focus on crosslinked networks of semiflexible biopolymers, either within cells---they are in this case part of the \emph{cytoskeleton}---or external to them---then part of the \emph{extracellular matrix} (ECM). These networks can be remodelled and crosslinked by other proteins which have out-of-equilibrium dynamics, fuelled by active processes. We will show how contraction and growth of these networks govern {shapes and deformations of subcellular compartments, cells and fibrous tissue}. Within the cytoskeleton, one network of particular interest is the one formed by \emph{actin} and the crosslinker (and molecular motor) \emph{myosin}.

Finally, in section 5, we will describe how prestress affects the shape of cellularised tissue. Here, {the material is formed by} cells of regulated volume {and} mechanically connected through \emph{adhesions}, molecular bonds joining their \emph{plasma membranes}. The adhesions are formed of transmembrane complexes, allowing tension to be transmitted between the cytoskeletons of neighbouring cells. {
In this section, the elements experiencing growth and contraction interact with these other elements, making the material heterogeneous and bringing additional effects. }
Prestress can also be built through topological changes of the {cell contours}. We refer to this type of prestress as {\emph{topological prestress}. In particular, we suggest that, from a mechanical point of view, morphogenesis through cell proliferation is conveniently described if one distinguishes within the effect of cell division between growth prestress (increase of volume of cells%linked with  water and nutrient intake and protein synthesis
) and topological prestress (due to the apprearance of new
cell--cell junctions following cytokinesis events).}

\section{\label{sec:Concept-of-prestress}Concept of prestress in living systems}

In engineering, generally, ``pre'' in ``prestress'' refers to the fact that it is due to an operation done before establishing the system, for instance imposing boundary conditions such as tension on a structure such as wires, called tendons, before putting them in parallel with a compression-bearing material, e.g.\ by casting concrete.
This results in a system whose reference configuration is not compatible with the reference configuration of each of its components, which are thus prestressed. In systems of linear elastic components, one can equivalently consider that they are prestrained, the prestress and prestrain being simply related by the elastic modulus.

The source of prestress is not necessarily an externally imposed force applied to a component of the system, but can also be due to an internal change in the system. Prestress can thus arise in a system which is already connected and is originally stress-free, if one of the components, which we will call the active component, changes its equilibrium configuration. As an example, a porous material (passive component)  whose pores are occupied by a liquid will become prestressed if that liquid (active component) changes volume due to freezing or crystallisation. In that case, ``pre'' is not understood anymore as referring to a process in time, but rather as making reference to the fact that prestrain corresponds to the deformation between the initial configuration and a virtual configuration where the active component has assumed its new equilibrium shape. However, this stress-free configuration is virtual because the shapes of the active and passive components are not compatible anymore. 

The actual configuration in the absence of external load results from the mechanical balance between the active and the passive components, neither of which will be stress-free: the observed stress field in the absence of external loads is called residual stress \citep{fung2013biomechanics,taber1995biomechanics,goriely2017mathematics}. While residual stress has this narrow definition, the term prestress is somewhat broader. In the engineering community, prestress is typically due to external loads \citep{gower2017new}. The prestressed configuration is then used as a reference configuration, from which another elastic deformation (for instance the wave propagation in the prestressed body) is studied \citep{parnell2012nonlinear}.  In biophysical models, prestress can also be of active origin, for instance due to a morphogenetic event \citep{amar2018assessing}. We take advantage from this relative freedom to define prestress so that it corresponds to the notion of \emph{active stress} which was defined in the context of actomyosin systems \citep{Juelicher+.2007.1}. Indeed, if one chooses to describe the prestressed material with respect to its original shape, that is, the configuration it would have in the absence of prestress, as a reference configuration, then one finds that there is now a stress field associated to it, which for this choice is a \emph{prestress field} \citep{salenccon1994mecanique}.
Since this prestress does not need to preexist the system, it can be adjusted at any time by nonmechanical processes, such as biological pathways, so that it can drive dynamics.

\begin{figure}
    \centering
    \includegraphics[width=0.75\textwidth]{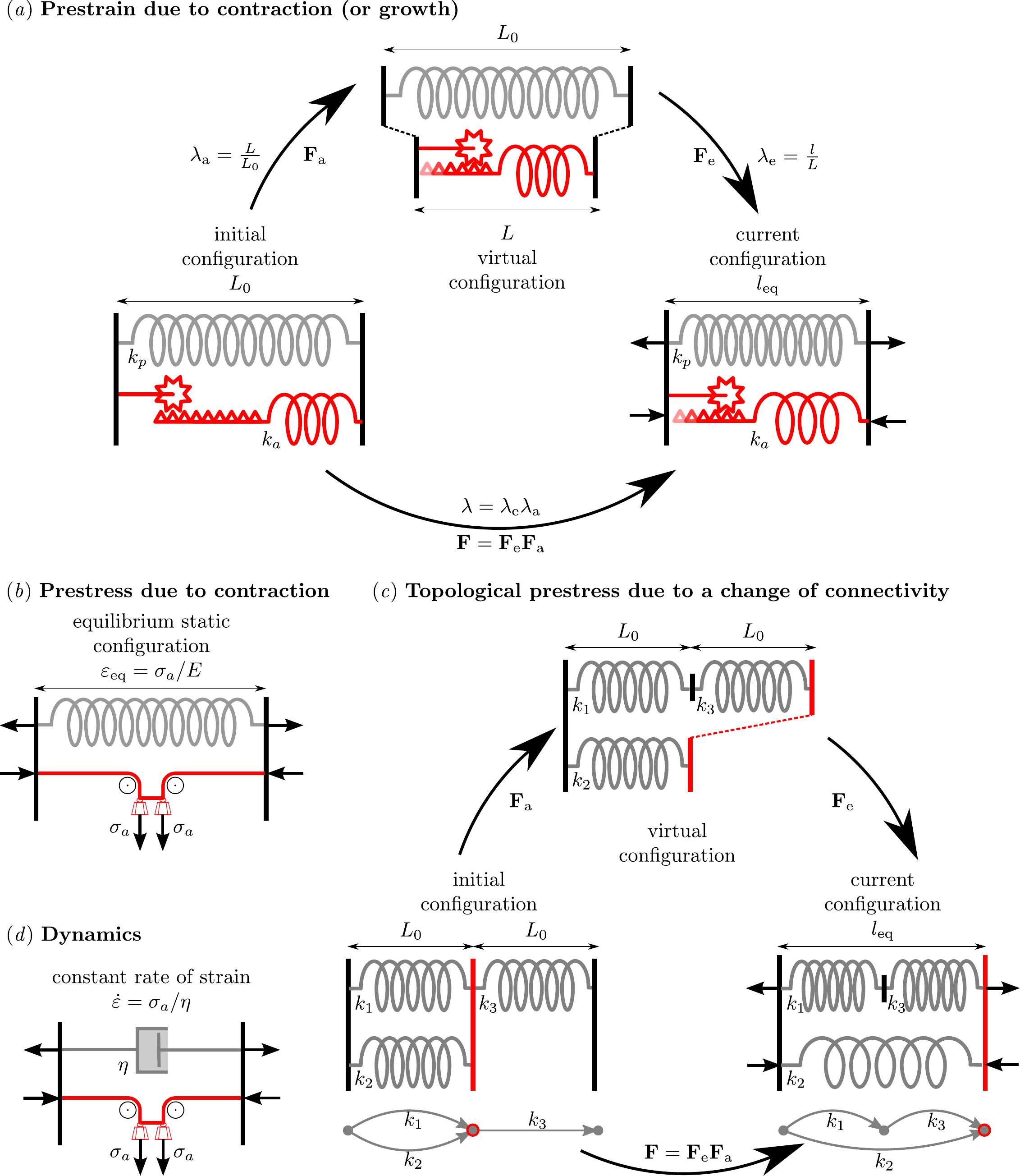}
    \caption{
    Prestrain and prestress in simple 1D systems.
    (\textit{a}) 
     Illustration of deformation gradient decomposition for contractile prestrain.
     An active element (red) and a passive one (gray) are put in parallel in between  force-bearing walls (black vertical lines).
     The active element is composed of a spring whose length is actively decreased (or increased) from $L_0$ to $L$, with an anelastic stretch, or \emph{prestrain}, $\lambda_{{a}}$ (here $\lambda_{{a}}<1$) imposed via a crank. In the virtual configuration, both elements remain stress-free but the system's topology (dashed line connections) is not respected. In the current configuration, even at equilibrium (no net force on the walls), the  structure is under stress. Operating the crank the other way, $\lambda_a>1$, gives the effect of growth prestrain. 
     (\textit{b}) A system equivalent to the one in \textit{a} can be obtained by replacing the active crank and spring element by a stress generator element (pulleys and weight system) whose magnitude is the \emph{prestress}. 
     (\textit{c}) For topological prestress, no active spring is necessary; the activity consists of {disconnecting} and reconnecting elements into a new network. Initially, springs $k_1$ and $k_2$ are in parallel, and the pair is connected in series with $k_3$. The topological change reconnects the springs, such that $k_1$ and $k_3$ are in series, this pair connected in parallel with $k_2$. Due to the change of topology (see directed graph insets) the initially stress-free structure becomes prestressed; spring $k_2$ is in tension, springs $k_1$ and $k_3$ are in compression.
     (\textit{d}) A viscous passive element (dashpot) in parallel with a stress generator gives a permanent regime of contraction at a constant strain rate.
     }
    \label{fig:decomposition}
\end{figure}

%\JE{I've updated for consistency (esp the actomyosin bit) and for these cycles things, rereading welcome.}
%\JF{that works. only thing left : there is twice the reference to active stress.}\JE{done}

%\JE[
%The concept of pre-stress differs somewhat from the related concept of \textit{residual stress}. The stress that remains in the absence of external loads is called residual stress {\citep{fung2013biomechanics,taber1995biomechanics,goriely2017mathematics}}. Pre-stress, on the other hand, can also arise due to external forces {\citep{gower2017new,parnell2012nonlinear,salenccon1994mecanique}}. The concept of pre-stress is therefore broader than, and encompasses, residual stress. We take this relative freedom and define, for the purpose of this article, pre-stress to be the stress caused by active internal re-arrangements, provided that the surface of the object is constrained to the initial stress-free configuration.  
%]{I have made suggestions to include this in the discussion above}

We now illustrate the concept using elementary mechanical elements. {%Note that a variety of descriptions exist when it comes to modelling pre-stress. 
In order to describe the behaviour of individual cells within tissue, a leading approach is to use cell-based discrete models, which can capture cell--cell interactions and dynamical changes in topology. % and topological stress. 
On the other hand, continuum descriptions offer the benefit of %being close to experimentally measurable mechanical 
giving access to quantities such as the Young's modulus or Poisson's ratio, and offer a vast array of tools for simplification through mathematical analysis %. The two types of modelling approaches are contrasted in great detail 
\citep{jones2012modeling}. For the sake of simplicity of presentation, we present ideas in terms of continuum models when possible, and discrete models when necessary.}

\subsection{Static description\label{sec:static}}

We illustrate the concept of a virtual configuration in \fig{fig:decomposition}\textit{a} based on a 1D example, meaning that we only consider deformations in the horizontal direction. An active element (red) is connected in parallel to a passive element (gray). The vertical black lines represent force-bearing walls.  In the initial configuration, both elements have length $L_0$, and the system is unstressed. A contraction reduces the length of the active element to $L$, acting effectively like a spring that reduces its rest length through an active process. We refer to this active stretch as $\lambda_a=L/L_0$. In this example, $\lambda_a<1$, since an active contraction occurs. In the virtual configuration, the system is still stress-free, but incompatible, since the two elements now have different rest lengths making their existing connection impossible  \citep{eckart1948thermodynamics}. Compatibility is restored through building stress: in the current configuration, both elements have length $l$, which is the new equilibrium length. The elastic stretch of the active element is $\lambda_e=l/L$. In this new equilibrium configuration, the active element is in tension ($\lambda_e>1$), and the passive spring is in compression (its elastic stretch is $l/L_0<1$). 
%\JF{'In this new equilibrium configuration, the elastic element is in tension', I would write 'In this new equilibrium configuration, the active element is in tension', to avoid confusion \JE{done}} 
In this framework, growth can be treated the same way, with one crucial difference: the active element increases, rather than decreases, its rest length, so that the active stretch is $\lambda_a>1$.

%\JE[In experiments, pre-strains are generally very difficult to access, as they can only be retrieved through destructive experiments (through cutting of arteries {\citep{chuong1986residual}}, cutting of solid tumors {\citep{stylianopoulos2012causes}}, and laser ablation of epithelia {\citep{bonnet2012mechanical}}). It is therefore useful to link the prestrain $\lambda_a$ to a prestress field, which we shall call $\sigma_a$.]{Unsure that this is the right motivation: prestress are not easy to monitor either!} 
It is useful to understand the relationship between prestrain and prestress, since it allows to make a link with the work on active stress in the context of contractile networks \citep{Juelicher+.2007.1}.
To understand this relationship, let us consider one active element in isolation, with initial length $L_0$, virtual length $L$, and current length $l$. The Hookean stress--strain relationship of this element is 
\begin{equation}
    \sigma = E\left(\frac{l}{L_0}\lambda_a^{-1} -1\right),
    \label{eq:stress-strain-spring}
\end{equation} 
where $E$ is Young's modulus. 
If we now deform the spring to the formerly stress-free state ($l=L_0$), we will be met with resisting stress $\sigma$ equal to the prestress $\sigma=\sigma_a=E(\lambda_a^{-1}-1)$.
{In this simple case, there is thus an explicit relation that can be written between the prestress field $\sigma_a$ and the active stretch that characterises prestrain $\lambda_a=L/L_0$.}
How can this be made sense of in the context of the stress--strain relationship of the spring, Eq.~\eqref{eq:stress-strain-spring}? It is possible to split \eqref{eq:stress-strain-spring} as follows:
\begin{equation}
    \sigma=\Tilde{E}\left(\frac{l}{L_0}-1\right)+\sigma_a\,.
     \label{eq:stress-strain-spring-transformed}
\end{equation}
This shows a Hookean behaviour near the initial, formerly stress-free state, with the modified Young's modulus $\Tilde{E}=E+\sigma_a$. At the formerly stress-free configuration $l=L_0$, we recover the prestress $\sigma=\sigma_a$. Up to the change of elastic modulus, we thus see the equivalence between a system in which the active element is a growing ($\sigma_a<0$) or contracting ($\sigma_a>0$) elastic material and a system in which the active element is a "stress generator" of magnitude $\sigma_a$, see \fig{fig:decomposition}\textit{b}.

% \JF{I find it interesting to write it like : 
% \begin{equation}
%     \sigma=\left(\sigma_a + E\right)\left(\frac{l}{L_0}-1\right)+\sigma_a
% \end{equation}
% I measured in suspended monolayers that E scales like $\sigma_a$
% \JE{Yes, very good like this, I've put that in the def of $\Tilde{E}$ so that it introduces one idea after the other.}
% }
%\JE{actually, we can make systems of fig 1a and b strictly equivalent for finite stiffness of both red and gray springs, so there's no "change of elastic modulus" then. It could be written out in a short section if we have a supplementary text?}
%\AE{I'm not following the new Fig 1b. What is the actual length, and rest length, of the red spring?}
%\JE{It's needed in order to make the 1a and 1b strictly equivalent,  not sure we want to keep it. I'll write details in appendix}

% \JE{Maybe we should give more details of what the "growing bar" is in terms of mechanics. I think it can be represented in your case as a crank, and the stress generator with a weight:\\
% \includegraphics[width=0.4\textwidth]{prestress_basics.pdf}
% Note that the crank needs to be in series with another element, because it is a "strain generator" which defines no force, whereas the weight element needs to be in parallel with some other because it defines no length.
% }

\begin{figure}
    \centering
    \includegraphics[width=0.75\textwidth]{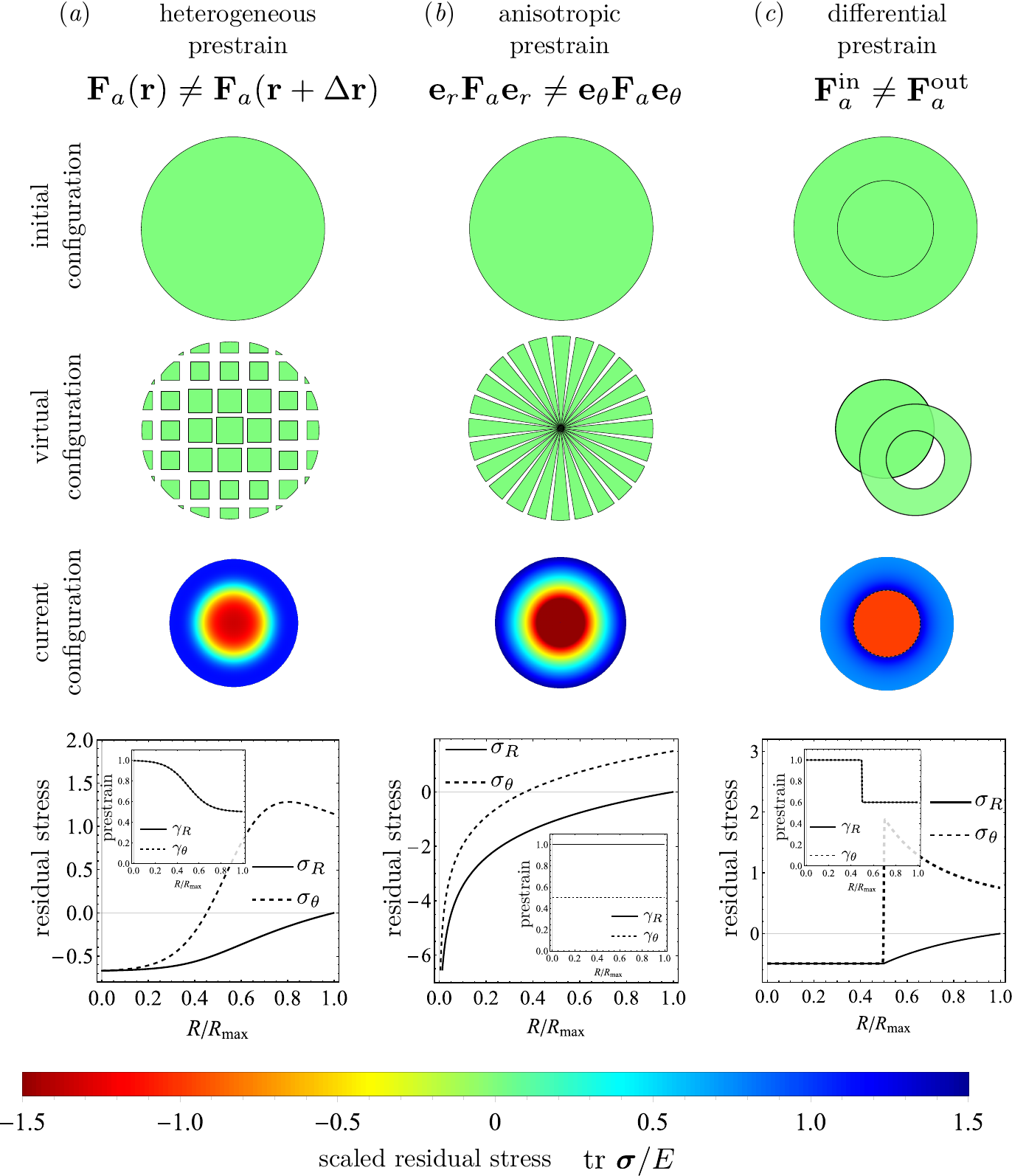}
    \caption{Creating residual stress from different patterns of {prestrain} in the anelastic framework for a disk. The disk is incompressible ($\det{\mathbf{F}_e}=1$) and of neo-Hookean material. The boundary conditions are no traction, $\boldsymbol{\sigma}\mathbf{e}_R=\mathbf{0}$. We denote the components of the {prestrain} in polar coordinates $\mathbf{F}_a=\text{diag}(\gamma_R, \gamma_\theta)$. Here we illustrate contraction, $\gamma<1$, however equivalent situations are found for growth, $\gamma>1$.  (\textit{a}) We consider spatially heterogeneous prestrain which is isotropic ($\gamma:=\gamma_R=\gamma_\theta$). The initial configuration shows the undeformed, uncontracted, stress-free disk. In the virtual configuration, most contraction occurs towards the periphery, leading to an incompatible body. The prestrain is explicitly shown in the inset, where $\gamma=1$ (no prestrain) at the center and $\gamma=0.5$ (contraction) at the boundary of the disk. The result in the current configuration is a residually stressed body, with tensile hoop stress ($\sigma_\theta>0$) at the disk boundary, and compressive stress ($\sigma_R<0$, $\sigma_\theta<0$) at the disk center. Due to the tensile hoop stress at the boundary, the disk would open if incised in the periphery.  (\textit{b}) We consider the case of anisotropic ($\gamma_R\neq \gamma_\theta$) but spatially homogeneous ($\mathrm{d}\gamma_R/\mathrm{d}R=\mathrm{d}\gamma_\theta/\mathrm{d}R=0$) prestrain. This corresponds to pizza slice shaped pieces being cut out in the virtual configuration. The resulting stress field is qualitatively the same as in the heterogeneous case.  (\textit{c}) We consider the case in which the outer part of the disk has a different prestrain than the inner part, prestrain being isotropic. The scenario is a discrete version of the heterogeneous case. The hoop stress is discontinuous. }
    \label{fig:morphoelastic}
\end{figure}

In order to describe more generally mechanical systems with prestress, and to allow the prestrain to be either due to growth or active contraction, we use a framework of anelasticity \citep{rodriguez1994stress,lubarda2004constitutive,epstein2012elements}. At its core is the  decomposition of the deformation gradient as $\mathbf{F}=\mathbf{F}_{e}\mathbf{F}_{a}$, where $\mathbf{F}$ is decomposed into an anelastic part $\mathbf{F}_{a}$ and an elastic part $\mathbf{F}_{e}$, see Fig.\ \ref{fig:decomposition}. 
The elastic deformation is taken to be hyperelastic and isotropic, for example neo-Hookean, captured by a strain-energy density $W=W\left(\mathbf{F}_{e}\right)$. 
The anelastic deformation is associated with an irreversible process that in some way modifies the microstructure: $\mathbf{F}_{a}$ can mean active contraction, like energy-consuming contraction due to myosin, in which mass is conserved or reduced (i.e. transferred from the solid phase to an extracellular reservoir), $\det\mathbf{F}_{a}\leq1$. 
Alternatively, it could mean growth, in which mass is added into the solid phase from an extracellular reservoir, $\det\mathbf{F}_{a}\geq1$. 
Cauchy stress is then defined as $\boldsymbol{\sigma}=\left(\det\mathbf{F}_{e}\right)^{-1}\left(\partial W/\partial\mathbf{F}_{e}\right)\mathbf{F}_{e}^{\mathsf{T}}$ and mechanical equilibrium is $\mathrm{div}\,\boldsymbol{\sigma}=0$ in the absence of body forces and respecting that anelasticity (active contraction or growth) occurs at time scales much larger than elastodynamics. In this view, prestress will exist in the system due to the incompatibility \citep{eckart1948thermodynamics,skalak1997kinematics,aharoni2016internal,truskinovsky2019nonlinear,lee2021geometry} of the anelastic strain  $\mathbf{F}_{a}$: for instance, if the system is composed of multiple components with different incompatible reference configurations ($\mathbf{F}_{a}$ in layer 1 is different from $\mathbf{F}_{a}$ in layer 2), or anisotropy ($\mathbf{F}_{a}$ in, e.g., radial direction does not match $\mathbf{F}_{a}$ in hoop direction), or some heterogeneity in the anelastic strain (say, $\mathbf{F}_{a}\left(\mathbf{x}\right)\neq\mathbf{F}_{a}\left(\mathbf{x}+\Delta\mathbf{x}\right)$), which could capture a spatial gradient in growth or active contraction. Fig.\ \ref{fig:morphoelastic} illustrates how  residual stress from different patterns of prestress can be created from the anelastic point of view.

A distinctly different possibility for building prestress is through changes in the microstructure due to rearrangements in the network. An example are T1 transitions, which are neighbour exchanges between cells in epithelial tissues, see section \ref{sec:tissue:topology}. {We offer to name this type of prestress \emph{topological prestress}.}
{We define it as prestress that is added to or removed from an interconnected network of mechanical elements by breaking and reconnecting network elements rather than prestressing individual elements. For instance, a passive T1 transition relaxes the stress in the system purely by exchanging which cell is connected to which, and not through modifying the reference configuration of any of the cells. }
%\AE{I slightly rephrased the sentence from the Discussion and added a sentence to connect back to the T1 transition example, OK?}
%\JF{good idea the added sentence at this stage !}
%\JE{Perfect!}
%Such transitions can be actively induced through myosin contraction \citep{nestor2022adhesion}, or purely passive rearrangements leading to a locally lower state of elastic energy, resembling viscoelastic dissipation \citep{farhadifar2007influence}.
%\JF{' Such transitions can be actively induced through myosin contraction...' Not sure this sentence is really useful at this stage. I would keep it for 5.4.}

The concept of a microstructural rearrangement leading to topological prestress is illustrated in Fig.~\ref{fig:decomposition}\textit{c}. A network of three springs is presented in the initial configuration, but  breaking and reconnecting bonds changes the network topology in the virtual configuration (topology refers to the connectivity of a network). Such changes of connectivity, which are meant to describe networks of  discrete elements such as cell adhesions or polymer crosslinks, are challenging to describe with a continuum field like $\mathbf{F}_a$ that is meant to describe the larger tissue scale. For example, discrete deformations like slip lines in crystal plasticity have been successfully described with the continuum framework $\mathbf{F}=\mathbf{F}_e \mathbf{F}_a$ where $\mathbf{F}_a$ describes the macroscopic  plastic deformation, $\mathbf{F}_e$ the elastic deformation, and $\mathbf{F}$ the total deformation \citep{reina2014kinematic,reina2016derivation}. But for biological tissue, which is generally amorphic and has no crystalline structure, the definition of macroscopic topological prestrain has not fully been achieved yet, although it is an active area of research \citep{chenchiah2014energy,murisic2015discrete,erlich2020role,kupferman2020continuum}.

\subsection{Dynamic description\label{sec:theory:dynamic}}

Both the active and passive elements of the system can exhibit a time-dependent behaviour.
The growth rate and actomyosin contractility both depend on some nonmechanical processes, such as protein synthesis, nutrient intake, or ATP hydrolysis.
All these are tightly regulated by biochemical pathways in physiological conditions, and in a large part the dynamics of the mechanical systems can be enslaved to biochemical clocks \citep{michaux2018excitable,Nishikawa+Grill.2017.1,Heer+Martin.2017.1,Blanchard+Gorfinkiel.2018.1}.
Active elements are also sensitive to the mechanical context. For instance, individual molecular motors are known to stall beyond some maximal load \citep{Liepelt-Lipowsky.2009.1}. 
In the context of growth, the timescale is rather large, since it is {the one of} the cycle of cell division which takes place over hours or days. Therefore, the passive elements are considered to be always at equilibrium.

On the other hand, regimes of {cell} motility often rely on the dynamics of the passive component with a constant prestress.
%\JF{I think this is wrong at the scale of the organism where muscle contraction needs varying prestress. }\JE{good point -- of \emph{cell} motility? Some morphogenetic movements such as germband extension are like that too, but I don't think it's necessary to mention}
This may be required to achieve movements which are faster than the rates at which prestress can be created. This is the case e.g.\ for carnivorous plants \citep{Forterre+Mahadevan.2005.1}  where an elastic instability is being used to suddenly release elastic energy that had been stored by the slow build-up of prestress. 

While the full complexity of interacting timescales between the active prestress one and the passive viscoelastic ones is encountered in some cases, we will focus here on models that describe cases where they are sufficiently well separated. In the case of growth, the system is often considered to be purely elastic and the models thus focus on the timescale of the evolution of prestrain. In the case of contractile networks, viscous dissipation in the microstructure is often important for the observed dynamics and sets their timescale, whereas the active stress is often assumed to be slowly varying.
% \JF{These paragraphs contain a lot of info that could be in the 'experimental' part of the review. Why not keeping only the first sentence : 'Both the active...' and the last paragraph here, and move the rest to experimental parts?}
% \JE{I think a few very specific examples can be good to illustrate the 2 extreme situations where dynamics is only in the active or only in the passive component, but I agree we should not list systems if it's not for a specific illustration.}

\subsubsection{\label{sec:dynamics-theory}Dynamics governed by the active component: the example of growth laws}

%When considering growth, it is typical to  focus on the characterisation of long-term behaviour of growth dynamics, neglecting the much faster inertial and viscous effects. This approach is common for the study of biological materials and is justified when observing the time scales of elastodynamics (inertial effects) and viscous effects vs.\ growth dynamics (growth-mechanical feedback). 
The timescale of elastodynamics in soft tissue  (i.e.\ wave propagation in soft elastic media)  is on the order of milliseconds, and a typical viscous time scale due to internal friction is on the order of seconds to minutes. The timescale of growth, on the other hand, varies from minutes (doubling time of \textit{Escherichia coli}) to years (slow growing tumours). Growth is thus a case where the separation of timescales is generally sufficient to consider that the passive components are instantaneously reaching their equilibrium configuration \citep{goriely2017mathematics}. 

%In morphogenesis, living tissues change shape and size very rapidly. \JF{'In morphogenesis, living tissues change shape and size very rapidly'. rapidly with respect to what other timescale ?}
%Morphogenetic events include spectacular shape changes such as self-inversion \citep{hohn2015dynamics}, looping as in the case of the heart tube \citep{ramasubramanian2008modeling}, branching such as in lungs, kidney and vascular networks \citep{erlich2019physical}. As these examples illustrate, morphogenesis involves complex dynamical interactions between growth, no-linear mechanics, shape and size. The determination of the appropriate form of evolution equations for growth and shape change has been the focus of much research (\citet{epstein2000thermomechanics,lubarda2002mechanics,dicarlo2002growth,ambrosi2007growth,ganghoffer2010mechanical}). 
%{With limited experimental data available, this research has employed thermodynamic arguments based on the entropy inequality or a dissipation principle to motivate appropriate forms \citep{epstein2000thermomechanics,lubarda2002mechanics,dicarlo2002growth,ambrosi2007growth,ganghoffer2010mechanical}.} The reasoning is typically to assume a single constituent theory in which the free energy depends on elastic deformation. 
%\JF{Just for you to know, I was not able to understand that bit : 'With limited experimental data... elastic deformation'}

The dynamics is then governed by a law that prescribes the evolution of the active component's prestrain as a function of the current configuration. Thermodynamic arguments based on the entropy inequality or a dissipation principle motivate appropriate forms of such a law \citep{epstein2000thermomechanics,lubarda2002mechanics,dicarlo2002growth,ambrosi2007growth,ganghoffer2010mechanical}.
By following a standard set of arguments and derivations, one arrives at a variant of the growth law
\begin{equation}
	\dot{\mathbf{F}}_{a}\mathbf{F}_{a}^{-1}=\mathbf{K}\left(\boldsymbol{\sigma}^{*}_{E}-\boldsymbol{\sigma}_{E}\right),
	\label{eq:growth-law-generic}
\end{equation}
where $\boldsymbol{\sigma}_{E}=\left(W\mathds{21}-\det\left(\mathbf{F}_{e}\right)\mathbf{F}_{e}^{\mathsf{T}}\boldsymbol{\sigma}\mathbf{F}_{e}^{-\mathsf{T}}\right)/\rho_{r}$ is the Eshelby stress, $W$ is the strain-energy density, and $\rho_r$ the density in the virtual configuration. The homeostatic Eshelby stress is  $\boldsymbol{\sigma}_{E}^{*}$ and $\mathbf{K}$ is a positive-definite coefficient matrix. The principle of homeostasis states that organisms have the ability to self-regulate some of their properties so as to optimise function in a physiological state, such as the ability of mammals to maintain a constant body temperature. In the context of mechanics, homeostasis can be understood as a living tissue's ability to grow and remodel to accommodate a preferred (homeostatic) stress, i.e. to reshape itself to reduce the difference between its actual stress and the a priori known or genetically encoded homeostatic stress \citep{dicarlo2002growth,erlich2019homeostatic}. 
Growth laws of the type \eqref{eq:growth-law-generic} which employ a homeostasis mechanism have been applied to morphogenesis problems, like sea urchin gastrulation \citep{taber2009towards}, the formation of ribs in Ammonite's seashells \citep{erlich2018mechanical}, and the intestinal crypt \citep{almet2021role}, as well as other applications such as wound healing \citep{Bowden2015wound,taber2009towards} and discrete networks such as plant cell networks \citep{erlich2020role}.

\subsubsection{Dynamics governed by the passive component: the example of contractile networks}

On the other hand, there are systems in which the limiting rate of strain is set by the passive component. 
Large amplitude motion, such as muscle contraction or intracellular retrograde flow, could not take place in a purely elastic medium. 
%\JF{At this stage, you could mention active gel theory \JE{You mean, lower down in this section? Yes, that's true, I was kept away from it by the derivation from transiently reticulated rubbers}}\JE{done}
Indeed, in the elastic models of \fig{fig:decomposition}\textit{a}, an obvious upper bound for the amplitude of strain is the magnitude of the prestrain. However, if the passive spring is replaced by a viscous dashpot (\fig{fig:decomposition}\textit{d}), then a constant rate of strain is achieved. This modelling is consistent for instance with the theory of sliding filaments for muscles, as delineated by \citep{Huxley.1957.1}, where the regime of maximum contraction speed is explained in terms of a balance between active stress and a sliding friction.
This compound element is thus governed by an equation of the form:
\begin{align}
    \sigma - \sigma_a = \eta\dot{\varepsilon}.
    \label{eq:viscoactive}
\end{align}
Here we find striking similarity with \eq{eq:growth-law-generic}, with the difference that the strain rate of the whole system appears directly. In both cases, as long as the stress differs from a given homeostatic stress ($\sigma^*$) or active stress ($\sigma_a$), an internal length, the virtual length $L$ or a length of `telescoping' of filaments, is being adjusted at a rate set by a parameter which has the dimensions of a viscosity.
We will come back to the molecular-scale understanding of the sliding friction in section \ref{sec:networks:dynamics}. In effect, this sliding friction can be likened to the fluidisation of any viscoelastic liquid beyond its relaxation time. In transiently reticulated networks, such as the actomyosin network, this relaxation time is related to the residence time of crosslinkers \citep{Larson.1999.1} and thus the maximum speed of actomyosin contraction can be related to these crosslinker dynamics \citep{Etienne+Asnacios.2015.1}, yielding a Maxwell model for the passive component:
\begin{equation}
\tau_{a}\overset{\triangledown}{\boldsymbol{\sigma}}+\boldsymbol{\sigma}-2\tau_{a}E\dot{\boldsymbol{\varepsilon}}=\boldsymbol{\sigma}_{a}\,,
\label{eq:cross-linked-network}
\end{equation}
where $\dot{\boldsymbol{\varepsilon}}$ is the rate of strain tensor, $E$ the elastic modulus of the crosslinked actin network, and $\tau_{a}$ a characteristic relaxation time.
The time derivative $\overset{\triangledown}{\boldsymbol{\sigma}}$ has to be an objective time derivative of the stress tensor $\boldsymbol{\sigma}$, starting from rubber elasticity theory one obtains an upper-convected Maxwell derivative \citep{Yamamoto.1956.1,Etienne+Asnacios.2015.1}. This can be related to the fact that the network structure is based on linear elements under stretch deformation \citep{Hinch-Harlen.2021.1}. Note that corotational derivatives are widely used in the field. 

This constitutive relation is consistent with the general framework of active gels  \citep{Juelicher+.2007.1}, which provides a thermodynamic framework relating the active prestress $\boldsymbol{\sigma}_a$ to the chemical potential difference associated with the myosin activity. At the molecular scale, the corresponding continuous injection of energy drives these system out-of-equilibrium and is at the origin of a spectacular violation of the fluctuation--dissipation relation \citep{Mizuno+MacKintosh.2007.1}, although effective equilibrium descriptions can be restored at higher scales \citep{o2022time}.

%\JF{Here the transition with density description is a bit abrupt, we could detail it a bit more. }\JE{Offer below}
{Contrary to growing systems, contractile ones generally deform while keeping a constant mass.}
In order to sustain a deformation rate that will in general not be volume-preserving, the density $\rho$ of the network needs to be actively regulated to a constant value $\rho_0$ by a reaction term, which in its simplest expression writes, in one dimension:
\begin{align}
    \tau_n \left( \dot{\rho} + \rho\partial_x v\right) = \rho_0-\rho,
    \label{eq:mass_balance}
\end{align}
where $v$ is the velocity in $x$ direction.
Here $\tau_n^{-1}$ provides another bound for the maximum rate of sustained flow. On the other hand, the reaction term in the mass balance can itself be a source of growth-related prestress. Assuming a density-dependent rheology of the material, such as $\sigma = -E {(\rho-\rho^*)}/{\rho^*}$ in its simplest form, with $\rho^*$ an equilibrium density. When density is close to this equilibrium, $\rho \simeq \rho^*$, we find again \eq{eq:viscoactive} with $\eta=\tau_n E$ and $\sigma_a = E(\rho^*-\rho_0)/\rho^*$ \citep{Putelat+Truskinovsky.2018.1}.

Finally, one situation which is encountered in several living systems is a stationary system size emerging from an enduring permanent internal flow regime. It can be easily seen e.g.\ that a closed system governed by \eq{eq:mass_balance}, but with different regulation densities $\rho_0^a < \rho_0^b$ in different geometrical regions $a$ and $b$, will establish a flow from the region $b$ to $a$. The system total size will adjust as the combination of the local growth and shrinkage, and there can be geometries and parameters for which this balances yields a constant total size. This sort of dynamic equilibrium will be exemplified below in actomyosin networks and cell spheroids.

\section{Measuring prestress and prestrain in living systems}
\label{sec:measure}

Obtaining detailed and reliable expositions of the prestresses which shape living matter has presented a great technical and conceptual challenge. A major difficulty is the large number of components which are in fact represented by the simplified components of Fig.\ \ref{fig:decomposition}. In every cell and tissue, numerous stress-bearing and -generating elements are mechanically coupled in complex (and often unknown) arrangements. Current techniques typically allow us to probe only small subsets of those components at once, so we are liable to overlook the many connected parts which remain invisible. On top of this, of course, the reference configuration of the components of Fig.\ \ref{fig:decomposition} cannot be deduced from the reference configuration of the ensemble alone, they can only be revealed through perturbations. But living matter has a great propensity to react and adapt to the perturbations we introduce in order to measure, so that there is always a real risk of measuring artefacts.

In order to face these challenges, in recent years, a multiplicity of experimental methods have been developed by biologists and physicists to study dynamic prestress. The most commonly used approaches are:

\begin{itemize}
\item live imaging of molecular actors generating prestress and subsequent strain of the biological material;
\item biological perturbation of prestress generators via drugs or molecular loss of function;
\item prestress release by cutting and ablation followed by measurement of resulting strains;
\item insertion of or embedding into stress-sensing deformable elements, functioning from the tissue scale down to the subcellular scale.
\end{itemize} 
These developments have been reviewed in detail elsewhere for both cell \citep{polacheck2016measuring} and tissue scale measurements \citep{Gomez.2020.1} and will be described when required in the sections below.

\section{Prestress in biopolymer networks}
\label{sec:networks}

\begin{figure}
    \centering
    \includegraphics[width=.75\textwidth]{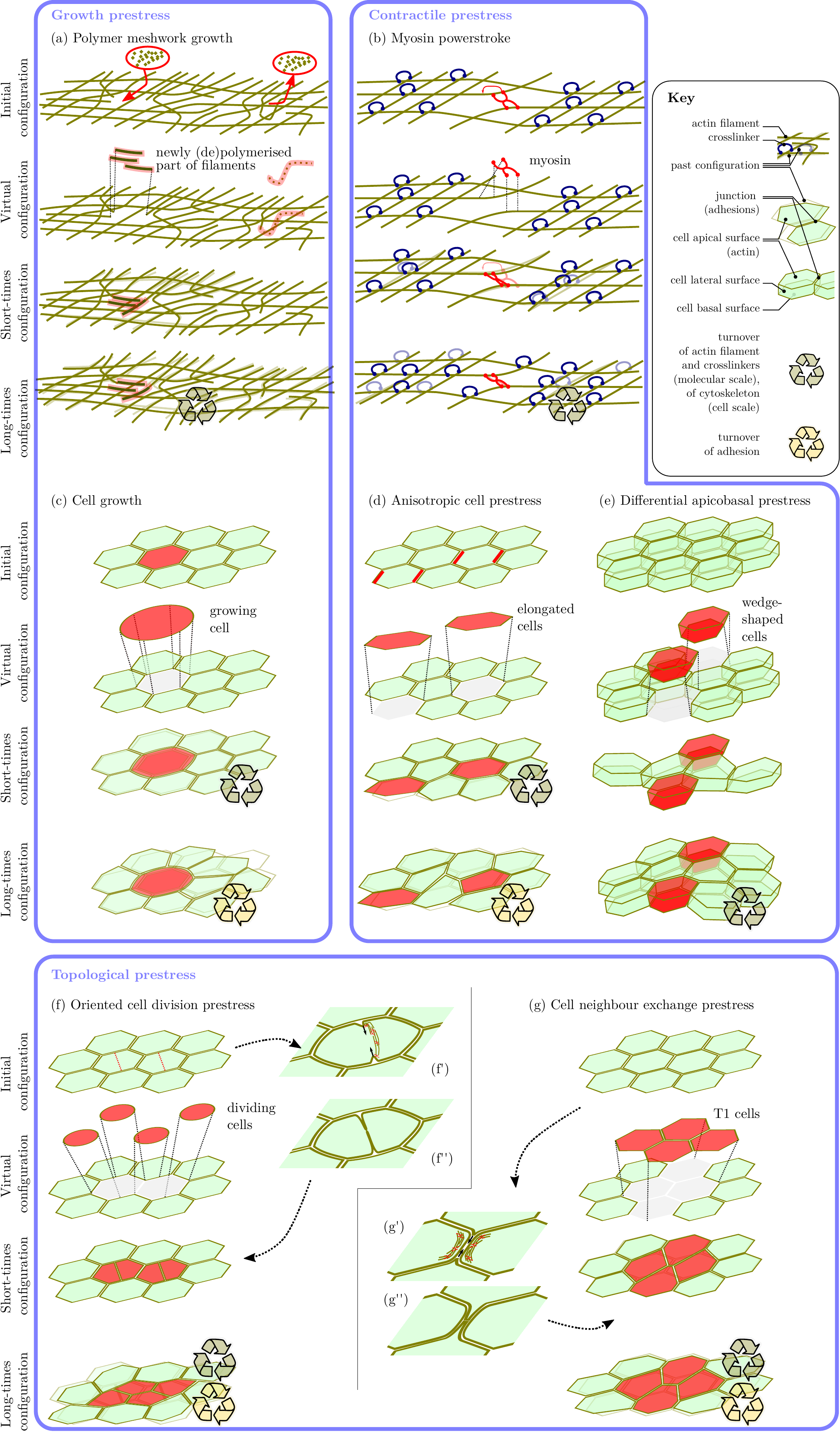}
    \caption{Examples of (\textit{a},\textit{b}) subcellular and (\textit{c}-\textit{g}) tissue scale deformations due to active prestress related (\textit{a}-\textit{e}) to microscale rest shape change or (\textit{f},\textit{g}) to a change of connectivity, which we refer to as topological prestress. (\textit{f}',\textit{f}'') and (\textit{g}',\textit{g}'') represent how the topological change in the tissue can be obtained by a subcellular active process, involving myosin prestress, however this level of detail can usefully be ignored when modelling tissue-scale deformations using the concept of topological prestress.
    }
    \label{fig:cell_tissue}
\end{figure}

\subsection{Prestress in the actin cytoskeleton}

{
The cytoskeleton is made of three categories of dynamic filaments (namely actin, microtubules and intermediate filaments) but only actin and microtubules are found to interact with molecular motors, which are major actors in prestress generation. We focus here on the actin cytoskeleton, although the framework defined above can be applied to the microtubule network, for instance to understand the force-balance within the mitotic spindle \citep{gay2012stochastic}. 

The actin cytoskeleton is made of polar semiflexible filaments composed of G-actin monomers, which turnover within filaments in time-scales ranging from seconds to minutes depending on cell types and actin structures considered \citep{amato1986probing, elkhatib2014fascin, Fritzsche+Charras.2013.1, saha2016determining, Clement+Lenne.2017.1}. Growth rate and geometry of the actin networks are regulated by a variety of actin-binding proteins which can either nucleate, elongate or sever actin filaments, or cap their ends \citep{pollard_actin_2016}.

Actin also binds to a specific type of crosslinker, the myosin molecular motors, which walk along actin filaments by using ATP hydrolysis (\fig{fig:cell_tissue}\textit{b}). Myosin~II filaments in particular can attach to two actin filaments thanks to head domains at their two ends. Actin and myosins form together a zoology of network structures which range from the crystalline structure found in muscle sarcomeres to the less ordered actomyosin cortex, a thin actin gel lying underneath the cell membrane. Intermediate in terms of organisation are linear bundles of actin enriched in myosin, such as the so-called stress fibres \citep{burnette_contractile_2014}} and junctional cortex in epithelia \citep{Bertet+Lecuit.2004.1}. The signalling pathways controlling the formation of these respective organisations is beyond the scope of this review and has been reviewed elsewhere \citep{tojkander_actin_2012}.

\subsubsection{Mechanical balance of prestressed actomyosin}
\paragraph{Contractile prestress.} 

{
Importantly, myosin generates contractile prestress within actin networks. The contractile nature of stress fibres was demonstrated by measuring the rate and amplitude of the viscoelastic recoil of individual stress fibres after laser ablation. These two quantities were shown to be reduced when Myosin II activity was inhibited by drug treatment \citep{kumar_viscoelastic_2006}. 
Furthermore, dose-dependent treatments of Blebbistatin (an inhibitor of Myosin II contractile activity) on single cells isolated in a parallel plates traction force apparatus revealed that cell-scale traction exerted by the actomyosin cortex is proportional to Myosin II ATPase activity, indicating that myosin is the main generator of contractile prestress in the cell cortex \citep{mitrossilis_single-cell_2009}. It can thus be established that the actomyosin meshwork is exerting a contractile active stress, proportional to the chemical potential of myosin \citep{Juelicher+.2007.1}, which can be understood as a prestress. 

How is it that contraction at the cell scale dominates over expansion despite the disordered nature of these networks? Various hypotheses have been proposed. The most documented one posits that because actin filaments buckle under compression,  myosin activity will result only in a tensile contribution \citep{lenz2014geometrical, belmonte2017theory, koenderink2018architecture}.

This contractile prestress can be balanced by the mechanical resistance of three types of other mechanical elements: the cell environment, the other cytoskeletal networks and the fluid component of the cytoplasm (or cytosol).  

\begin{figure}[t]
    \centering
    \includegraphics[width=\linewidth]{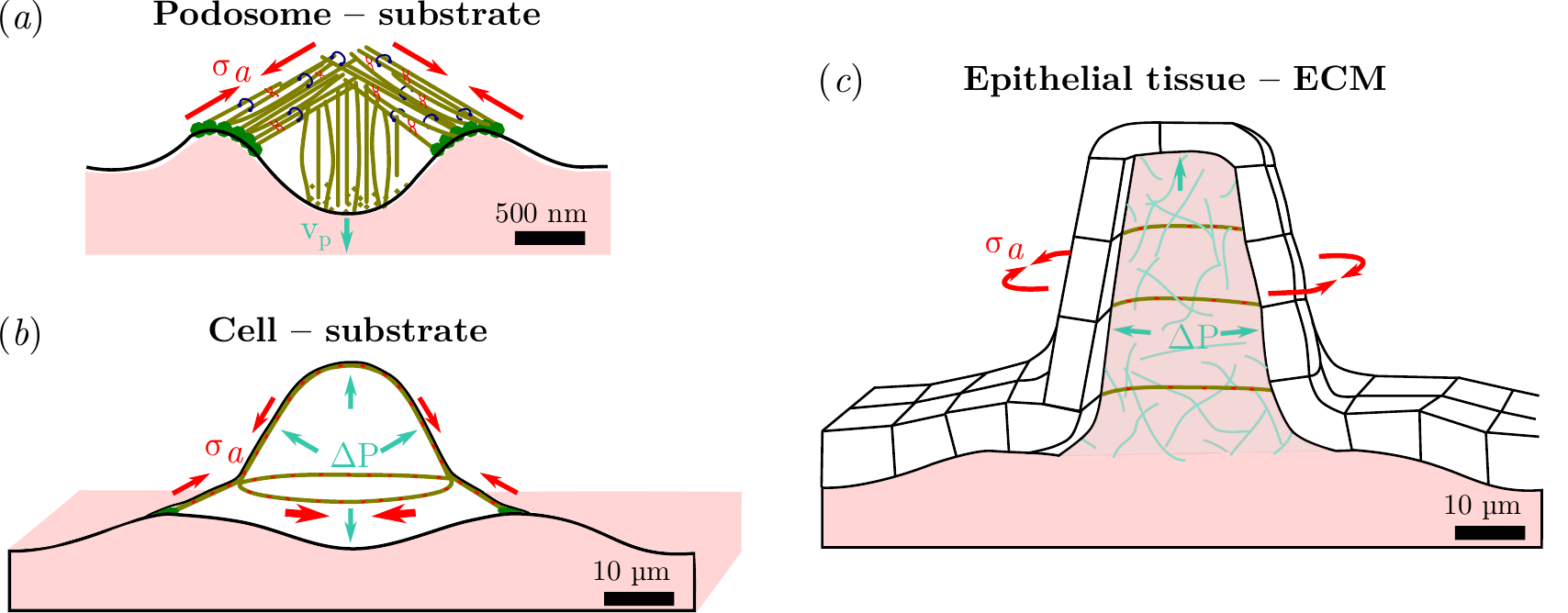}
    \caption{Growth and/or contractile prestress governs the shape of subcellular, cellular and tissue-scale structures through mechanical connection between active and passive elements. (\textit{a}) In podosomes, protruding forces applied by the actin core (growing at a rate $v_p$) onto the substrate are balanced by a contractile actomyosin network of prestress ($\sigma_a$), organised as a dome and attached to the substrate at the periphery via adhesion proteins \citep{labernadie_protrusion_2014}. (\textit{b}) In adherent cells, actomyosin prestress ($\sigma_a$) is balanced by cytosol pressure ($\Delta$P) and substrate deformation. Cell shape is further refined by anisotropic and heterogeneous actomyosin network contraction. Here, orthoradial stress fibres are connected to radial stress fibres, which are attached to the substrate via adhesion proteins  at the cell periphery \citep{burnette_contractile_2014}. (\textit{c}) In the zebrafish semicircular canal, pressure ($\Delta$P) is generated within the ECM via synthesis of hyaluronan, pumping in interstitial fluid. This deforms the overlying epithelium which is further shaped by an anisotropic prestress ($\sigma_a$) generated by actin- and cadherin-rich protrusions \citep{munjal_extracellular_2021}.
    }
    \label{fig:balance}
\end{figure}

The tension--compression balance between tensile actomyosin and the external environment to which cells adhere was revealed thanks to the development of deformable substrates (elastomers, hydrogels, or micro-posts arrays). It was shown that adherent cells seemingly “at rest” apply tensile stresses radially directed towards the cell center \citep{Harris+.1981.1, pelham_cell_1997, Dembo+.1999.1, tan_cells_2003}. This stress is transmitted to the substrate at the sites of focal adhesions, mechano-sensitive protein aggregates connecting cells to the ECM \citep{tan_cells_2003, burridge2016focal}. Larger deformations are observed in the direction parallel to stress fibres \citep{mandal_cell_2014}. In line with this, the recoil of the cell substrate away from the site of incision after ablation of stress fibres demonstrated the mechanical connection between those dense actomyosin fibres and the ECM \citep{kumar_viscoelastic_2006}.  

Actomyosin tensile stress could also be balanced by compression of other cytoskeletal components. Among them, the microtubule network was suggested as a major mechanical actor bearing actomyosin tensile prestress, forming a biological illustration of the tensegrity model \citep{ingber_cellular_1993, ingber2014tensegrity}. Depolymerising microtubules using a drug increases traction forces on the substrate, suggesting that the compression borne by microtubules is transferred to the substrate \citep{Stamenovic+Wang.2002.1}. Actomyosin pretension can however also be biochemically affected by microtubule depolymerisation \citep{rape_microtubule_2011}.

Finally, the tension developed within the actomyosin cortex can be balanced by the cytosol, an incompressible fluid which permeates the whole cytoplasm of the cell \citep{salbreux_actin_2012}. The cytosol is restricted from escaping by the cell plasma membrane to which the actomyosin cortex is adhered via the ERM family of proteins \citep{mangeat1999erm}. This mechanical balance is spectacularly broken when this cortex or its adhesion with the plasma membrane is locally ruptured: in such occasions, a high-curvature spherical protrusion, called a bleb, forms at the wounded site and inflates with cytosol from the cell body \citep{charras_life_2008, tinevez_role_2009}. Reducing myosin activity, e.g. with the fittingly-named drug Blebbistatin, decreases the volume and rate of expansion of blebs, evidencing that the pressure driving the flow is due to the prestressed actomyosin cortex. 

The actomyosin cortex and stress fibres constitute a continuous network and both contribute to cell-scale prestress \citep{labouesse_cell_2015, vignaud_stress_2021}, although their mechanical properties and regulatory pathways are different \citep{labouesse_cell_2015}. The spatial arrangement of these prestressed structures in equilibrium with the passive elements described above give rise, for instance, to the typical three-dimensional shape of crawling cells. In these cells, a flat compartment is formed at the cell front and an inversion of curvature of the cell profile is observed at the junction between this compartment and the dorsal cortex (see \fig{fig:balance}\emph{b}). This specific shape results from the presence of orthoradial fibres whose pretension opposes cytosol pressure. These fibres are then connected to radial stress fibres which are attached to the substrate near the cell edge via focal adhesions. Depolymerising the orthoradial fibres via biochemical treatment restores a constant curvature along the cell profile \citep{burnette_contractile_2014}.

\paragraph{Growth prestress.}

The actin cytoskeleton can also exert growth prestress. This was shown in vitro where actin networks nucleated by Arp2/3 under an AFM cantilever \citep{bieling_force_2016} or at the surface of spaced magnetic cylinders \citep{bauer2017new} were shown to generate compression within the network. This effect can be conceptualised as in \eq{eq:mass_balance} (also \fig{fig:cell_tissue}\textit{a}), although the growth is often localised at the boundary. In vitro and in vivo, the mechanical activity of network growth is mechanosensitive as shown by a force--velocity relationship in line with an increase of the network density in response to load \citep{bieling_force_2016,mueller_load_2017}.

The generation of such growth prestress is involved in various biological functions. First, it is at the origin of the motility of the Listeria pathogen \citep{theriot_rate_1992}. Here, like on spherical beads immersed in actin in vitro \citep{marcy_forces_2004}, growth occurs first homogeneously at the surface, which generates residual stress at the periphery of the network. The network ultimately fails via an elastic instability \citep{john_nonlinear_2008}. This symmetry breaking thus forms an anisotropic gel at the surface of the object resulting in a directional movement.

In mammalian cells, a length increase of actin filaments can increase cortical thickness and more importantly counteract tension generated by myosin  in the network \citep{chugh_actin_2017}. The best known example of a cell function in which actin network growth is involved is cell protrusivity, where actin can form a variety of structures pushing the cell membrane forward. This topic is a field of research in itself and we refer the reader to reviews treating it specifically \citep{blanchoin_actin_2014}. 
}

\paragraph{Example: podosomes are shaped by growth and contractile prestress.}

An example of a prestressed structure combining growth and tensile prestress at the subcellular scale is the podosome, whose mechanics have recently been clarified (see \fig{fig:balance}\textit{a}). Podosomes are micron-scale structures present in various cell types (macrophages, cancer cells, endothelial cells) known to probe cell substrate mechanical properties and to be the site of ECM digestion. They are formed of a dense core of actin filaments, oriented normally to the substrate, and bound to a corona of radially-oriented filaments which are tangential to the substrate and adhere to it thanks to focal adhesion proteins \citep{luxenburg_involvement_2012, van_den_dries_probing_2019}. Protruding forces generated by actin polymerisation within the podosome's core were proposed to be a major contributor to the compression of the substrate \citep{labernadie_protrusion_2014}. While this growth-related prestress remains a possible player, recent findings show that the mechanical balance is dominated by the peripheral actomyosin filaments, which are exerting tensile forces between the tip of the core and the substrate, and hence press the core into the substrate \citep{jasnin_elasticity_2021}. This system shows that despite their complexity, {the mechanical equilibrium of biological structures can be unravelled by combining force measurements, imaging and careful biological perturbation experiments}. 

\subsubsection{Dynamics}
\label{sec:networks:dynamics}
%\JF{I am in favor of putting this section 4.3 at the end of 4.1, just because it only discusses dynamics of actin csk. Also I think we could retitle 4.1 'Prestress in acto-myosin cytoskeleton', because we only touch this network of the csk. I could also add a sentence in 4.2, saying that we don't know much about dynamics. Those little reorganisations may help reviewer 2.}
%\JE{Fine for me. Do you want to have subsectioning into 4.1.1 and 4.1.2 Dynamics? For 4.1.1 then maybe rather than "statics" it could be "mechanical balance of prestress actomyosin"? }

As mentioned in section \ref{sec:theory:dynamic}, the dynamics of the system can come either from an evolution of the prestress or from a passive relaxation of the material. With some notable exceptions \citep{Fierling+Etienne+Rauzi.2022.1}, the evolution of the prestress is generally slower than the relaxation of the material.
%
%\JE[
%Cells live in dynamic mechanical environment. Because of visco-elastic and non-linear mechanical properties of the ECM, stiffness felt by cells can be modified at rather short time-scales. Equally, mechanical stresses exerted by surrounding tissues (growth, contraction,...), induce cell strains at seconds to minutes time-scales, time-scales which are often close to those of turnover of actin and myosin filaments and focal adhesions. ]{move to the cellularised section?}
%
%In the above examples, the prestressed network is seen as having reached an equilibrium. 
%The prestress is then revealed by the fact that its mechanical environment is being strained by the network, while a corresponding strain occurs within the network itself.
%
%
%\JE[
%At very short times (up to a fraction of a second after a step strain), fluxes of water that permeate  the network have been shown to dominate the force response {\citep{Moeendarbary+Charras.2013.1}}. 
%At intermediate timescales however,]{Move to cell-scale?}
%
Indeed, biopolymer networks are in general very dynamic as they are being constantly remodelled by processes of (de)reticulation and (de)polymerisation. While the time scales of ECM proteolysis and synthesis remain largely unknown, the rapid turnover time of actin filaments %in many cell types ranges from seconds to barely more than ten seconds \citep{Fritzsche+Charras.2013.1, saha2016determining, Clement+Lenne.2017.1}. %\JE{Y a t il d'autres refs depuis...? on dirait pas...}
%This rapid turnover of the network itself 
entails its long times liquid-like behaviour \citep{Kruse+.2006.1}. Its effective viscosity scales like the product of the network short-times elastic modulus $E$ and its characteristic turnover time $\tau_a$, see \eq{eq:cross-linked-network}.
In the muscle, the only crosslinkers between actin and thick filaments are the myosin heads themselves. In order to function as sliding filaments, myosin needs to cycle from attached to detached from the actin in order to perform repeated steps \citep{Caruel-Truskinovsky.2018.1}, and the dissipation associated with maximal contraction velocity is due to an internal friction associated with the rate of detachment \citep{Huxley.1957.1}.
This is also the case in most motile and tissue cells, where the turnover time of myosin %II has been measured \JE[in the cortex of epithelial cells]{only? what about fibroblasts e.g.?}, and 
was found to be close to that of actin monomers in actin filaments and of $\alpha$-actinin (one of the major actin crosslinkers) \citep{Khalilgharibi+Charras.2019.1}.
This explains persistent flows such as the retrograde flow observed in migrating and spreading cells, whose maximal rate of strain is thus set by the ratio of the prestress $\sigma_{{a}}$ to the viscosity, as in \eq{eq:viscoactive} \citep{Kruse+.2006.1,Etienne+Asnacios.2015.1}.
This is also what sets the rate at which actomyosin-rich subcellular components or tissues straighten after having been buckled by external compressive forces \citep{tofangchi2016mechanism, wyatt_actomyosin_2020}.

It is also of interest to consider the case in which macroscopic-scale deformation of the network is prevented by the boundary conditions. In those conditions, forces at the boundaries will need to balance the internal tension of the network. {The} energy input {from myosin motors} will then be dissipated internally by a microscopic scale creep, corresponding to the elastic energy loss incurred when elements in the network detach while under tension \citep{Huxley.1957.1,Etienne+Asnacios.2015.1}. In this case, the stress measured at the boundaries will be equal to the prestress.
Note that in energetic terms, there is work being continuously performed internally by the myosin in both the cases of zero external load or zero contraction of the system, the corresponding energy being respectively dissipated by internal friction and by internal creep \citep{Etienne+Asnacios.2015.1}.
%\JE{Which brings me to note that it shouldn't be the case in theories which assume that myosin stall under high stress, then what you measure is the stalling stress, not the prestress. Is that a contradiction in active gel theory papers?? I have to check this.}
These observations demonstrate a peculiarity of actomyosin networks where prestress and dissipation possess common molecular origins. 
At the microscopic scale, active processes increase the intensity of fluctuations in the medium. This has been evidenced both for the out-of-equilibrium (de)polymerisation of cytoskeletal filaments \citep{robert2010vivo} and for myosin activity  \citep{Mizuno+MacKintosh.2007.1}, by comparing the fluctuations to those that would be expected from thermal forces only.  
%Microrheology, which is based on tracking the displacement of microscopic tracers embedded in the biopolymer network, has allowed to describe the rheology of such systems \citep{Mason-Weitz.1995.1,Tseng+Wirtz.2002.1, gambini2012micro}. In active microrheology, tracers are being submitted to an external force field, allowing to measure the storage and loss modulus associated with the drag force. Passive microrheology, on the other hand, is performed without external force and is thus sensitive to fluctuations within the system. Notably, comparing passive and active microrheology allows to measure the respective contributions of actin and microtubules networks to fluctuations \citep{robert2010vivo}, and especially the contribution of myosin \citep{Mizuno+MacKintosh.2007.1}.

The actomyosin cortex and myosin contractility also participate in dynamic equilibration of prestress when substrate stiffness is varied. Cell traction assays in parallel plate geometry performed in various contractile cell types revealed that the loading rate during traction increased with substrate stiffness, with the same trend as the maximum traction force \citep{mitrossilis_single-cell_2009, lam_mechanics_2011}.
{While the models most often evoked to explain the variations of cell tractions in response to changes in substrate stiffness at long time-scale rely on mechano-transduction and subsequent changes in biochemical activity,
it was found that the loading rate adapted to real-time changes of cantilever stiffness in a sub-second time-scale \citep{mitrossilis_real-time_2010, crow2012contractile}, making purely mechanical explanations appealing \citep{fouchard_acto-myosin_2011}. Independently of the adaptation of myosin prestress to mechanical cues,} a simple model such as \eq{eq:cross-linked-network} already predicts a biphasic behaviour with a maximal traction for very large stiffness of the exterior and a traction proportional to that stiffness when it is below a threshold \citep{Etienne+Asnacios.2015.1}. When the actomyosin cortex is continuously adhering to a deformable substrate, the interplay of this system with the elastic length scale of the substrate yields complex interactions which translate into a biphasic behaviour of the cell crawling velocity as a function of the elastic modulus of the substrate \citep{Chelly+Recho.2022.1}.
    
The combination of growth at the leading edge of cells and active contraction to the back is the hallmark of cell crawling on a flat substrate \citep{Mitchison-Cramer.1996.1,Kruse+.2006.1,Recho-Truskinovsky.2013.1}, these two effects giving rise to a dynamic equilibrium setting the size of the system \citep{Etienne+Asnacios.2015.1,Ambrosi-Zanzottera.2016.1}.

\subsection{Prestress in fibrous tissues}

The ECM is a biopolymer network which is made of assemblies of filamentous proteins (the most abundant being collagen I), proteoglycans and glycosaminoglycans (GAGs). It is the main component of fibrous tissues---also called soft connective tissues---and of the basement membrane on which bidimensional epithelial tissues lie. In most tissues, matrix content is dominated by collagen (especially type I collagen), which has the ability to self-organise into fibrils and fibres through physical bonds \citep{giraud2008liquid}. These fibres can measure up to dozens of microns and often form a crosslinked gel in vivo. 
In soft fibrous tissues, the ECM is synthesised by cells of the fibroblast family which are embedded {and mechanically connected to the matrix}. Here, both cells and ECM lie in an interstitial fluid. Mechanical interactions between these three phases generate a large variety of architectures \citep{wershof_matrix_2019} and mechanical properties \citep{levental_soft_2007}, which vary from organ to organ and according to patho-physiological conditions.

The difficulty of performing live imaging in these three-dimensional systems, combined with their physical and biological complexity, has so far impaired the understanding of their active mechanical properties. Nevertheless, recent data show how prestress can be generated in fibrous tissues and affects tissue development and pathology. {Despite being largely composed of inert protein}, the ECM is not mechanically passive. 

\paragraph{Contractile prestress.}

First, contraction of the ECM can result from variation in water content within the interstitial fluid. In particular, collagen molecules, known to resist tensile stresses in fibrous tissues, change conformation with decreasing water content of the surrounding medium. This effect induces high contraction of the network, which could be important for the function of the load-bearing tendon \citep{masic2015osmotic, Bertinetti+Fratzl.2015.1}. 

But the ECM is also made contractile through the traction forces that fibroblasts exert within it. Since fibroblasts are polarised mechanically and bound to the ECM through focal adhesions, they apply force dipoles on the collagenous network, acting like active crosslinkers. Isometric contraction of fibroblast assemblies self-organising in collagenous matrix could be measured in between parallel cantilevers and was shown to be dependent on myosin activity \citep{delvoye_measurement_1991, legant_microfabricated_2009}. In these systems, like in the actomyosin cortex, the nonlinear mechanical properties of collagenous networks resulting from the semiflexible nature of the fibres are thought to play a major role in the propagation and amplification of contractile stress at the tissue scale \citep{ronceray_fiber_2016, han_cell_2018}. 
%\JF{new sentence for everyone to check below}
{Because cell traction forces are sensitive to the stiffness of the extracellular environment \citep{discher2005tissue,mitrossilis_single-cell_2009}, stiffening of the ECM fibres generated by cell traction forces  triggers a positive feedback loop amplifying stiffening in fibrotic reactions \citep{calvo2013mechanotransduction} or during directed cell migration \citep{van2018mechanoreciprocity}.}

This mechanical activity of fibroblasts contracting the ECM influences tissue function during developement and in adulthood. First, in confirmation of a long-standing hypothesis \citep{harris_generation_1984}, recent work shows that the patterning of multicellular aggregates in the chick dermis initiating feather follicles depends on fibroblast cell contractility, preceding differentiation of the epidermis \citep{shyer_emergent_2017}. This indicates that fibroblast contractile activity can indeed act as an organising factor of fibrous tissues during development. Second, during lymph node physiological function, immune cells signal their arrival to the fibroblast reticular cells and tune their contractility in order to relax the tissue. This is thought to help maintaining lymph node integrity while the lymph node expands \citep{acton2014dendritic}. Finally, cancer-associated fibroblasts, which shape the fibrous tissue of the tumour microenvironment and are more contractile than normal fibroblasts \citep{sahai_framework_2020}, could also apply active stresses at the global scale of the tumour. Along this line, it was shown that their collective contraction concomitant with an orthoradial assembly around tumour aggregates compress tumour compartments in vivo, as well as cylindrical micropillars in vitro \citep{barbazan2021cancer}.

\paragraph{Growth prestress.}

The interstitial fluid can generate growth prestress within the ECM. Indeed, the osmotic pressure within the ECM can be tuned in particular by the presence of GAGs and proteoglycans, thanks to their long, negatively charged chains. This property is used during developmental morphogenesis where localised synthesis of hyaluronan (a common GAG) contributes to the bulging of an epithelial monolayer laying on top of the swelling ECM \citep{munjal_extracellular_2021} (see \fig{fig:balance}\textit{c}). Such water influx is then balanced by the matrix fibres under tensile load \citep{ehret_inverse_2017}. Thus, in contrast to the poro-elastic behaviour observed in the cell cortex \citep{moeendarbary_cytoplasm_2013}, tension relaxation correlates with expelled interstitial fluid \citep{ehret_inverse_2017}. In the context of cancer, the deregulated fibrous tissue defining the mechanical properties of the tumour stroma (i.e the tumour micro-environment) is also affected by an increase in hyaluronan synthesis and subsequent elevated interstitial fluid pressure. In pancreatic adeno-carcinoma, this effect generates collapse of blood vessels, which could participate in organ loss of function and impairs delivery of therapeutic agents \citep{provenzano_enzymatic_2012}.

%\JE{Cell migration?}

%\JF{Is the differential pre-stress fundamentally different from heterogeneous prestress?} \JE{Discussed on May 12}

%\JE{The stuff below is probably for cell/tissue scale}
%One should note that, although actomyosin is restricted to the cell cytoplasm, it can form supracellular structures by means of adhesion molecules that connect mechanically one subcellular network to the one of a neighbouring cell. In this way, one-dimensional supra cellular cables \citep{Monier+Sanson.2010.1} or two-dimensional meshworks \citep{Chanet+Martin.2017.1} have been shown to exert stress at a tissue scale. \JE{Adhesion turnover: is it said elsewhere?} \JE{Muscles and syncytial tissue...? Too biological?}
%Finally, there is a nascent recognition that prestress is also a property of other biopolymer networks, although in a much less versatile and pervasive manner. Indeed, variations in osmolarity can modify the rest length of extra-cellular matrix filaments. It has been shown that such variations can occur in physiological conditions \citep{Bertinetti+Fratzl.2015.1}. \JE{Also mention the swelling of Megason et al \ref{fig:balance}C ?}

\section{Prestress in cellularised tissue}
At the tissue scale, a new structural unit becomes fundamental: the cell. From a purely mechanical view point, one crucial aspect is that cells tessellate the space occupied by a tissue into units of regulated volume, which have interactions via adhesion molecules. Being functional units, cells may have individual biochemical activity which can translate into mechanical activity, in turn giving rise to strains that can alter the tissue geometry. This biochemical activity can be patterned at the scale of single cells. For example, in epithelial tissues, which typically form thin sheets of a single cell layer, \emph{apical} (cell `top' surface), \emph{basal} (`bottom') and \emph{lateral} (where cells are in contact) surfaces are, to some extent, independent biological and hence mechanical units. 

Since tight junctions between cells allow tissues to form impermeable layers, cells which actively and directionaly pump ions can in this way impose a pressure difference between apical and basal surfaces, which can generate and maintain a liquid-filled cavity called \emph{lumen}. Lumens, like neighbouring tissues and the ECM network synthesised by cells, in turn impact tissue mechanics by  boundary conditions that vary in space and time. 

Cells also define a tissue topology via the neighbour relations created by cell--cell adhesion. This topology can evolve over the course of time as cell--cell junctions are assembled and disassembled, which can in turn give rise to further tissue-scale strain that can be modelled as the result of a topological prestress, as opposed to contractile or growth prestress. 

With these differences in microstructure come processes with new timescales \citep{khalilgharibi_dynamic_2016}. While a single actin filament may turn-over in seconds, a cell–cell junction requires at least minutes to disassemble and reassemble when a cell changes neighbour \citep{tlili_migrating_2020, Clement+Lenne.2017.1}. The creation of a new junction during cell division similarly occurs over minutes. However, the full timescale of the cell cycle, which we may consider to encompass the prestress changes associated with cell growth and cytokinesis, varies greatly between animal cell types, from minutes to years. These cell-level processes may be synchronised throughout a tissue, generating tissue-level deformations at a similar timescale, or they may be asynchronous and so add up gradually over longer timescales.

\subsection{Contractile prestress}
\subsubsection{Supracellular scale prestressed networks in interaction with their environment} % should be shortened? "prestressed networks in interaction with their environment"?

Although actomyosin is restricted to the cell cytoplasm, it can form supracellular structures by means of adhesion molecules that mechanically connect one subcellular network to that of a neighbouring cell, forming multicellular structures under tension
\citep{Fernandez+Eaton-Zallen.2009.1,Chanet+Martin.2017.1}. The emergence of such a mechanical continuum is illustrated during the reformation of a dissociated epithelial monolayer in vitro by an increase of apparent tissue stiffness, which is coincident with the development of cell--cell junctions and dependent on actomyosin activity \citep{harris_formation_2014}. The tissue-level prestress generated by the continuous actomyosin network can be measured directly in in vitro epithelial monolayers devoid of ECM and suspended between the arms of a force cantilever \citep{wyatt_actomyosin_2020}. Here, a ramp of compressive strain imposed at the tissue boundary results in a linear decrease of tissue stress. As would be the case for a prestressed thin elastic plate, the tissue then buckles when it reaches a compressed state. Strikingly, the dynamics of stress recovery upon a rapidly applied compressive strain match those of an isolated actomyosin network, indicating that, in this case, the cellularised structure has little impact on the overall mechanics \citep{wyatt_actomyosin_2020}. Such a response, consistent with a continuous model of an epithelium, has also been observed in vivo, in several Drosophila epithelia, where anisotropies of prestress were revealed by the recoil of circular regions of tissue after laser-cutting \citep{bonnet_mechanical_2012}. 

Just as epithelia can change length and generate prestress through their actomyosin networks, they can also be connected to an active element and play a passive role. Their challenge is then to bear the stress generated in order to maintain epithelial integrity \citep{bonfanti_fracture_2022}. This is illustrated by the blisters that epithelia form under active ion pumping directed towards their basal side. In analogy with the pressure in the cell cytoplasm, the pressure within the blister is balanced by the tension in the actomyosin network. Under increased stress, cell deformations can reach hundreds of percents. When the actin pool is exhausted so that the filament network can no longer cover the cell surface area, the keratin network (an intermediate filament network connected through an other type of cell--cell junctions called desmosomes) takes over to resist mechanical stress \citep{latorre2018active}. 

While the passive response of an epithelium leads to a dome shape in the system above, an active participation of the epithelium is required to generate a tubular shape, as is the case in the zebrafish inner ear \citep{munjal_extracellular_2021}. Here, anisotropic multicellular cables which are both contractile and adhesive form in the direction orthoradial to the cylinder axis, and so by breaking the symmetry of prestress allows an anisotropic shape to arise. Notably in this example, the element driving growth is an increase of osmotic pressure in the ECM actively regulated by cell synthesis of hyaluronan.

\subsubsection{Spatially patterned prestress and tissue bending}

The generation of anisotropic prestress, as in the zebrafish inner ear, is one example of a common strategy of spatially organising prestress in order to control tissue shape. The most studied example is perhaps epithelial tissues where prestress, tangential to the plane of the tissue, varies along its transverse direction and drives bending via differential prestress. This aspect is particularly important during animal development and detailed reviews of this process have been made by others \citep{pearl_cellular_2017, tozluoglu_folding_2020}. For the sake of this review, note that a variety of fine-tuned prestress regulation have been documented, including increase of prestress in the apical domain \citep{martin2009pulsed}, increase or decrease of prestress in the basal domain \citep{sidhaye_concerted_2017, krueger_downregulation_2018}, and increase in lateral prestress \citep{brodland_video_2010, gracia_mechanical_2019}. 

This differential prestress can be revealed by laser ablation in cultured suspended epithelia by measuring the spontaneous curvature generated orthogonal to tissue plane at a newly created free edge. This too allows for measurement of the out-of-plane forces involved by unfurling the curled tissue with a force cantilever \citep{fouchard_curling_2020}.   

Three-dimensional geometries other than linear folds can be produced through the same mechanism, via patterning of differential prestress throughout the plane of an epithelium. For instance in 2D-cultured gut organoids, three cell types are organised within the tissue plane into concentric circular islands. Traction force microscopy and laser ablations revealed that these regions display different mechanical behaviours, with apical--basal myosin polarisation in the central region leading to the doming of crypts (cup-shaped structures of the digestive tracts) \citep{perez2021mechanical}. In 3D gut organoids, in which cells embedded in ECM form cysts, experiments revealed that crypt morphogenesis is also driven by membrane transporters which cause liquid transfer from the crypt cavity to the tissue \citep{Yang.2021.1}. Interestingly, the apical--basal polarisation of the crypt region was found to be large enough in comparison to that outside of the crypt region for the overall morphology to be robust to organoid volume changes \citep{Yang.2021.1}.

Recent advances in optogenetics (optical activation of engineered biomolecules) have also allowed for the experimental control of this patterned differential prestress. For example, at the stage preceding gastrulation in Drosophila, localised apical activation of RhoA, an activator of myosin contractility, was shown to be sufficient to initiate folding in a variety of directions and locations where invagination does not normally occur \citep{izquierdo_guided_2018}. 
In vitro measurements showed that the spontaneous curvature generated by active differential prestress is so high (on the order of the inverse of tissue thickness) that a competition between in-plane elastic energy and bending energy takes place. In this way, tissue folding is continuously modulated by external tension and reciprocally \citep{recho_tug--war_2020}.  
Nevertheless, differential prestress is not the only way to achieve folding. Recent data from the ventral furrow formation in Drosophila embryo showed the importance of the global ellipsoidal geometry of the embryo for an elongated ventral patch of myosin to achieve a fold along its long axis \citep{Chanet+Martin.2017.1}. Indeed, heterogeneneous prestress at the surface of a thin shell respecting this geometry leads to surface buckling initiating folding with the correct pattern of strain and dynamics \citep{Fierling+Etienne+Rauzi.2022.1}.

\subsection{\label{sec:Prestress-via-growth} Growth prestress}

%A key aspect that plays an important role in many living tissues is residual stress. As defined in Section \ref{sec:Concept-of-prestress}, residual stress is an internal stress that remains when all external loads of an originally unloaded configuration are removed. Tissues actively build these internal stresses both during morphogenesis (when they rapidly change shape and add mass) and in the adult physiological state (when mass and volume changes serve the purpose of maintenance and are comparatively small). It

Physiological growth in living tissues often results in material added (or lost) in a non-uniform manner, forcing neighbouring tissue to accommodate the newly added material through elastic deformation. These nonuniformities, as illustrated in Fig.\ \ref{fig:morphoelastic}, include heterogeneous, anisotropic and differential prestrain $\mathbf{F}_a$. 
As discussed in Section \ref{sec:Concept-of-prestress}, this nonuniform prestrain is revealed by residual stress: an internal stress that remains when all external loads of an originally unloaded configuration are removed.
Tissues actively build these internal stresses both during morphogenesis (when they rapidly change shape and add mass) and in the adult physiological state (when mass and volume changes serve the purpose of maintenance and are comparatively small).

A classic example is the residual stress in arteries, which has been theoretically and experimentally described by Fung and others \citep{fung1991residual,vaishnav1987residual}. The observation is that arteries, when radially cut, open up due to compressive stress built in the hoop (circumferential) direction. The opening angle can be used to quantify residual stress \citep{fung1991residual}. Experiments also suggest that arteries are residually stressed in the axial direction \citep{goriely2010mechanical}. In a series of seminal studies, Fung and co-workers demonstrated residual stress in other cardiovascular systems such as the heart \citep{omens1990residual}, veins \citep{xie1991zero}, and the trachea \citep{han1991residual}. {Residual stress was also identified in other physiological tissues and organs such as the brain \citep{budday2014mechanical} and bones \citep{yamada2011residual}, as well as morphogenetic systems such as the optic cup \citep{oltean2016tissue} and the developing embryo \citep{beloussov2003morphomechanics}.} It was also found in pathological tissue such as solid tumours \citep{ambrosi2004role}. Proliferation also produces compression within the core of tumours and orthoradial tension at the periphery as revealed by cutting of an excised tumour along its radius \citep{stylianopoulos2012causes}. In this context, growth-related prestress is referred to as 'solid stress'.
%\AE{I updated some references here -- not sure if this is worth highlighting in read since the reviewer did not ask for these changes. }\JE{that's ok I'm sure}
%This list of examples of residual stress in physiological tissue is not exhaustive, and the plant kingdom also offers a rich pool of examples of residually stressed systems... [Jensen and Fozard, Dervaux and Ben Amar, ...]

% \AE{Add Alessandri, and Giovanni, possibly Dolega and Pierre's paper, as further examples of how stress is measured. }

Apart from physiological systems, growth-induced prestress has been measured in a multitude of ways in cultured systems. Stress in growing multicellular spheroids has been measured with great precision: the growing cells were encapsulated inside permeable, elastic, hollow microspheres which deformed as the spheroids grew inside of them, allowing to reconstruct the traction forces \citep{alessandri2013cellular}. The external pressure applied on the spheroid by the elastic coating leads to a steady-state size of the spheroid, in which there is an equilibrium between a necrotic core and a proliferating rim. The existence of a pressure at which spheroids reach a stationary size was theoretically proposed by \citet{basan2009homeostatic} and independently experimentally addressed by applying osmotic pressure to the spheroid \citep{montel2011stress}.

A similar technique for measuring growth-induced stress, by the elastic deformation of the environment, was demonstrated for yeast cell colonies (which do not form cell--cell junctions) \citep{delarue2016self}. A distinctly different method was used in \citep{cheng2009micro}, where  three-dimensional aggregates of tumour cells were co-embeded with fluorescent micro-beads in agarose gels. The displacement of the beads allows a reconstruction of the spatial stress distribution. 

Prestress generated by proliferation can also be observed in two-dimensional tissues, for example during the early development of the {\textit{Drosophila}} wing disc.
In a 3D finite element simulation of the wing disc as one heterogeneous layer, the growth rate and mechanical coupling to the elastic basement membrane of the tissue provoke the formation of spatially regulated folds \citep{tozluoglu_planar_2019}. {The doming of a part of the wing disc, the wing pouch, was recently explained through a combination of differential growth and differential growth anisotropy between tissue layers \citep{harmansa2022growth}.}  Similarly, differential growth rates between adhered tissues has been shown to regulate the looping of the chick gut \citep{Savin.2011.1}, the formation of villi in the chick gut \citep{Shyer.2013.1} and the gyrification of the brain \citep{Tallinen.2016.1}.

Until here, we considered growth to be dictated by proliferation rate, but an increase of cell density can occur independently to proliferation. For instance, during the formation of the zebrafish optic cup, migration of cells from the outside of the organ is analogue to a local tissue growth. This process increases tissue curvature and is required to produce correct bending of the organ, in parallel to differential contractility \citep{sidhaye_concerted_2017}.

Finally, the removal of cells, through cellular processes such as apoptosis, can be seen as a reverse growth and can play equally important roles during tissue morphogenesis. For example, in a transient extra-embryonic epithelial tissue named the amnioserosa, which sits between two embryonic epithelia during Drosophila embryogenesis, the gradual apoptosis of amnioserosa cells drives the final stage of dorsal closure in which those two sheets are brought together \citep{Pasakarnis.2016.1}.

\subsection{Topological prestress}
\label{sec:tissue:topology}

The cellular microstructure of tissues offers a further mechanism through which to generate or relax prestress via changing the topology through cell--cell neighbour exchange events, termed intercalations (or T1 transitions in foam literature) \citep{Blanchard+Adams.2009.1,Blanchard.2017.1}. At the subcellular scale, this process requires the coordinated action of many biomolecular players to disassemble and reassemble cell--cell junctions and has often been found to depend on the active generation of stress to shrink or expand junctions \citep{Bertet+Lecuit.2004.1,nestor2022adhesion}. The resulting T1 defines an orientation, as specified by the orientation of the removed and of the added junctions. In some tissues, T1 transitions occur throughout tissues with no preferred orientation, in which case they relieve local cell packing stresses, allowing tissue ordering \citep{Curran.2017.1} or facilitating flow during migration \citep{tlili_migrating_2020}. The transition from such a liquid-like state to a `jammed' state has been shown theoretically to relate to cell density, junctional tension and fluctuations \citep{LawsonKeister-Manning.2021.1}. However, the most dramatic examples occur when intercalations are globally aligned throughout a tissue, such as during Drosophila germband extension \citep{Zallen-Wieschaus.2004.1,Tetley-Blanchard+Sanson.2016.1}. This results in creating neighbour relations which, in the initial configuration of the tissue, strain the cells in the transverse direction: their relaxation thus entails a tissue-scale deformation in the longitudinal direction \citep{Collinet+Lecuit.2015.1}, see \fig{fig:cell_tissue}g. Similarly, multilayered epithelia can expand (here isotropically in the plane) through `radial intercalation’ in which cells in lower layers intercalate into upper layers creating a precompression which is relaxed by expansion \citep{Szabo.2016.1}. Since these are relaxation processes following out-of-equilibrium topological changes, they can be conceptualised as topological prestress. A major challenge in the analysis of such deformations is to disentangle this process from boundary stresses that may also act on the tissue \citep{Butler+Kabla+Mahadevan+.2009.1,Lye+Sanson.2015.1,Collinet+Lecuit.2015.1}.

Although epithelial topology is often represented as a two-dimensional network, a full three-dimensional treatment revealed different connectivity in the apical and basal domains \citep{gomez-galvez_scutoids_2018}. Topological transitions were indeed observed along the apico-basal axis in a range of in vivo tissues, resulting in a three-dimensional cell shape named a scutoid. This solution is favourable when tissues are curved by different amounts with respect to their principle planar axes (i.e.\ tubular rather than spherical) unless the radius of curvature is large compared with the tissue thickness. It remains to be found whether these intercalations could drive bending itself, rather than simply relax stress.

A separate class of topological change which affects prestress is the introduction of a new junction into the network, which occurs during cytokinesis, see \fig{fig:cell_tissue}\emph{f}. Again this is an oriented process, as the degree of freedom is the orientation of the new interface. Indeed, much has been discovered about the cell- and tissue-level signals read by a cell when choosing a division orientation. These signals include biochemical cues such as tissue polarisation \citep{Gong.2004.1, Gho.1998.1}, as well as mechanical and geometrical signals such as tissue stress \citep{Niwayama.2019.1, Fink.2011.1, Scarpa.2018.1} and cell shape \citep{wyatt_emergence_2015, Bosveld.2016.1}. 
In turn, the choice of division orientation alters prestress. A prestrained mother cell, depending on the choice of division orientation, could be divided into two daughter cells with either an increased or reduced anisotropy in shape. In tissues, the latter choice is usually observed, so as to homeostatically regulate cell shape anisotropy \citep{Mao+Tapon.2013.1,wyatt_emergence_2015,Xiong.2014.1}. In the zebrafish, this alignment of division with cell shape (which coincides with the principal axis of tissue stress), was shown using laser ablations to reduce the stress that builds up as a cell layer migrates over the embryo’s surface \citep{Campinho+Heisenberg.2013.1}. Similarly, alignment of cytokineses along a given axis can contribute to driving directional growth \citep{daSilva-Vincent.2007.1, Li.2014.1}.

\medskip

\begin{figure}[t]
    \centering
    \includegraphics[width=\linewidth]{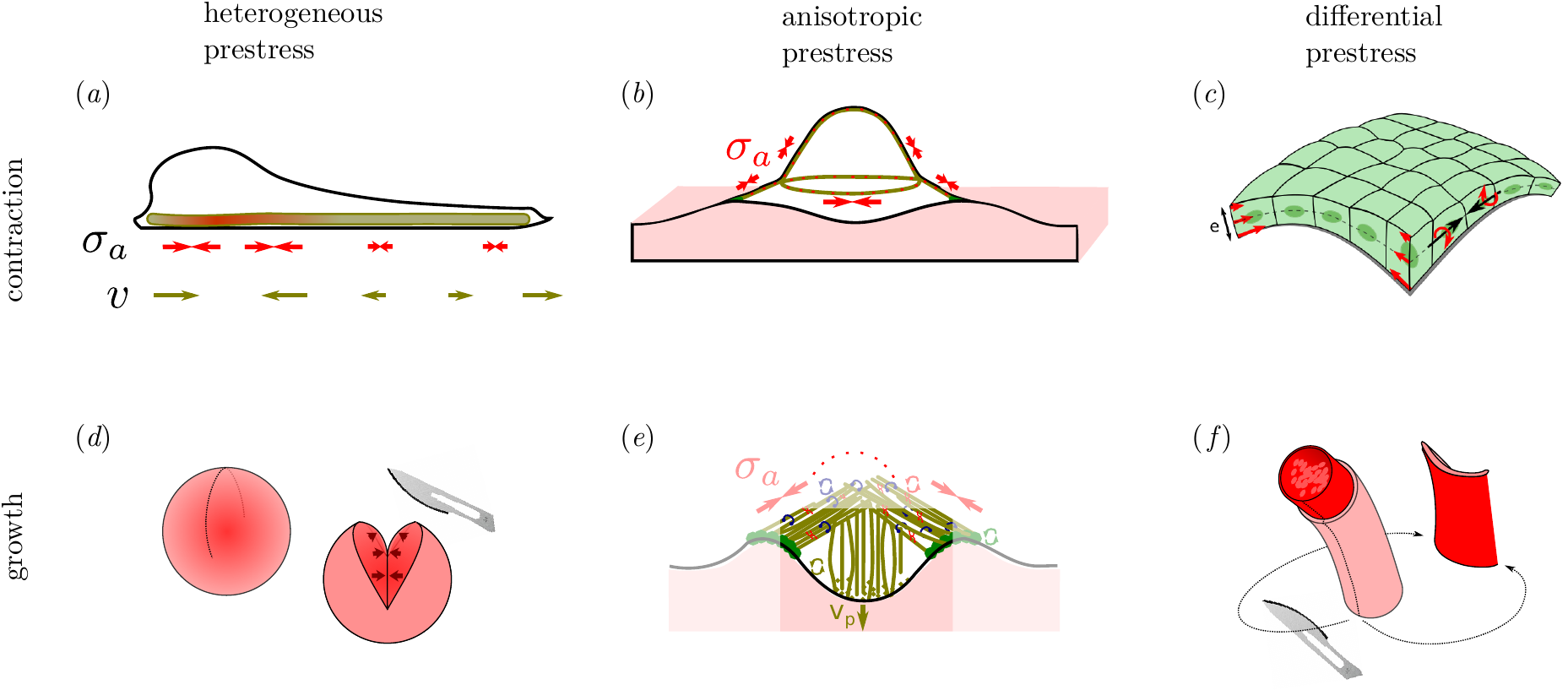}
    \caption{
    Example biophysical systems where mechanics is governed by heterogeneous, anisotropic or differential prestress of either sign, corresponding to contraction or growth.
    (\textit{a}) Contractile actomyosin with local accumulation, in parallel with a length-regulating element and in continuous adhesion with a substrate, generates a friction pattern that enables motility \citep{Recho+.2013.1}.
    (\textit{b}) Anisotropic pretension of the apical surface of cells regulates their shape \citep{burnette_contractile_2014}.
    (\textit{c}) Differential prestress between the weakly contractile apical (top) surface and strongly contractile basal (bottom) surface causes tissue curling \citep{recho_tug--war_2020} (\textit{d}) Residual stress due to heterogeneous growth is characterised by cutting experiments in tumours, revealing tensile hoop stress at the periphery \citep{stylianopoulos2012causes} (\textit{e}) The core of podosomes grows within a confined space, generating anisotropic prestress \citep{labernadie_protrusion_2014}
    (\textit{f}) Arteries change curvature if cut, this is believed to be caused by differential growth and remodelling of concentric layers \citep{goriely2010mechanical}
    }
    \label{fig:heterogeneities}
\end{figure}

Altogether, a complete description of tissue morphogenesis can be achieved by a linear combination of the active and passive mechanical modules we have described so far (cell stretching, oriented cell division, T1 transitions), provided that one has a good knowledge of the boundary conditions, which can often be dynamic in living systems \citep{etournay2015interplay}.

\section{\label{sec:Conclusions}Conclusions and future directions}

Living organisms have evolved a large number of controllable processes %\JE[phenomena]{better word?} 
able to drive deformations from within the system itself. In this review, we have tentatively brought together the understanding of those systems under the generic banner of prestressed materials. Beyond the convenience of relating these working principles with a known engineering concept, this allows us to offer a classification into heterogeneous, anisotropic or differential prestress systems driven by either contractile or growth activity. Figure \ref{fig:heterogeneities} provides an example for each of the six typologies that thus emerge. 

Additionally, %we propose that some active deformations can be understood as resulting from topological prestress, which is neither growth nor contraction of subelements but consists of a change of connectivity between them. 
{we define topological prestress as the prestress that is added to, or removed from, an interconnected network of mechanical elements not through prestressing individual elements, but purely by breaking and creating connections between the network elements.}
This latter type of prestress corresponds to a higher level of phenomenology, since it does not in itself describe by which (active) process the connectivity is being changed, see the examples in \fig{fig:cell_tissue}\textit{f,g}. It also has the property that it is always providing anisotropic prestress.

%\TW{I'm not sure, the radial intercalations that I wrote about are heterogeneous prestress no?
%\JE{Well, true, it's not so clear cut simple. It is always anisotropic, but on top of that it can be heterogeneous and/or differential (active scutoids would, eg). I guess we can say with more assurance that it is "always providing anisotropic prestress", rather than providing "only anisotropic prestress" which I had written. Else if you feel this is still a bit shortcutting the difficulty we can remove the sentence altogether and refer directly to fig 3f,g.}} (\fig{fig:cell_tissue}\textit{f,g}). 

\medskip

%The potential of this research field \JE[bla bla]{...?}.

From an experimental point of view, characterising living systems as active materials is highly challenging. One difficulty is linked with the necessity to test systems while they are maintained in a  state of function as close as possible to physiological conditions. For this, recent developments in organoid systems present great opportunities, since elements of tissue morphogenesis can now be recapitulated in an in vitro setting which is much simplified and much more amenable to experimental pertubations.

Observing the system simultaneously at different scales, especially combining global-scale stress and strain measurements with very local measurements could allow a better characterisation of how the microstructure dynamics give rise to emergent properties. For instance, tools measuring strain at the molecular scale, like Förster resonance energy transfer (FRET) \citep{borghi2012cadherin}, or at the meso-scale, like micro-magnets \citep{laplaud2021pinching}, micro-droplets \citep{mongera2018fluid} or optical tweezers \citep{han_cell_2018}, could be coupled to cell-scale techniques, like parallel plate rheometry or TFM, to decipher the subcellular contributions to cell shape changes. On the other hand, cell-scale in parallel to tissue-scale mechanical testing appears necessary to understand the cellular origins of tissue flows \citep{moisdon2022mapping}. 

In all cases, the interpretation of force--displacement relation measurements must be supported by careful modelling, as exemplified by the violation of the fluctuation--dissipation relation \citep{o2022time}. This sheds light on the vital need for combining any experimental approach with a theoretical understanding and of testing this by employing multiple modalities. 
%\JF{The bit from 'Beyond this' above could be somewhere else, connected to the theory paragraph for instance} 
%The theoretical models that can describe these systems should thus continue to be developed so as to provide frameworks for this understanding.
Notably, extending the existing models towards geometric or material nonlinearities is a necessity given the large deformations that are commonly encountered in real systems. How the prestrain and prestress fields relate in a nonlinear context also remains to be clarified. Solving mechanical models coupling different biological structures, in relevant geometries in three dimensions also remains an important challenge: for example, a full understanding of the three-dimensional mechanical balance of single or tissue cells, or of the dynamics of fibrous tissues which is governed both by cells and ECM are still lacking.

{Another} challenging theoretical task is to describe topological prestrain and prestress in a continuum framework. A missing intermediate  step is the linking between the cellular and tissue scales. At the scale of several cells,  vertex models \citep{farhadifar2007influence} are a highly studied family of models for biological tissue, both for their strength at capturing various tissue properties \citep{alt2017vertex,cheddadi2019coupling,latorre2018active} and for their interesting physical behaviour \citep{bi2015density,farhadifar2007influence,schoetz2013glassy}. However, the coarse-graining of vertex models to continuum descriptions, such as anelasticity, remains highly challenging, and is being tackled with approaches based on nearly periodic lattices \citep{murisic2015discrete,chenchiah2014energy,kupferman2020continuum} and discrete calculus \citep{jensen2020force,nestor2018PRE}. These efforts may in the long term lead to the possibility of encoding topological transitions, like active or passive neighbour exchanges (T1 transitions), into a continuum field usable, for instance, in anelasticity approaches. 

{Seventy percent of the total cell volume is water, and there is growing evidence in favour of a coupling between cell mechanics and osmotic gradients controlling volume \citep{xie2018controlling,cadart2019physics}. Cells maintain a prestress inside the membrane by carefully controlling the flow of water: water mobility in and out of cells relies on the permeation of water through the
plasma membrane, which can be regulated by aquaporin
channels  \citep{kedem1958thermodynamic},
as well as ion pumps which actively consume energy. Therefore, there has been recently considerable interest in the modelling of water mobility. This has been approached via a fluid phase in the framework of poroelasticity \citep{ambrosi2017solid,fraldi2018cells,xue2016biochemomechanical}, as well as by explicitly tracking water fluxes in a vertex model \citep{cheddadi2019coupling}. In parallel, there are new developments coupling the electrochemistry of ion fluxes, mechanics of cell volume regulation, and active pumping \citep{cadart2019physics,deshpande2021chemo}. Water mobility should ultimately add contributions to coarse-grained models such as anelasticity. From the experimental side, this requires the development of non-perturbative pressure sensors which is an important challenge for the future. }


\begin{thebibliography}{215}
\providecommand{\natexlab}[1]{#1}
\providecommand{\url}[1]{\texttt{#1}}
\expandafter\ifx\csname urlstyle\endcsname\relax
  \providecommand{\doi}[1]{doi: #1}\else
  \providecommand{\doi}{doi: \begingroup \urlstyle{rm}\Url}\fi

\bibitem[Acton et~al.(2014)Acton, Farrugia, Astarita, Mour{\~a}o-S{\'a},
  Jenkins, Nye, Hooper, Van~Blijswijk, Rogers, Snelgrove,
  et~al.]{acton2014dendritic}
Sophie~E Acton, Aaron~J Farrugia, Jillian~L Astarita, Diego Mour{\~a}o-S{\'a},
  Robert~P Jenkins, Emma Nye, Steven Hooper, Janneke Van~Blijswijk, Neil~C
  Rogers, Kathryn~J Snelgrove, et~al.
\newblock Dendritic cells control fibroblastic reticular network tension and
  lymph node expansion.
\newblock \emph{Nature}, 514\penalty0 (7523):\penalty0 498--502, 2014.

\bibitem[Aharoni et~al.(2016)Aharoni, Kolinski, Moshe, Meirzada, and
  Sharon]{aharoni2016internal}
Hillel Aharoni, John~M Kolinski, Michael Moshe, Idan Meirzada, and Eran Sharon.
\newblock Internal stresses lead to net forces and torques on extended elastic
  bodies.
\newblock \emph{Physical review letters}, 117\penalty0 (12):\penalty0 124101,
  2016.

\bibitem[Alessandri et~al.(2013)Alessandri, Sarangi, Gurchenkov, Sinha,
  Kie{\ss}ling, Fetler, Rico, Scheuring, Lamaze, Simon,
  et~al.]{alessandri2013cellular}
K{\'e}vin Alessandri, Bibhu~Ranjan Sarangi, Vasily~Val{\'e}r{\"\i}{\'e}vitch
  Gurchenkov, Bidisha Sinha, Tobias~Reinhold Kie{\ss}ling, Luc Fetler, Felix
  Rico, Simon Scheuring, Christophe Lamaze, Anthony Simon, et~al.
\newblock Cellular capsules as a tool for multicellular spheroid production and
  for investigating the mechanics of tumor progression in vitro.
\newblock \emph{Proceedings of the National Academy of Sciences}, 110\penalty0
  (37):\penalty0 14843--14848, 2013.

\bibitem[Almet et~al.(2021)Almet, Byrne, Maini, and Moulton]{almet2021role}
Axel~A Almet, Helen~M Byrne, Philip~K Maini, and Derek~E Moulton.
\newblock The role of mechanics in the growth and homeostasis of the intestinal
  crypt.
\newblock \emph{Biomechanics and Modeling in Mechanobiology}, 20\penalty0
  (2):\penalty0 585--608, 2021.

\bibitem[Alt et~al.(2017)Alt, Ganguly, and Salbreux]{alt2017vertex}
Silvanus Alt, Poulami Ganguly, and Guillaume Salbreux.
\newblock Vertex models: from cell mechanics to tissue morphogenesis.
\newblock \emph{Philosophical Transactions of the Royal Society B: Biological
  Sciences}, 372\penalty0 (1720):\penalty0 20150520, 2017.

\bibitem[Amato and Taylor(1986)]{amato1986probing}
Philip~A Amato and D~Lansing Taylor.
\newblock Probing the mechanism of incorporation of fluorescently labeled actin
  into stress fibers.
\newblock \emph{The Journal of cell biology}, 102\penalty0 (3):\penalty0
  1074--1084, 1986.

\bibitem[Ambrosi and Guillou(2007)]{ambrosi2007growth}
D~Ambrosi and A~Guillou.
\newblock Growth and dissipation in biological tissues.
\newblock \emph{Continuum Mechanics and Thermodynamics}, 19\penalty0
  (5):\penalty0 245--251, 2007.

\bibitem[Ambrosi and Mollica(2004)]{ambrosi2004role}
D~Ambrosi and F~Mollica.
\newblock The role of stress in the growth of a multicell spheroid.
\newblock \emph{Journal of mathematical biology}, 48\penalty0 (5):\penalty0
  477--499, 2004.

\bibitem[Ambrosi and Zanzottera(2016)]{Ambrosi-Zanzottera.2016.1}
D.~Ambrosi and A.~Zanzottera.
\newblock Mechanics and polarity in cell motility.
\newblock \emph{Physica D}, 2016.

\bibitem[Ambrosi et~al.(2017)Ambrosi, Pezzuto, Riccobelli, Stylianopoulos, and
  Ciarletta]{ambrosi2017solid}
D~Ambrosi, Simone Pezzuto, Davide Riccobelli, T~Stylianopoulos, and Pasquale
  Ciarletta.
\newblock Solid tumors are poroelastic solids with a chemo-mechanical feedback
  on growth.
\newblock \emph{Journal of Elasticity}, 129\penalty0 (1):\penalty0 107--124,
  2017.

\bibitem[Barbazan et~al.(2021)Barbazan, P{\'e}rez-Gonz{\'a}lez,
  G{\'o}mez-Gonz{\'a}lez, Dedenon, Richon, Latorre, Serra, Mariani, Descroix,
  Sens, et~al.]{barbazan2021cancer}
Jorge Barbazan, Carlos P{\'e}rez-Gonz{\'a}lez, Manuel G{\'o}mez-Gonz{\'a}lez,
  Mathieu Dedenon, Sophie Richon, Ernest Latorre, Marco Serra, Pascale Mariani,
  St{\'e}phanie Descroix, Pierre Sens, et~al.
\newblock Cancer-associated fibroblasts actively compress cancer cells and
  modulate mechanotransduction.
\newblock \emph{BioRxiv}, 2021.

\bibitem[Basan et~al.(2009)Basan, Risler, Joanny, Sastre-Garau, and
  Prost]{basan2009homeostatic}
Markus Basan, Thomas Risler, Jean-Fran{\c{c}}ois Joanny, Xavier Sastre-Garau,
  and Jacques Prost.
\newblock Homeostatic competition drives tumor growth and metastasis
  nucleation.
\newblock \emph{HFSP journal}, 3\penalty0 (4):\penalty0 265--272, 2009.

\bibitem[Bau{\"e}r et~al.(2017)Bau{\"e}r, Tavacoli, Pujol, Planade, Heuvingh,
  and Du~Roure]{bauer2017new}
Pierre Bau{\"e}r, Joseph Tavacoli, Thomas Pujol, Jessica Planade, Julien
  Heuvingh, and Olivia Du~Roure.
\newblock A new method to measure mechanics and dynamic assembly of branched
  actin networks.
\newblock \emph{Scientific reports}, 7\penalty0 (1):\penalty0 1--11, 2017.

\bibitem[Belmonte et~al.(2017)Belmonte, Leptin, and
  N{\'e}d{\'e}lec]{belmonte2017theory}
Julio~M Belmonte, Maria Leptin, and Fran{\c{c}}ois N{\'e}d{\'e}lec.
\newblock A theory that predicts behaviors of disordered cytoskeletal networks.
\newblock \emph{Molecular systems biology}, 13\penalty0 (9):\penalty0 941,
  2017.

\bibitem[Beloussov and Grabovsky(2003)]{beloussov2003morphomechanics}
Lev~V Beloussov and Vassily~I Grabovsky.
\newblock Morphomechanics: goals, basic experiments and models.
\newblock \emph{International Journal of Developmental Biology}, 50\penalty0
  (2-3):\penalty0 81--92, 2003.

\bibitem[Ben~Amar et~al.(2018)Ben~Amar, Qiuyang-Qu, Vuong-Brender, and
  Labouesse]{amar2018assessing}
Martine Ben~Amar, Paul Qiuyang-Qu, Thanh Thi~Kim Vuong-Brender, and Michel
  Labouesse.
\newblock Assessing the contribution of active and passive stresses in {\emph{c
  elegans}} elongation.
\newblock \emph{Physical Review Letters}, 121\penalty0 (26):\penalty0 268102,
  2018.

\bibitem[Bertet et~al.(2004)Bertet, Sulak, and Lecuit]{Bertet+Lecuit.2004.1}
C.~Bertet, L.~Sulak, and T.~Lecuit.
\newblock Myosin-dependent junction remodelling controls planar cell
  intercalation and axis elongation.
\newblock \emph{Nature}, 429:\penalty0 667, 2004.

\bibitem[Bertinetti et~al.(2015)Bertinetti, Masic, Schuetz, Barbetta, Seidt,
  Wagermaier, and Fratzl]{Bertinetti+Fratzl.2015.1}
Luca Bertinetti, Admir Masic, Roman Schuetz, Aurelio Barbetta, Britta Seidt,
  Wolfgang Wagermaier, and Peter Fratzl.
\newblock Osmotically driven tensile stress in collagen-based mineralized
  tissues.
\newblock \emph{Journal of the Mechanical Behavior of Biomedical Materials},
  52:\penalty0 14--21, 2015.
\newblock \doi{10.1016/j.jmbbm.2015.03.010}.

\bibitem[Bi et~al.(2015)Bi, Lopez, Schwarz, and Manning]{bi2015density}
Dapeng Bi, JH~Lopez, Jennifer~M Schwarz, and M~Lisa Manning.
\newblock A density-independent rigidity transition in biological tissues.
\newblock \emph{Nature Physics}, 11\penalty0 (12):\penalty0 1074--1079, 2015.

\bibitem[Bieling et~al.(2016)Bieling, Li, Weichsel, McGorty, Jreij, Huang,
  Fletcher, and Mullins]{bieling_force_2016}
Peter Bieling, Tai-De Li, Julian Weichsel, Ryan McGorty, Pamela Jreij,
  Bo~Huang, Daniel~A. Fletcher, and R.~Dyche Mullins.
\newblock Force {Feedback} {Controls} {Motor} {Activity} and {Mechanical}
  {Properties} of {Self}-{Assembling} {Branched} {Actin} {Networks}.
\newblock \emph{Cell}, 164\penalty0 (1):\penalty0 115--127, January 2016.
\newblock ISSN 0092-8674.
\newblock \doi{10.1016/j.cell.2015.11.057}.
\newblock URL
  \url{https://www.sciencedirect.com/science/article/pii/S0092867415015767}.

\bibitem[Blanchard et~al.(2009)Blanchard, Kabla, Schultz, Butler, Sanson,
  Gorfinkiel, Mahadevan, and Adams]{Blanchard+Adams.2009.1}
G.~B. Blanchard, A.J. Kabla, N.L. Schultz, L.C. Butler, B.~Sanson,
  N.~Gorfinkiel, L.~Mahadevan, and R.J. Adams.
\newblock Tissue tectonics: morphogenetic strain rates, cell shape change and
  intercalation.
\newblock \emph{Nature Methods}, 6:\penalty0 458--464, 2009.

\bibitem[Blanchard(2017)]{Blanchard.2017.1}
Guy~B. Blanchard.
\newblock Taking the strain: quantifying the contributions of all cell
  behaviours to changes in epithelial shape.
\newblock \emph{Phil. Trans. R. Soc. B}, 372:\penalty0 20150513, 2017.
\newblock \doi{10.1098/rstb.2015.0513}.

\bibitem[Blanchard et~al.(2018)Blanchard, \'Etienne, and
  Gorfinkiel]{Blanchard+Gorfinkiel.2018.1}
Guy~B Blanchard, Jocelyn \'Etienne, and Nicole Gorfinkiel.
\newblock From pulsatile apicomedial contractility to effective epithelial
  mechanics.
\newblock \emph{Current Opinion in Genetics \& Development}, 51:\penalty0
  78--87, 2018.
\newblock \doi{10.1016/j.gde.2018.07.004}.

\bibitem[Blanchoin et~al.(2014)Blanchoin, Boujemaa-Paterski, Sykes, and
  Plastino]{blanchoin_actin_2014}
Laurent Blanchoin, Rajaa Boujemaa-Paterski, Cécile Sykes, and Julie Plastino.
\newblock Actin dynamics, architecture, and mechanics in cell motility.
\newblock \emph{Physiological reviews}, 94\penalty0 (1):\penalty0 235--263,
  2014.
\newblock Publisher: American Physiological Society Bethesda, MD.

\bibitem[Bonfanti et~al.(2022)Bonfanti, Duque, Kabla, and
  Charras]{bonfanti_fracture_2022}
Alessandra Bonfanti, Julia Duque, Alexandre Kabla, and Guillaume Charras.
\newblock Fracture in living tissues.
\newblock \emph{Trends in Cell Biology}, 2022.
\newblock Publisher: Elsevier.

\bibitem[Bonnet et~al.(2012)Bonnet, Marcq, Bosveld, Fetler, Bellaïche, and
  Graner]{bonnet_mechanical_2012}
Isabelle Bonnet, Philippe Marcq, Floris Bosveld, Luc Fetler, Yohanns
  Bellaïche, and François Graner.
\newblock Mechanical state, material properties and continuous description of
  an epithelial tissue.
\newblock \emph{Journal of The Royal Society Interface}, 9\penalty0
  (75):\penalty0 2614--2623, 2012.
\newblock Publisher: The Royal Society.

\bibitem[Borghi et~al.(2012)Borghi, Sorokina, Shcherbakova, Weis, Pruitt,
  Nelson, and Dunn]{borghi2012cadherin}
Nicolas Borghi, Maria Sorokina, Olga~G Shcherbakova, William~I Weis, Beth~L
  Pruitt, W~James Nelson, and Alexander~R Dunn.
\newblock E-cadherin is under constitutive actomyosin-generated tension that is
  increased at cell--cell contacts upon externally applied stretch.
\newblock \emph{Proceedings of the National Academy of Sciences}, 109\penalty0
  (31):\penalty0 12568--12573, 2012.

\bibitem[Bosveld et~al.(2016)Bosveld, Markova, Guirao, Martin, Wang, Pierre,
  Balakireva, Gaugue, Ainslie, Christophorou, et~al.]{Bosveld.2016.1}
Floris Bosveld, Olga Markova, Boris Guirao, Charlotte Martin, Zhimin Wang,
  Ana{\"e}lle Pierre, Maria Balakireva, Isabelle Gaugue, Anna Ainslie, Nicolas
  Christophorou, et~al.
\newblock Epithelial tricellular junctions act as interphase cell shape sensors
  to orient mitosis.
\newblock \emph{Nature}, 530\penalty0 (7591):\penalty0 495--498, 2016.

\bibitem[Bowden et~al.(2015)Bowden, Byrne, Maini, and Moulton]{Bowden2015wound}
L~G Bowden, H~M Byrne, P~K Maini, and D~E Moulton.
\newblock {A morphoelastic model for dermal wound closure}.
\newblock \emph{Biomechanics and modeling in mechanobiology}, pages 1--19,
  August 2015.

\bibitem[Brodland et~al.(2010)Brodland, Conte, Cranston, Veldhuis, Narasimhan,
  Hutson, Jacinto, Ulrich, Baum, and Miodownik]{brodland_video_2010}
G.~Wayne Brodland, Vito Conte, P.~Graham Cranston, Jim Veldhuis, Sriram
  Narasimhan, M.~Shane Hutson, Antonio Jacinto, Florian Ulrich, Buzz Baum, and
  Mark Miodownik.
\newblock Video force microscopy reveals the mechanics of ventral furrow
  invagination in {Drosophila}.
\newblock \emph{Proceedings of the National Academy of Sciences}, 107\penalty0
  (51):\penalty0 22111--22116, 2010.
\newblock Publisher: National Acad Sciences.

\bibitem[Budday et~al.(2014)Budday, Raybaud, and Kuhl]{budday2014mechanical}
Silvia Budday, Charles Raybaud, and Ellen Kuhl.
\newblock A mechanical model predicts morphological abnormalities in the
  developing human brain.
\newblock \emph{Scientific reports}, 4\penalty0 (1):\penalty0 1--7, 2014.

\bibitem[Burnette et~al.(2014)Burnette, Shao, Ott, Pasapera, Fischer, Baird,
  Der~Loughian, Delanoe-Ayari, Paszek, and Davidson]{burnette_contractile_2014}
Dylan~T. Burnette, Lin Shao, Carolyn Ott, Ana~M. Pasapera, Robert~S. Fischer,
  Michelle~A. Baird, Christelle Der~Loughian, Helene Delanoe-Ayari, Matthew~J.
  Paszek, and Michael~W. Davidson.
\newblock A contractile and counterbalancing adhesion system controls the {3D}
  shape of crawling cells.
\newblock \emph{Journal of Cell Biology}, 205\penalty0 (1):\penalty0 83--96,
  2014.
\newblock Publisher: The Rockefeller University Press.

\bibitem[Burridge and Guilluy(2016)]{burridge2016focal}
Keith Burridge and Christophe Guilluy.
\newblock Focal adhesions, stress fibers and mechanical tension.
\newblock \emph{Experimental cell research}, 343\penalty0 (1):\penalty0 14--20,
  2016.

\bibitem[Butler et~al.(2009)Butler, Blanchard, Kabla, Lawrence, Welchman,
  Mahadevan, Adams, and Sanson]{Butler+Kabla+Mahadevan+.2009.1}
L.~C. Butler, G.~B. Blanchard, A.~J. Kabla, N.~J. Lawrence, D.~P. Welchman,
  L.~Mahadevan, R.~J. Adams, and B.~Sanson.
\newblock Cell shape changes indicate a role for extrinsic tensile forces in
  drosophila germ-band extension.
\newblock \emph{Nature Cell Biol.}, 11:\penalty0 859--864, 2009.

\bibitem[Cadart et~al.(2019)Cadart, Venkova, Recho, Lagomarsino, and
  Piel]{cadart2019physics}
Clotilde Cadart, Larisa Venkova, Pierre Recho, Marco~Cosentino Lagomarsino, and
  Matthieu Piel.
\newblock The physics of cell-size regulation across timescales.
\newblock \emph{Nature Physics}, 15\penalty0 (10):\penalty0 993--1004, 2019.

\bibitem[Calvo et~al.(2013)Calvo, Ege, Grande-Garcia, Hooper, Jenkins,
  Chaudhry, Harrington, Williamson, Moeendarbary, Charras,
  et~al.]{calvo2013mechanotransduction}
Fernando Calvo, Nil Ege, Araceli Grande-Garcia, Steven Hooper, Robert~P
  Jenkins, Shahid~I Chaudhry, Kevin Harrington, Peter Williamson, Emad
  Moeendarbary, Guillaume Charras, et~al.
\newblock Mechanotransduction and yap-dependent matrix remodelling is required
  for the generation and maintenance of cancer-associated fibroblasts.
\newblock \emph{Nature cell biology}, 15\penalty0 (6):\penalty0 637--646, 2013.

\bibitem[Campinho et~al.(2013)Campinho, Behrndt, Ranft, Risler, Minc, and
  Heisenberg]{Campinho+Heisenberg.2013.1}
Pedro Campinho, Martin Behrndt, Jonas Ranft, Thomas Risler, Nicolas Minc, and
  Carl-Philipp Heisenberg.
\newblock Tension-oriented cell divisions limit anisotropic tissue tension in
  epithelial spreading during zebrafish epiboly.
\newblock \emph{Nat Cell Biol}, 15:\penalty0 1405--1414, 2013.
\newblock \doi{10.1038/ncb2869}.

\bibitem[Caruel and Truskinovsky(2018)]{Caruel-Truskinovsky.2018.1}
M~Caruel and L~Truskinovsky.
\newblock Physics of muscle contraction.
\newblock \emph{Rep. Prog. Phys.}, 81:\penalty0 036602, 2018.
\newblock \doi{10.1088/1361-6633/aa7b9e}.

\bibitem[Chanet et~al.(2017)Chanet, Miller, Vaishnav, Ermentrout, Davidson, and
  Martin]{Chanet+Martin.2017.1}
Soline Chanet, Callie~J. Miller, Eeshit~Dhaval Vaishnav, Bard Ermentrout,
  Lance~A. Davidson, and Adam~C. Martin.
\newblock Actomyosin meshwork mechanosensing enables tissue shape to orient
  cell force.
\newblock \emph{Nat Comms}, 8:\penalty0 15014, 2017.
\newblock \doi{10.1038/ncomms15014}.

\bibitem[Charras et~al.(2008)Charras, Coughlin, Mitchison, and
  Mahadevan]{charras_life_2008}
Guillaume~T. Charras, Margaret Coughlin, Timothy~J. Mitchison, and
  L.~Mahadevan.
\newblock Life and {Times} of a {Cellular} {Bleb}.
\newblock \emph{Biophysical Journal}, 94\penalty0 (5):\penalty0 1836--1853,
  March 2008.
\newblock ISSN 0006-3495.
\newblock \doi{10.1529/biophysj.107.113605}.
\newblock URL
  \url{https://www.sciencedirect.com/science/article/pii/S0006349508706222}.

\bibitem[Cheddadi et~al.(2019)Cheddadi, G{\'e}nard, Bertin, and
  Godin]{cheddadi2019coupling}
Ibrahim Cheddadi, Michel G{\'e}nard, Nadia Bertin, and Christophe Godin.
\newblock Coupling water fluxes with cell wall mechanics in a multicellular
  model of plant development.
\newblock \emph{PLoS computational biology}, 15\penalty0 (6):\penalty0
  e1007121, 2019.

\bibitem[Chelly et~al.(2022)Chelly, Jahangiri, Mireux, {{\'E}}tienne, Dysthe,
  Verdier, and Recho]{Chelly+Recho.2022.1}
H.~Chelly, A.~Jahangiri, M.~Mireux, J.~{{\'E}}tienne, D.K. Dysthe, C.~Verdier,
  and P.~Recho.
\newblock Cell crawling on a compliant substrate: A biphasic relation with
  linear friction.
\newblock \emph{International Journal of Non-Linear Mechanics}, 139:\penalty0
  103897, 2022.
\newblock \doi{10.1016/j.ijnonlinmec.2021.103897}.

\bibitem[Chenchiah and Shipman(2014)]{chenchiah2014energy}
Isaac~Vikram Chenchiah and Patrick~D Shipman.
\newblock An energy-deformation decomposition for morphoelasticity.
\newblock \emph{Journal of the Mechanics and Physics of Solids}, 67:\penalty0
  15--39, 2014.

\bibitem[Cheng et~al.(2009)Cheng, Tse, Jain, and Munn]{cheng2009micro}
Gang Cheng, Janet Tse, Rakesh~K Jain, and Lance~L Munn.
\newblock Micro-environmental mechanical stress controls tumor spheroid size
  and morphology by suppressing proliferation and inducing apoptosis in cancer
  cells.
\newblock \emph{PLoS one}, 4\penalty0 (2):\penalty0 e4632, 2009.

\bibitem[Chugh et~al.(2017)Chugh, Clark, Smith, Cassani, Dierkes, Ragab, Roux,
  Charras, Salbreux, and Paluch]{chugh_actin_2017}
Priyamvada Chugh, Andrew~G. Clark, Matthew~B. Smith, Davide~AD Cassani, Kai
  Dierkes, Anan Ragab, Philippe~P. Roux, Guillaume Charras, Guillaume Salbreux,
  and Ewa~K. Paluch.
\newblock Actin cortex architecture regulates cell surface tension.
\newblock \emph{Nature cell biology}, 19\penalty0 (6):\penalty0 689--697, 2017.
\newblock Publisher: Nature Publishing Group.

\bibitem[Cl\'ement et~al.(2017)Cl\'ement, Dehapiot, Collinet, Lecuit, and
  Lenne]{Clement+Lenne.2017.1}
R.~Cl\'ement, B.~Dehapiot, Claudio Collinet, Thomas Lecuit, and
  Pierre-Fran\c{c}ois Lenne.
\newblock Viscoelastic dissipation stabilizes cell shape changes during tissue
  morphogenesis.
\newblock \emph{Current Biology}, 2017.
\newblock \doi{10.1016/j.cub.2017.09.005}.

\bibitem[Collinet et~al.(2015)Collinet, Rauzi, Lenne, and
  Lecuit]{Collinet+Lecuit.2015.1}
C.~Collinet, M.~Rauzi, P.-F. Lenne, and T.~Lecuit.
\newblock Local and tissue-scale forces drive oriented junction growth during
  tissue extension.
\newblock \emph{Nature Cell Biol.}, 17:\penalty0 1247--1258, 2015.

\bibitem[Crow et~al.(2012)Crow, Webster, Hohlfeld, Ng, Geissler, and
  Fletcher]{crow2012contractile}
Ailey Crow, Kevin~D Webster, Evan Hohlfeld, Win~Pin Ng, Phillip Geissler, and
  Daniel~A Fletcher.
\newblock Contractile equilibration of single cells to step changes in
  extracellular stiffness.
\newblock \emph{Biophysical journal}, 102\penalty0 (3):\penalty0 443--451,
  2012.

\bibitem[Curran et~al.(2017)Curran, Strandkvist, Bathmann, {de Gennes}, Kabla,
  Salbreux, and Baum]{Curran.2017.1}
Scott Curran, Charlotte Strandkvist, Jasper Bathmann, Marc {de Gennes},
  Alexandre Kabla, Guillaume Salbreux, and Buzz Baum.
\newblock Myosin ii controls junction fluctuations to guide epithelial tissue
  ordering.
\newblock \emph{Developmental Cell}, 43\penalty0 (4):\penalty0 480--492.e6,
  2017.
\newblock ISSN 1534-5807.
\newblock \doi{https://doi.org/10.1016/j.devcel.2017.09.018}.
\newblock URL
  \url{https://www.sciencedirect.com/science/article/pii/S1534580717307712}.

\bibitem[da~Silva and Vincent(2007)]{daSilva-Vincent.2007.1}
Sara~Morais da~Silva and Jean-Paul Vincent.
\newblock Oriented cell divisions in the extending germband of drosophila.
\newblock \emph{Development}, 134:\penalty0 3049--3054, 2007.
\newblock \doi{10.1242/dev.004911}.

\bibitem[Delarue et~al.(2016)Delarue, Hartung, Schreck, Gniewek, Hu,
  Herminghaus, and Hallatschek]{delarue2016self}
Morgan Delarue, J{\"o}rn Hartung, Carl Schreck, Pawel Gniewek, Lucy Hu, Stephan
  Herminghaus, and Oskar Hallatschek.
\newblock Self-driven jamming in growing microbial populations.
\newblock \emph{Nature physics}, 12\penalty0 (8):\penalty0 762--766, 2016.

\bibitem[Delvoye et~al.(1991)Delvoye, Wiliquet, Levêque, Nusgens, and
  Lapière]{delvoye_measurement_1991}
Pierre Delvoye, Philippe Wiliquet, Jean-Luc Levêque, Betty~V. Nusgens, and
  Charles~M. Lapière.
\newblock Measurement of mechanical forces generated by skin fibroblasts
  embedded in a three-dimensional collagen gel.
\newblock \emph{Journal of Investigative Dermatology}, 97\penalty0
  (5):\penalty0 898--902, 1991.
\newblock Publisher: Elsevier.

\bibitem[Dembo and Wang(1999)]{Dembo+.1999.1}
M.~Dembo and Y.-L. Wang.
\newblock Stresses at the cell-to-substrate interface during locomotion of
  fibroblasts.
\newblock \emph{Biophys. J.}, 76:\penalty0 2307--2316, 1999.

\bibitem[Deshpande et~al.(2021)Deshpande, DeSimone, McMeeking, and
  Recho]{deshpande2021chemo}
Vikram Deshpande, Antonio DeSimone, Robert McMeeking, and Pierre Recho.
\newblock Chemo-mechanical model of a cell as a stochastic active gel.
\newblock \emph{Journal of the Mechanics and Physics of Solids}, 151:\penalty0
  104381, 2021.

\bibitem[DiCarlo and Quiligotti(2002)]{dicarlo2002growth}
Antonio DiCarlo and Sara Quiligotti.
\newblock Growth and balance.
\newblock \emph{Mechanics Research Communications}, 29\penalty0 (6):\penalty0
  449--456, 2002.

\bibitem[Discher et~al.(2005)Discher, Janmey, and Wang]{discher2005tissue}
Dennis~E Discher, Paul Janmey, and Yu-li Wang.
\newblock Tissue cells feel and respond to the stiffness of their substrate.
\newblock \emph{Science}, 310\penalty0 (5751):\penalty0 1139--1143, 2005.

\bibitem[Eckart(1948)]{eckart1948thermodynamics}
Carl Eckart.
\newblock The thermodynamics of irreversible processes. iv. the theory of
  elasticity and anelasticity.
\newblock \emph{Physical Review}, 73\penalty0 (4):\penalty0 373, 1948.

\bibitem[Ehret et~al.(2017)Ehret, Bircher, Stracuzzi, Marina, Zündel, and
  Mazza]{ehret_inverse_2017}
Alexander~E. Ehret, Kevin Bircher, Alberto Stracuzzi, Vita Marina, Manuel
  Zündel, and Edoardo Mazza.
\newblock Inverse poroelasticity as a fundamental mechanism in biomechanics and
  mechanobiology.
\newblock \emph{Nature communications}, 8\penalty0 (1):\penalty0 1--10, 2017.
\newblock Publisher: Nature Publishing Group.

\bibitem[Elkhatib et~al.(2014)Elkhatib, Neu, Zensen, Schmoller, Louvard,
  Bausch, Betz, and Vignjevic]{elkhatib2014fascin}
Nadia Elkhatib, Matthew~B Neu, Carla Zensen, Kurt~M Schmoller, Daniel Louvard,
  Andreas~R Bausch, Timo Betz, and Danijela~Matic Vignjevic.
\newblock Fascin plays a role in stress fiber organization and focal adhesion
  disassembly.
\newblock \emph{Current Biology}, 24\penalty0 (13):\penalty0 1492--1499, 2014.

\bibitem[Epstein(2012)]{epstein2012elements}
Marcelo Epstein.
\newblock \emph{The elements of continuum biomechanics}.
\newblock John Wiley \& Sons, 2012.

\bibitem[Epstein and Maugin(2000)]{epstein2000thermomechanics}
Marcelo Epstein and G{\'e}rard~A Maugin.
\newblock Thermomechanics of volumetric growth in uniform bodies.
\newblock \emph{International Journal of Plasticity}, 16\penalty0
  (7-8):\penalty0 951--978, 2000.

\bibitem[Erlich et~al.(2018)Erlich, Howell, Goriely, Chirat, and
  Moulton]{erlich2018mechanical}
Alexander Erlich, Rowan Howell, Alain Goriely, R{\'e}gis Chirat, and
  DE~Moulton.
\newblock Mechanical feedback in seashell growth and form.
\newblock \emph{The ANZIAM Journal}, 59\penalty0 (4):\penalty0 581--606, 2018.

\bibitem[Erlich et~al.(2019)Erlich, Moulton, and
  Goriely]{erlich2019homeostatic}
Alexander Erlich, Derek~E Moulton, and Alain Goriely.
\newblock Are homeostatic states stable? dynamical stability in
  morphoelasticity.
\newblock \emph{Bulletin of mathematical biology}, 81\penalty0 (8):\penalty0
  3219--3244, 2019.

\bibitem[Erlich et~al.(2020)Erlich, Jones, Tisseur, Moulton, and
  Goriely]{erlich2020role}
Alexander Erlich, Gareth~W Jones, Fran{\c{c}}oise Tisseur, Derek~E Moulton, and
  Alain Goriely.
\newblock The role of topology and mechanics in uniaxially growing cell
  networks.
\newblock \emph{Proceedings of the Royal Society A}, 476\penalty0
  (2233):\penalty0 20190523, 2020.

\bibitem[\'Etienne et~al.(2015)\'Etienne, Fouchard, Mitrossilis, Bufi,
  Durand-Smet, and Asnacios]{Etienne+Asnacios.2015.1}
Jocelyn \'Etienne, Jonathan Fouchard, D\'emosth\`ene Mitrossilis, Nathalie
  Bufi, Pauline Durand-Smet, and Atef Asnacios.
\newblock Cells as liquid motors: Mechanosensitivity emerges from collective
  dynamics of actomyosin cortex.
\newblock \emph{Proc Natl Acad Sci USA}, 112:\penalty0 2740--2745, 2015.
\newblock \doi{10.1073/pnas.1417113112}.

\bibitem[Etournay et~al.(2015)Etournay, Popovi{\'c}, Merkel, Nandi, Blasse,
  Aigouy, Brandl, Myers, Salbreux, J{\"u}licher, et~al.]{etournay2015interplay}
Rapha{\"e}l Etournay, Marko Popovi{\'c}, Matthias Merkel, Amitabha Nandi,
  Corinna Blasse, Beno{\^\i}t Aigouy, Holger Brandl, Gene Myers, Guillaume
  Salbreux, Frank J{\"u}licher, et~al.
\newblock Interplay of cell dynamics and epithelial tension during
  morphogenesis of the drosophila pupal wing.
\newblock \emph{Elife}, 4:\penalty0 e07090, 2015.

\bibitem[Farhadifar et~al.(2007)Farhadifar, Röper, Aigouy, Eaton, and
  J{\"u}licher]{farhadifar2007influence}
Reza Farhadifar, Jens-Christian Röper, Benoit Aigouy, Suzanne Eaton, and Frank
  J{\"u}licher.
\newblock The influence of cell mechanics, cell-cell interactions, and
  proliferation on epithelial packing.
\newblock \emph{Current Biology}, 17\penalty0 (24):\penalty0 2095--2104, 2007.

\bibitem[Fernandez-Gonzalez et~al.(2009)Fernandez-Gonzalez, de~Matos~Simoes,
  Röper, Eaton, and Zallen]{Fernandez+Eaton-Zallen.2009.1}
R.~Fernandez-Gonzalez, S.~de~Matos~Simoes, J.-C. Röper, S.~Eaton, and J.~A.
  Zallen.
\newblock Myosin ii dynamics are regulated by tension in intercalating cells.
\newblock \emph{Dev. Cell}, 17:\penalty0 736--743, 2009.

\bibitem[Fierling et~al.(2022)Fierling, John, Delorme, Torzynski, Blanchard,
  Lye, Popkova, Malandain, Sanson, {{\'E}}tienne, Marmottant, Quilliet, and
  Rauzi]{Fierling+Etienne+Rauzi.2022.1}
Julien Fierling, Alphy John, Barth{{\'e}}l{{\'e}}my Delorme, Alexandre
  Torzynski, Guy~B. Blanchard, Claire~M. Lye, Anna Popkova, Gr{{\'e}}goire
  Malandain, B{{\'e}}n{{\'e}}dicte Sanson, Jocelyn {{\'E}}tienne, Philippe
  Marmottant, Catherine Quilliet, and Matteo Rauzi.
\newblock Embryo-scale epithelial buckling forms a propagating furrow that
  initiates gastrulation.
\newblock \emph{Nat Commun}, 13:\penalty0 859864, 2022.
\newblock \doi{10.1038/s41467-022-30493-3}.

\bibitem[Fink et~al.(2011)Fink, Carpi, Betz, B{\'e}tard, Chebah, Azioune,
  Bornens, Sykes, Fetler, Cuvelier, et~al.]{Fink.2011.1}
Jenny Fink, Nicolas Carpi, Timo Betz, Angelique B{\'e}tard, Meriem Chebah,
  Ammar Azioune, Michel Bornens, Cecile Sykes, Luc Fetler, Damien Cuvelier,
  et~al.
\newblock External forces control mitotic spindle positioning.
\newblock \emph{Nature cell biology}, 13\penalty0 (7):\penalty0 771--778, 2011.

\bibitem[Forterre et~al.(2005)Forterre, Skotheim, Dumais, and
  Mahadevan]{Forterre+Mahadevan.2005.1}
Yo{\"e}l Forterre, Jan~M. Skotheim, Jacques Dumais, and L.~Mahadevan.
\newblock How the venus flytrap snaps.
\newblock \emph{Nature}, 433:\penalty0 421--425, 2005.
\newblock \doi{10.1038/nature03185}.

\bibitem[Fouchard et~al.(2011)Fouchard, Mitrossilis, and
  Asnacios]{fouchard_acto-myosin_2011}
Jonathan Fouchard, Démosthène Mitrossilis, and Atef Asnacios.
\newblock Acto-myosin based response to stiffness and rigidity sensing.
\newblock \emph{Cell Adhesion \& Migration}, 5\penalty0 (1):\penalty0 16--19,
  January 2011.
\newblock ISSN 1933-6918, 1933-6926.
\newblock \doi{10.4161/cam.5.1.13281}.
\newblock URL \url{http://www.tandfonline.com/doi/abs/10.4161/cam.5.1.13281}.

\bibitem[Fouchard et~al.(2020)Fouchard, Wyatt, Proag, Lisica, Khalilgharibi,
  Recho, Suzanne, Kabla, and Charras]{fouchard_curling_2020}
Jonathan Fouchard, Tom P.~J. Wyatt, Amsha Proag, Ana Lisica, Nargess
  Khalilgharibi, Pierre Recho, Magali Suzanne, Alexandre Kabla, and Guillaume
  Charras.
\newblock Curling of epithelial monolayers reveals coupling between active
  bending and tissue tension.
\newblock \emph{Proceedings of the National Academy of Sciences}, 117\penalty0
  (17):\penalty0 9377--9383, April 2020.
\newblock ISSN 0027-8424, 1091-6490.
\newblock \doi{10.1073/pnas.1917838117}.
\newblock URL \url{https://pnas.org/doi/full/10.1073/pnas.1917838117}.

\bibitem[Fraldi and Carotenuto(2018)]{fraldi2018cells}
Massimiliano Fraldi and Angelo~R Carotenuto.
\newblock Cells competition in tumor growth poroelasticity.
\newblock \emph{Journal of the Mechanics and Physics of Solids}, 112:\penalty0
  345--367, 2018.

\bibitem[Fritzsche et~al.(2013)Fritzsche, Lewalle, Duke, Kruse, and
  Charras]{Fritzsche+Charras.2013.1}
M.~Fritzsche, A.~Lewalle, T.~Duke, K.~Kruse, and G.~Charras.
\newblock Analysis of turnover dynamics of the submembranous actin cortex.
\newblock \emph{Mol. Biol. Cell}, 24:\penalty0 757--767, 2013.
\newblock \doi{10.1091/mbc.E12-06-0485}.

\bibitem[Fung(1991)]{fung1991residual}
Yuan~Cheng Fung.
\newblock What are the residual stresses doing in our blood vessels?
\newblock \emph{Annals of biomedical engineering}, 19\penalty0 (3):\penalty0
  237--249, 1991.

\bibitem[Fung(2013)]{fung2013biomechanics}
Yuan-cheng Fung.
\newblock \emph{Biomechanics: motion, flow, stress, and growth}.
\newblock Springer Science \& Business Media, 2013.

\bibitem[Ganghoffer(2010)]{ganghoffer2010mechanical}
Jean-Fran{\c{c}}ois Ganghoffer.
\newblock Mechanical modeling of growth considering domain variation—part ii:
  volumetric and surface growth involving eshelby tensors.
\newblock \emph{Journal of the Mechanics and Physics of Solids}, 58\penalty0
  (9):\penalty0 1434--1459, 2010.

\bibitem[Gay et~al.(2012)Gay, Courtheoux, Reyes, Tournier, and
  Gachet]{gay2012stochastic}
Guillaume Gay, Thibault Courtheoux, C{\'e}line Reyes, Sylvie Tournier, and
  Yannick Gachet.
\newblock A stochastic model of kinetochore--microtubule attachment accurately
  describes fission yeast chromosome segregation.
\newblock \emph{Journal of Cell Biology}, 196\penalty0 (6):\penalty0 757--774,
  2012.

\bibitem[Gho and Schweisguth(1998)]{Gho.1998.1}
Michel Gho and Fran{\c{c}}ois Schweisguth.
\newblock Frizzled signalling controls orientation of asymmetric sense organ
  precursor cell divisions in drosophila.
\newblock \emph{Nature}, 393\penalty0 (6681):\penalty0 178--181, 1998.

\bibitem[Giraud-Guille et~al.(2008)Giraud-Guille, Mosser, and
  Belamie]{giraud2008liquid}
Marie~Madeleine Giraud-Guille, Gervaise Mosser, and Emmanuel Belamie.
\newblock Liquid crystallinity in collagen systems in vitro and in vivo.
\newblock \emph{Current Opinion in Colloid \& Interface Science}, 13\penalty0
  (4):\penalty0 303--313, 2008.

\bibitem[G{\'o}mez-Gonz{\'a}lez et~al.(2020)G{\'o}mez-Gonz{\'a}lez, Latorre,
  Arroyo, and Trepat]{Gomez.2020.1}
Manuel G{\'o}mez-Gonz{\'a}lez, Ernest Latorre, Marino Arroyo, and Xavier
  Trepat.
\newblock Measuring mechanical stress in living tissues.
\newblock \emph{Nature Reviews Physics}, 2\penalty0 (6):\penalty0 300--317,
  2020.

\bibitem[Gong et~al.(2004)Gong, Mo, and Fraser]{Gong.2004.1}
Ying Gong, Chunhui Mo, and Scott~E Fraser.
\newblock Planar cell polarity signalling controls cell division orientation
  during zebrafish gastrulation.
\newblock \emph{Nature}, 430\penalty0 (7000):\penalty0 689--693, 2004.

\bibitem[Goriely(2017)]{goriely2017mathematics}
Alain Goriely.
\newblock \emph{The mathematics and mechanics of biological growth}, volume~45.
\newblock Springer, 2017.

\bibitem[Goriely and Vandiver(2010)]{goriely2010mechanical}
Alain Goriely and Rebecca Vandiver.
\newblock On the mechanical stability of growing arteries.
\newblock \emph{IMA journal of applied mathematics}, 75\penalty0 (4):\penalty0
  549--570, 2010.

\bibitem[Gower et~al.(2017)Gower, Shearer, and Ciarletta]{gower2017new}
Artur~L Gower, Tom Shearer, and Pasquale Ciarletta.
\newblock A new restriction for initially stressed elastic solids.
\newblock \emph{The Quarterly Journal of Mechanics and Applied Mathematics},
  70\penalty0 (4):\penalty0 455--478, 2017.

\bibitem[Gracia et~al.(2019)Gracia, Theis, Proag, Gay, Benassayag, and
  Suzanne]{gracia_mechanical_2019}
Mélanie Gracia, Sophie Theis, Amsha Proag, Guillaume Gay, Corinne Benassayag,
  and Magali Suzanne.
\newblock Mechanical impact of epithelial- mesenchymal transition on epithelial
  morphogenesis in {Drosophila}.
\newblock \emph{Nature communications}, 10\penalty0 (1):\penalty0 1--17, 2019.
\newblock Publisher: Nature Publishing Group.

\bibitem[Gómez-Gálvez et~al.(2018)Gómez-Gálvez, Vicente-Munuera, Tagua,
  Forja, Castro, Letrán, Valencia-Expósito, Grima, Bermúdez-Gallardo,
  Serrano-Pérez-Higueras, Cavodeassi, Sotillos, Martín-Bermudo, Márquez,
  Buceta, and Escudero]{gomez-galvez_scutoids_2018}
Pedro Gómez-Gálvez, Pablo Vicente-Munuera, Antonio Tagua, Cristina Forja,
  Ana~M. Castro, Marta Letrán, Andrea Valencia-Expósito, Clara Grima, Marina
  Bermúdez-Gallardo, Óscar Serrano-Pérez-Higueras, Florencia Cavodeassi, Sol
  Sotillos, María~D. Martín-Bermudo, Alberto Márquez, Javier Buceta, and
  Luis~M. Escudero.
\newblock Scutoids are a geometrical solution to three-dimensional packing of
  epithelia.
\newblock \emph{Nature Communications}, 9\penalty0 (1):\penalty0 2960, July
  2018.
\newblock ISSN 2041-1723.
\newblock \doi{10.1038/s41467-018-05376-1}.
\newblock URL \url{https://www.nature.com/articles/s41467-018-05376-1.}
\newblock Number: 1 Publisher: Nature Publishing Group.

\bibitem[Han and Fung(1991)]{han1991residual}
HC~Han and YC~Fung.
\newblock Residual strains in porcine and canine trachea.
\newblock \emph{Journal of biomechanics}, 24\penalty0 (5):\penalty0 307--315,
  1991.

\bibitem[Han et~al.(2018)Han, Ronceray, Xu, Malandrino, Kamm, Lenz, Broedersz,
  and Guo]{han_cell_2018}
Yu~Long Han, Pierre Ronceray, Guoqiang Xu, Andrea Malandrino, Roger~D. Kamm,
  Martin Lenz, Chase~P. Broedersz, and Ming Guo.
\newblock Cell contraction induces long-ranged stress stiffening in the
  extracellular matrix.
\newblock \emph{Proceedings of the National Academy of Sciences}, 115\penalty0
  (16):\penalty0 4075--4080, 2018.
\newblock Publisher: National Acad Sciences.

\bibitem[Harmansa et~al.(2022)Harmansa, Erlich, Eloy, Zurlo, and
  Lecuit]{harmansa2022growth}
Stefan Harmansa, Alexander Erlich, Christophe Eloy, Giuseppe Zurlo, and Thomas
  Lecuit.
\newblock Growth anisotropy of the extracellular matrix drives mechanics in a
  developing organ.
\newblock \emph{bioRxiv}, 2022.

\bibitem[Harris et~al.(1980)Harris, Wild, and Stopak]{Harris+.1981.1}
A.~K. Harris, P.~Wild, and D.~Stopak.
\newblock Silicone rubber substrata: a new wrinkle in the study of cell
  locomotion.
\newblock \emph{Science}, 208:\penalty0 177--179, 1980.

\bibitem[Harris et~al.(1984)Harris, Warner, and Stopak]{harris_generation_1984}
Albert~K. Harris, Patricia Warner, and David Stopak.
\newblock Generation of spatially periodic patterns by a mechanical
  instability: a mechanical alternative to the {Turing} model.
\newblock \emph{Development}, 1984.
\newblock Publisher: The Company of Biologists Limited.

\bibitem[Harris et~al.(2014)Harris, Daeden, and Charras]{harris_formation_2014}
Andrew~R. Harris, Alicia Daeden, and Guillaume~T. Charras.
\newblock Formation of adherens junctions leads to the emergence of a
  tissue-level tension in epithelial monolayers.
\newblock \emph{Journal of cell science}, 127\penalty0 (11):\penalty0
  2507--2517, 2014.
\newblock Publisher: The Company of Biologists Bidder Building, 140 Cowley
  Road, Cambridge, CB4 ….

\bibitem[Heer et~al.(2017)Heer, Miller, Chanet, Stoop, Dunkel, and
  Martin]{Heer+Martin.2017.1}
Natalie~C. Heer, Pearson~W. Miller, Soline Chanet, Norbert Stoop, J.~Dunkel,
  and Adam~C. Martin.
\newblock Actomyosin-based tissue folding requires a multicellular myosin
  gradient.
\newblock \emph{Development}, 144:\penalty0 1876--1886, 2017.
\newblock \doi{10.1242/dev.146761}.

\bibitem[Hinch and Harlen(2021)]{Hinch-Harlen.2021.1}
John Hinch and Oliver Harlen.
\newblock Oldroyd b, and not a?
\newblock \emph{Journal of Non-Newtonian Fluid Mechanics}, 298:\penalty0
  104668, 2021.
\newblock \doi{10.1016/j.jnnfm.2021.104668}.

\bibitem[Huxley(1957)]{Huxley.1957.1}
A.~F. Huxley.
\newblock Muscle structure and theories of contraction.
\newblock \emph{Prog. Biophys. Biophys. Chem.}, 7:\penalty0 255--318, 1957.

\bibitem[Ingber et~al.(2014)Ingber, Wang, and
  Stamenovi{\'c}]{ingber2014tensegrity}
Donald~E Ingber, Ning Wang, and Dimitrije Stamenovi{\'c}.
\newblock Tensegrity, cellular biophysics, and the mechanics of living systems.
\newblock \emph{Reports on Progress in Physics}, 77\penalty0 (4):\penalty0
  046603, 2014.

\bibitem[Ingber(1993)]{ingber_cellular_1993}
Donovan~E. Ingber.
\newblock Cellular tensegrity: defining new rules of biological design that
  govern the cytoskeleton.
\newblock \emph{Journal of cell science}, 104\penalty0 (3):\penalty0 613--627,
  1993.
\newblock Publisher: Citeseer.

\bibitem[Izquierdo et~al.(2018)Izquierdo, Quinkler, and
  De~Renzis]{izquierdo_guided_2018}
Emiliano Izquierdo, Theresa Quinkler, and Stefano De~Renzis.
\newblock Guided morphogenesis through optogenetic activation of {Rho}
  signalling during early {Drosophila} embryogenesis.
\newblock \emph{Nature communications}, 9\penalty0 (1):\penalty0 1--13, 2018.
\newblock Publisher: Nature Publishing Group.

\bibitem[Jasnin et~al.(2021)Jasnin, Hervy, Balor, Bouissou, Proag, Voituriez,
  Maridonneau-Parini, Baumeister, Dmitrieff, and
  Poincloux]{jasnin_elasticity_2021}
Marion Jasnin, Jordan Hervy, Stéphanie Balor, Anais Bouissou, Amsha Proag,
  Raphaël Voituriez, Isabelle Maridonneau-Parini, Wolfgang Baumeister, Serge
  Dmitrieff, and Renaud Poincloux.
\newblock Elasticity of dense actin networks produces nanonewton protrusive
  forces.
\newblock \emph{bioRxiv}, 2021.
\newblock Publisher: Cold Spring Harbor Laboratory.

\bibitem[Jensen et~al.(2020)Jensen, Johns, and Woolner]{jensen2020force}
Oliver~E Jensen, Emma Johns, and Sarah Woolner.
\newblock Force networks, torque balance and airy stress in the planar vertex
  model of a confluent epithelium.
\newblock \emph{Proceedings of the Royal Society A}, 476\penalty0
  (2237):\penalty0 20190716, 2020.

\bibitem[John et~al.(2008)John, Peyla, Kassner, Prost, and
  Misbah]{john_nonlinear_2008}
Karin John, Philippe Peyla, Klaus Kassner, Jacques Prost, and Chaouqi Misbah.
\newblock Nonlinear study of symmetry breaking in actin gels: implications for
  cellular motility.
\newblock \emph{Physical review letters}, 100\penalty0 (6):\penalty0 068101,
  2008.
\newblock Publisher: APS.

\bibitem[Jones and Chapman(2012)]{jones2012modeling}
Gareth~Wyn Jones and S~Jonathan Chapman.
\newblock Modeling growth in biological materials.
\newblock \emph{Siam review}, 54\penalty0 (1):\penalty0 52--118, 2012.

\bibitem[J{\"u}licher et~al.(2007)J{\"u}licher, Kruse, Prost, and
  Joanny]{Juelicher+.2007.1}
F.~J{\"u}licher, K.~Kruse, J.~Prost, and J.-F. Joanny.
\newblock Active behavior of the cytoskeleton.
\newblock \emph{Phys. Rep.}, 449:\penalty0 3--28, 2007.

\bibitem[Kedem and Katchalsky(1958)]{kedem1958thermodynamic}
Ora Kedem and Aharon Katchalsky.
\newblock Thermodynamic analysis of the permeability of biological membranes to
  non-electrolytes.
\newblock \emph{Biochimica et biophysica Acta}, 27:\penalty0 229--246, 1958.

\bibitem[Khalilgharibi et~al.(2016)Khalilgharibi, Fouchard, Recho, Charras, and
  Kabla]{khalilgharibi_dynamic_2016}
Nargess Khalilgharibi, Jonathan Fouchard, Pierre Recho, Guillaume Charras, and
  Alexandre Kabla.
\newblock The dynamic mechanical properties of cellularised aggregates.
\newblock \emph{Current Opinion in Cell Biology}, 42:\penalty0 113--120,
  October 2016.
\newblock ISSN 09550674.
\newblock \doi{10.1016/j.ceb.2016.06.003}.
\newblock URL
  \url{https://linkinghub.elsevier.com/retrieve/pii/S0955067416301119}.

\bibitem[Khalilgharibi et~al.(2019)Khalilgharibi, Fouchard, Asadipour,
  Barrientos, Duda, Bonfanti, Yonis, Harris, Mosaffa, Fujita, Kabla, Mao, Baum,
  Mu{\~{n}}oz, Miodownik, and Charras]{Khalilgharibi+Charras.2019.1}
Nargess Khalilgharibi, Jonathan Fouchard, Nina Asadipour, Ricardo Barrientos,
  Maria Duda, Alessandra Bonfanti, Amina Yonis, Andrew Harris, Payman Mosaffa,
  Yasuyuki Fujita, Alexandre Kabla, Yanlan Mao, Buzz Baum, Jos{\'e}~J
  Mu{\~{n}}oz, Mark Miodownik, and Guillaume Charras.
\newblock Stress relaxation in epithelial monolayers is controlled by the
  actomyosin cortex.
\newblock \emph{Nat. Phys.}, 15:\penalty0 839--847, 2019.
\newblock \doi{10.1038/s41567-019-0516-6}.

\bibitem[Koenderink and Paluch(2018)]{koenderink2018architecture}
Gijsje~H Koenderink and Ewa~K Paluch.
\newblock Architecture shapes contractility in actomyosin networks.
\newblock \emph{Current opinion in cell biology}, 50:\penalty0 79--85, 2018.

\bibitem[Krueger et~al.(2018)Krueger, Tardivo, Nguyen, and
  De~Renzis]{krueger_downregulation_2018}
Daniel Krueger, Pietro Tardivo, Congtin Nguyen, and Stefano De~Renzis.
\newblock Downregulation of basal myosin-{II} is required for cell shape
  changes and tissue invagination.
\newblock \emph{The EMBO journal}, 37\penalty0 (23):\penalty0 e100170, 2018.

\bibitem[Kruse et~al.(2006)Kruse, Joanny, J{\"u}licher, and
  Prost]{Kruse+.2006.1}
K.~Kruse, J.-F. Joanny, F.~J{\"u}licher, and J.~Prost.
\newblock Contractility and retrograde flow in lamellipodium motion.
\newblock \emph{Phys. Biol.}, 3:\penalty0 130--137, 2006.

\bibitem[Kumar et~al.(2006)Kumar, Maxwell, Heisterkamp, Polte, Lele, Salanga,
  Mazur, and Ingber]{kumar_viscoelastic_2006}
Sanjay Kumar, Iva~Z. Maxwell, Alexander Heisterkamp, Thomas~R. Polte, Tanmay~P.
  Lele, Matthew Salanga, Eric Mazur, and Donald~E. Ingber.
\newblock Viscoelastic {Retraction} of {Single} {Living} {Stress} {Fibers} and
  {Its} {Impact} on {Cell} {Shape}, {Cytoskeletal} {Organization}, and
  {Extracellular} {Matrix} {Mechanics}.
\newblock \emph{Biophysical Journal}, 90\penalty0 (10):\penalty0 3762--3773,
  May 2006.
\newblock ISSN 0006-3495.
\newblock \doi{10.1529/biophysj.105.071506}.
\newblock URL
  \url{https://www.sciencedirect.com/science/article/pii/S0006349506725565}.

\bibitem[Kupferman et~al.(2020)Kupferman, Maman, and
  Moshe]{kupferman2020continuum}
Raz Kupferman, Ben Maman, and Michael Moshe.
\newblock Continuum mechanics of a cellular tissue model.
\newblock \emph{Journal of the Mechanics and Physics of Solids}, 143:\penalty0
  104085, 2020.

\bibitem[Labernadie et~al.(2014)Labernadie, Bouissou, Delobelle, Balor,
  Voituriez, Proag, Fourquaux, Thibault, Vieu, and
  Poincloux]{labernadie_protrusion_2014}
Anna Labernadie, Anais Bouissou, Patrick Delobelle, Stéphanie Balor, Raphael
  Voituriez, Amsha Proag, Isabelle Fourquaux, Christophe Thibault, Christophe
  Vieu, and Renaud Poincloux.
\newblock Protrusion force microscopy reveals oscillatory force generation and
  mechanosensing activity of human macrophage podosomes.
\newblock \emph{Nature communications}, 5\penalty0 (1):\penalty0 1--10, 2014.
\newblock Publisher: Nature Publishing Group.

\bibitem[Labouesse et~al.(2015)Labouesse, Verkhovsky, Meister, Gabella, and
  Vianay]{labouesse_cell_2015}
Céline Labouesse, Alexander~B. Verkhovsky, Jean-Jacques Meister, Chiara
  Gabella, and Benoît Vianay.
\newblock Cell shape dynamics reveal balance of elasticity and contractility in
  peripheral arcs.
\newblock \emph{Biophysical journal}, 108\penalty0 (10):\penalty0 2437--2447,
  2015.
\newblock Publisher: Elsevier.

\bibitem[Lam et~al.(2011)Lam, Chaudhuri, Crow, Webster, Li, Kita, Huang, and
  Fletcher]{lam_mechanics_2011}
Wilbur~A. Lam, Ovijit Chaudhuri, Ailey Crow, Kevin~D. Webster, Tai-De Li,
  Ashley Kita, James Huang, and Daniel~A. Fletcher.
\newblock Mechanics and contraction dynamics of single platelets and
  implications for clot stiffening.
\newblock \emph{Nature Materials}, 10\penalty0 (1):\penalty0 61--66, January
  2011.
\newblock ISSN 1476-1122, 1476-4660.
\newblock \doi{10.1038/nmat2903}.
\newblock URL \url{http://www.nature.com/articles/nmat2903}.

\bibitem[Laplaud et~al.(2021)Laplaud, Levernier, Pineau, Roman, Barbier,
  S{\'a}ez, Lennon-Dum{\'e}nil, Vargas, Kruse, Du~Roure,
  et~al.]{laplaud2021pinching}
Valentin Laplaud, Nicolas Levernier, Judith Pineau, Mabel~San Roman, Lucie
  Barbier, Pablo~J S{\'a}ez, Ana-Maria Lennon-Dum{\'e}nil, Pablo Vargas,
  Karsten Kruse, Olivia Du~Roure, et~al.
\newblock Pinching the cortex of live cells reveals thickness instabilities
  caused by myosin ii motors.
\newblock \emph{Science Advances}, 7\penalty0 (27):\penalty0 eabe3640, 2021.

\bibitem[Larson(1999)]{Larson.1999.1}
R.~G. Larson.
\newblock \emph{The structure and rheology of complex fluids}.
\newblock Topics Chem. Engng. Oxford Univ. Press, 1999.

\bibitem[Latorre et~al.(2018)Latorre, Kale, Casares, G{\'o}mez-Gonz{\'a}lez,
  Uroz, Valon, Nair, Garreta, Montserrat, Del~Campo, et~al.]{latorre2018active}
Ernest Latorre, Sohan Kale, Laura Casares, Manuel G{\'o}mez-Gonz{\'a}lez,
  Marina Uroz, L{\'e}o Valon, Roshna~V Nair, Elena Garreta, Nuria Montserrat,
  Ar{\'a}nzazu Del~Campo, et~al.
\newblock Active superelasticity in three-dimensional epithelia of controlled
  shape.
\newblock \emph{Nature}, 563\penalty0 (7730):\penalty0 203--208, 2018.

\bibitem[Lawson-Keister and Manning(2021)]{LawsonKeister-Manning.2021.1}
Elizabeth Lawson-Keister and M.~Lisa Manning.
\newblock Jamming and arrest of cell motion in biological tissues.
\newblock \emph{Current Opinion in Cell Biology}, 72:\penalty0 146--155, 2021.
\newblock \doi{10.1016/j.ceb.2021.07.011}.

\bibitem[Lee et~al.(2021)Lee, Holland, Weickenmeier, Gosain, and
  Tepole]{lee2021geometry}
Taeksang Lee, Maria~A Holland, Johannes Weickenmeier, Arun~K Gosain, and
  Adrian~Buganza Tepole.
\newblock The geometry of incompatibility in growing soft tissues: Theory and
  numerical characterization.
\newblock \emph{Journal of the Mechanics and Physics of Solids}, 146:\penalty0
  104177, 2021.

\bibitem[Legant et~al.(2009)Legant, Pathak, Yang, Deshpande, McMeeking, and
  Chen]{legant_microfabricated_2009}
Wesley~R. Legant, Amit Pathak, Michael~T. Yang, Vikram~S. Deshpande, Robert~M.
  McMeeking, and Christopher~S. Chen.
\newblock Microfabricated tissue gauges to measure and manipulate forces from
  {3D} microtissues.
\newblock \emph{Proceedings of the National Academy of Sciences}, 106\penalty0
  (25):\penalty0 10097--10102, June 2009.
\newblock ISSN 0027-8424, 1091-6490.
\newblock \doi{10.1073/pnas.0900174106}.
\newblock URL \url{https://pnas.org/doi/full/10.1073/pnas.0900174106}.

\bibitem[Lenz(2014)]{lenz2014geometrical}
Martin Lenz.
\newblock Geometrical origins of contractility in disordered actomyosin
  networks.
\newblock \emph{Physical Review X}, 4\penalty0 (4):\penalty0 041002, 2014.

\bibitem[Levental et~al.(2007)Levental, Georges, and
  Janmey]{levental_soft_2007}
Ilya Levental, Penelope~C. Georges, and Paul~A. Janmey.
\newblock Soft biological materials and their impact on cell function.
\newblock \emph{Soft Matter}, 3\penalty0 (3):\penalty0 299--306, 2007.
\newblock ISSN 1744-683X, 1744-6848.
\newblock \doi{10.1039/B610522J}.
\newblock URL \url{http://xlink.rsc.org/?DOI=B610522J}.

\bibitem[Li et~al.(2014)Li, Naveed, Kachalo, Xu, and Liang]{Li.2014.1}
Yingzi Li, Hammad Naveed, Sema Kachalo, Lisa~X Xu, and Jie Liang.
\newblock Mechanisms of regulating tissue elongation in drosophila wing: Impact
  of oriented cell divisions, oriented mechanical forces, and reduced cell
  size.
\newblock \emph{PLoS One}, 9\penalty0 (2):\penalty0 e86725, 2014.

\bibitem[Liepelt and Lipowsky(2009)]{Liepelt-Lipowsky.2009.1}
S.~Liepelt and R.~Lipowsky.
\newblock Operation modes of the molecular motor kinesin.
\newblock \emph{Phys. Rev. E}, 79:\penalty0 105, 2009.
\newblock \doi{10.1103/PhysRevE.79.011917}.

\bibitem[Lubarda(2004)]{lubarda2004constitutive}
Vlado~A Lubarda.
\newblock Constitutive theories based on the multiplicative decomposition of
  deformation gradient: Thermoelasticity, elastoplasticity, and biomechanics.
\newblock \emph{Appl. Mech. Rev.}, 57\penalty0 (2):\penalty0 95--108, 2004.

\bibitem[Lubarda and Hoger(2002)]{lubarda2002mechanics}
Vlado~A Lubarda and Anne Hoger.
\newblock On the mechanics of solids with a growing mass.
\newblock \emph{International journal of solids and structures}, 39\penalty0
  (18):\penalty0 4627--4664, 2002.

\bibitem[Luxenburg et~al.(2012)Luxenburg, Winograd-Katz, Addadi, and
  Geiger]{luxenburg_involvement_2012}
Chen Luxenburg, Sabina Winograd-Katz, Lia Addadi, and Benjamin Geiger.
\newblock Involvement of actin polymerization in podosome dynamics.
\newblock \emph{Journal of cell science}, 125\penalty0 (7):\penalty0
  1666--1672, 2012.
\newblock Publisher: Company of Biologists.

\bibitem[Lye et~al.(2015)Lye, Blanchard, Naylor, Muresan, Huisken, Adams, and
  Sanson]{Lye+Sanson.2015.1}
C.~M. Lye, G.~B. Blanchard, H.~Naylor, L.~Muresan, J.~Huisken, R.~Adams, and
  B.~Sanson.
\newblock Mechanical coupling between endoderm invagination and axis extension
  in drosophila.
\newblock \emph{PLoS Biol.}, 13:\penalty0 e1002292, 2015.

\bibitem[Mandal et~al.(2014)Mandal, Wang, Vitiello, Orellana, and
  Balland]{mandal_cell_2014}
Kalpana Mandal, Irène Wang, Elisa Vitiello, Laura Andreina~Chacòn Orellana,
  and Martial Balland.
\newblock Cell dipole behaviour revealed by {ECM} sub-cellular geometry.
\newblock \emph{Nature communications}, 5\penalty0 (1):\penalty0 1--10, 2014.
\newblock Publisher: Nature Publishing Group.

\bibitem[Mangeat et~al.(1999)Mangeat, Roy, and Martin]{mangeat1999erm}
Paul Mangeat, Christian Roy, and Marianne Martin.
\newblock Erm proteins in cell adhesion and membrane dynamics.
\newblock \emph{Trends in cell biology}, 9\penalty0 (5):\penalty0 187--192,
  1999.

\bibitem[Mao et~al.(2013)Mao, Tournier, Hoppe, Kester, Thompson, and
  Tapon]{Mao+Tapon.2013.1}
Y.~Mao, A.~L. Tournier, A.~Hoppe, L.~Kester, B.~J. Thompson, and N.~Tapon.
\newblock Differential proliferation rates generate patterns of mechanical
  tension that orient tissue growth.
\newblock \emph{EMBO J.}, 32:\penalty0 2790--2803, 2013.

\bibitem[Marcy et~al.(2004)Marcy, Prost, Carlier, and Sykes]{marcy_forces_2004}
Yann Marcy, Jacques Prost, Marie-France Carlier, and Cécile Sykes.
\newblock Forces generated during actin-based propulsion: a direct measurement
  by micromanipulation.
\newblock \emph{Proceedings of the National Academy of Sciences}, 101\penalty0
  (16):\penalty0 5992--5997, 2004.
\newblock Publisher: National Acad Sciences.

\bibitem[Martin et~al.(2009)Martin, Kaschube, and Wieschaus]{martin2009pulsed}
Adam~C Martin, Matthias Kaschube, and Eric~F Wieschaus.
\newblock Pulsed contractions of an actin--myosin network drive apical
  constriction.
\newblock \emph{Nature}, 457\penalty0 (7228):\penalty0 495--499, 2009.

\bibitem[Masic et~al.(2015)Masic, Bertinetti, Schuetz, Chang, Metzger, Buehler,
  and Fratzl]{masic2015osmotic}
Admir Masic, Luca Bertinetti, Roman Schuetz, Shu-Wei Chang, Till~Hartmut
  Metzger, Markus~J Buehler, and Peter Fratzl.
\newblock Osmotic pressure induced tensile forces in tendon collagen.
\newblock \emph{Nature communications}, 6\penalty0 (1):\penalty0 1--8, 2015.

\bibitem[Michaux et~al.(2018)Michaux, Robin, McFadden, and
  Munro]{michaux2018excitable}
Jonathan~B Michaux, Fran{\c{c}}ois~B Robin, William~M McFadden, and Edwin~M
  Munro.
\newblock Excitable rhoa dynamics drive pulsed contractions in the early c.
  elegans embryo.
\newblock \emph{Journal of Cell Biology}, 217\penalty0 (12):\penalty0
  4230--4252, 2018.

\bibitem[Mitchison and Cramer(1996)]{Mitchison-Cramer.1996.1}
T.~J. Mitchison and L.~P. Cramer.
\newblock Actin-based cell motility and cell locomotion.
\newblock \emph{Cell}, 84:\penalty0 371--379, 1996.

\bibitem[Mitrossilis et~al.(2009)Mitrossilis, Fouchard, Guiroy, Desprat,
  Rodriguez, Fabry, and Asnacios]{mitrossilis_single-cell_2009}
Démosthène Mitrossilis, Jonathan Fouchard, Axel Guiroy, Nicolas Desprat,
  Nicolas Rodriguez, Ben Fabry, and Atef Asnacios.
\newblock Single-cell response to stiffness exhibits muscle-like behavior.
\newblock \emph{Proceedings of the National Academy of Sciences}, 106\penalty0
  (43):\penalty0 18243--18248, October 2009.
\newblock ISSN 0027-8424, 1091-6490.
\newblock \doi{10.1073/pnas.0903994106}.
\newblock URL \url{https://pnas.org/doi/full/10.1073/pnas.0903994106}.

\bibitem[Mitrossilis et~al.(2010)Mitrossilis, Fouchard, Pereira, Postic,
  Richert, Saint-Jean, and Asnacios]{mitrossilis_real-time_2010}
Démosthène Mitrossilis, Jonathan Fouchard, David Pereira, François Postic,
  Alain Richert, Michel Saint-Jean, and Atef Asnacios.
\newblock Real-time single-cell response to stiffness.
\newblock \emph{Proceedings of the National Academy of Sciences}, 107\penalty0
  (38):\penalty0 16518--16523, September 2010.
\newblock ISSN 0027-8424, 1091-6490.
\newblock \doi{10.1073/pnas.1007940107}.
\newblock URL \url{https://pnas.org/doi/full/10.1073/pnas.1007940107}.

\bibitem[Mizuno et~al.(2007)Mizuno, Tardin, Schmidt, and
  MacKintosh]{Mizuno+MacKintosh.2007.1}
D.~Mizuno, C.~Tardin, C.F. Schmidt, and F.C. MacKintosh.
\newblock Nonequilibrium mechanics of active cytoskeletal networks.
\newblock \emph{Science}, 315:\penalty0 370--373, 2007.

\bibitem[Moeendarbary et~al.(2013)Moeendarbary, Valon, Fritzsche, Harris,
  Moulding, Thrasher, Stride, Mahadevan, and
  Charras]{moeendarbary_cytoplasm_2013}
Emad Moeendarbary, Léo Valon, Marco Fritzsche, Andrew~R. Harris, Dale~A.
  Moulding, Adrian~J. Thrasher, Eleanor Stride, L.~Mahadevan, and Guillaume~T.
  Charras.
\newblock The cytoplasm of living cells behaves as a poroelastic material.
\newblock \emph{Nature materials}, 12\penalty0 (3):\penalty0 253--261, 2013.
\newblock Publisher: Nature Publishing Group.

\bibitem[Moisdon et~al.(2022)Moisdon, Seez, No{\^u}s, Molino, Marcq, and
  Gay]{moisdon2022mapping}
{\'E}tienne Moisdon, Pierre Seez, Camille No{\^u}s, Fran{\c{c}}ois Molino,
  Philippe Marcq, and Cyprien Gay.
\newblock Mapping cell cortex rheology to tissue rheology, and vice-versa.
\newblock \emph{arXiv preprint arXiv:2204.10907}, 2022.

\bibitem[Mongera et~al.(2018)Mongera, Rowghanian, Gustafson, Shelton,
  Kealhofer, Carn, Serwane, Lucio, Giammona, and Camp{\`a}s]{mongera2018fluid}
Alessandro Mongera, Payam Rowghanian, Hannah~J Gustafson, Elijah Shelton,
  David~A Kealhofer, Emmet~K Carn, Friedhelm Serwane, Adam~A Lucio, James
  Giammona, and Otger Camp{\`a}s.
\newblock A fluid-to-solid jamming transition underlies vertebrate body axis
  elongation.
\newblock \emph{Nature}, 561\penalty0 (7723):\penalty0 401--405, 2018.

\bibitem[Montel et~al.(2011)Montel, Delarue, Elgeti, Malaquin, Basan, Risler,
  Cabane, Vignjevic, Prost, Cappello, et~al.]{montel2011stress}
Fabien Montel, Morgan Delarue, Jens Elgeti, Laurent Malaquin, Markus Basan,
  Thomas Risler, Bernard Cabane, Danijela Vignjevic, Jacques Prost, Giovanni
  Cappello, et~al.
\newblock Stress clamp experiments on multicellular tumor spheroids.
\newblock \emph{Physical review letters}, 107\penalty0 (18):\penalty0 188102,
  2011.

\bibitem[Mueller et~al.(2017)Mueller, Szep, Nemethova, De~Vries, Lieber,
  Winkler, Kruse, Small, Schmeiser, and Keren]{mueller_load_2017}
Jan Mueller, Gregory Szep, Maria Nemethova, Ingrid De~Vries, Arnon~D. Lieber,
  Christoph Winkler, Karsten Kruse, J.~Victor Small, Christian Schmeiser, and
  Kinneret Keren.
\newblock Load adaptation of lamellipodial actin networks.
\newblock \emph{Cell}, 171\penalty0 (1):\penalty0 188--200, 2017.
\newblock Publisher: Elsevier.

\bibitem[Munjal et~al.(2021)Munjal, Hannezo, Tsai, Mitchison, and
  Megason]{munjal_extracellular_2021}
Akankshi Munjal, Edouard Hannezo, Tony Y.-C. Tsai, Timothy~J. Mitchison, and
  Sean~G. Megason.
\newblock Extracellular hyaluronate pressure shaped by cellular tethers drives
  tissue morphogenesis.
\newblock \emph{Cell}, 184\penalty0 (26):\penalty0 6313--6325, 2021.
\newblock Publisher: Elsevier.

\bibitem[Murisic et~al.(2015)Murisic, Hakim, Kevrekidis, Shvartsman, and
  Audoly]{murisic2015discrete}
Nebojsa Murisic, Vincent Hakim, Ioannis~G Kevrekidis, Stanislav~Y Shvartsman,
  and Basile Audoly.
\newblock From discrete to continuum models of three-dimensional deformations
  in epithelial sheets.
\newblock \emph{Biophysical journal}, 109\penalty0 (1):\penalty0 154--163,
  2015.

\bibitem[Nestor-Bergmann et~al.(2018)Nestor-Bergmann, Johns, Woolner, and
  Jensen]{nestor2018PRE}
Alexander Nestor-Bergmann, Emma Johns, Sarah Woolner, and Oliver~E Jensen.
\newblock Mechanical characterization of disordered and anisotropic cellular
  monolayers.
\newblock \emph{Physical Review E}, 97\penalty0 (5):\penalty0 052409, 2018.

\bibitem[Nestor-Bergmann et~al.(2022)Nestor-Bergmann, Blanchard, Hervieux,
  Fletcher, {\'E}tienne, and Sanson]{nestor2022adhesion}
Alexander Nestor-Bergmann, Guy~B Blanchard, Nathan Hervieux, Alexander~G
  Fletcher, Jocelyn {\'E}tienne, and B{\'e}n{\'e}dicte Sanson.
\newblock Adhesion-regulated junction slippage controls cell intercalation
  dynamics in an apposed-cortex adhesion model.
\newblock \emph{PLoS computational biology}, 18\penalty0 (1):\penalty0
  e1009812, 2022.

\bibitem[Nishikawa et~al.(2017)Nishikawa, Naganathan, J{\"u}licher, and
  Grill]{Nishikawa+Grill.2017.1}
Masatoshi Nishikawa, Sundar~Ram Naganathan, Frank J{\"u}licher, and Stephan~W
  Grill.
\newblock Controlling contractile instabilities in the actomyosin cortex.
\newblock \emph{eLife}, 6:\penalty0 058101, 2017.
\newblock \doi{10.7554/eLife.19595}.

\bibitem[Niwayama et~al.(2019)Niwayama, Moghe, Liu, Fabrèges, Buchholz, Piel,
  and Hiiragi]{Niwayama.2019.1}
Ritsuya Niwayama, Prachiti Moghe, Yan-Jun Liu, Dimitri Fabrèges, Frank
  Buchholz, Matthieu Piel, and Takashi Hiiragi.
\newblock A tug-of-war between cell shape and polarity controls division
  orientation to ensure robust patterning in the mouse blastocyst.
\newblock \emph{Developmental Cell}, 51\penalty0 (5):\penalty0 564--574.e6,
  2019.
\newblock ISSN 1534-5807.
\newblock \doi{https://doi.org/10.1016/j.devcel.2019.10.012}.
\newblock URL
  \url{https://www.sciencedirect.com/science/article/pii/S1534580719308561}.

\bibitem[Oltean et~al.(2016)Oltean, Huang, Beebe, and Taber]{oltean2016tissue}
Alina Oltean, Jie Huang, David~C Beebe, and Larry~A Taber.
\newblock Tissue growth constrained by extracellular matrix drives invagination
  during optic cup morphogenesis.
\newblock \emph{Biomechanics and modeling in mechanobiology}, 15\penalty0
  (6):\penalty0 1405--1421, 2016.

\bibitem[Omens and Fung(1990)]{omens1990residual}
Jeffrey~H Omens and Yuan-Cheng Fung.
\newblock Residual strain in rat left ventricle.
\newblock \emph{Circulation research}, 66\penalty0 (1):\penalty0 37--45, 1990.

\bibitem[O’Byrne et~al.(2022)O’Byrne, Kafri, Tailleur, and van
  Wijland]{o2022time}
J{\'e}r{\'e}my O’Byrne, Yariv Kafri, Julien Tailleur, and Fr{\'e}d{\'e}ric
  van Wijland.
\newblock Time irreversibility in active matter, from micro to macro.
\newblock \emph{Nature Reviews Physics}, pages 1--17, 2022.

\bibitem[Parnell(2012)]{parnell2012nonlinear}
William~J Parnell.
\newblock Nonlinear pre-stress for cloaking from antiplane elastic waves.
\newblock \emph{Proceedings of the Royal Society A: Mathematical, Physical and
  Engineering Sciences}, 468\penalty0 (2138):\penalty0 563--580, 2012.

\bibitem[Pasakarnis et~al.(2016)Pasakarnis, Frei, Caussinus, Affolter, and
  Brunner]{Pasakarnis.2016.1}
Laurynas Pasakarnis, Erich Frei, Emmanuel Caussinus, Markus Affolter, and
  Damian Brunner.
\newblock Amnioserosa cell constriction but not epidermal actin cable tension
  autonomously drives dorsal closure.
\newblock \emph{Nature cell biology}, 18\penalty0 (11):\penalty0 1161--1172,
  2016.

\bibitem[Pearl et~al.(2017)Pearl, Li, and Green]{pearl_cellular_2017}
Esther~J. Pearl, Jingjing Li, and Jeremy B.~A. Green.
\newblock Cellular systems for epithelial invagination.
\newblock \emph{Philosophical Transactions of the Royal Society B: Biological
  Sciences}, 372\penalty0 (1720):\penalty0 20150526, May 2017.
\newblock ISSN 0962-8436, 1471-2970.
\newblock \doi{10.1098/rstb.2015.0526}.
\newblock URL
  \url{https://royalsocietypublishing.org/doi/10.1098/rstb.2015.0526}.

\bibitem[Pelham and Wang(1997)]{pelham_cell_1997}
Robert~J. Pelham and Yu-li Wang.
\newblock Cell locomotion and focal adhesions are regulated by substrate
  flexibility.
\newblock \emph{Proceedings of the National Academy of Sciences}, 94\penalty0
  (25):\penalty0 13661--13665, December 1997.
\newblock ISSN 0027-8424, 1091-6490.
\newblock \doi{10.1073/pnas.94.25.13661}.
\newblock URL \url{https://pnas.org/doi/full/10.1073/pnas.94.25.13661}.

\bibitem[P{\'e}rez-Gonz{\'a}lez et~al.(2021)P{\'e}rez-Gonz{\'a}lez, Ceada,
  Greco, Matej{\v{c}}i{\'c}, G{\'o}mez-Gonz{\'a}lez, Castro, Menendez, Kale,
  Krndija, Clark, et~al.]{perez2021mechanical}
Carlos P{\'e}rez-Gonz{\'a}lez, Gerardo Ceada, Francesco Greco, Marija
  Matej{\v{c}}i{\'c}, Manuel G{\'o}mez-Gonz{\'a}lez, Natalia Castro, Anghara
  Menendez, Sohan Kale, Denis Krndija, Andrew~G Clark, et~al.
\newblock Mechanical compartmentalization of the intestinal organoid enables
  crypt folding and collective cell migration.
\newblock \emph{Nature cell biology}, 23\penalty0 (7):\penalty0 745--757, 2021.

\bibitem[Polacheck and Chen(2016)]{polacheck2016measuring}
William~J Polacheck and Christopher~S Chen.
\newblock Measuring cell-generated forces: a guide to the available tools.
\newblock \emph{Nature methods}, 13\penalty0 (5):\penalty0 415--423, 2016.

\bibitem[Pollard(2016)]{pollard_actin_2016}
Thomas~D. Pollard.
\newblock Actin and actin-binding proteins.
\newblock \emph{Cold Spring Harbor perspectives in biology}, 8\penalty0
  (8):\penalty0 a018226, 2016.
\newblock Publisher: Cold Spring Harbor Lab.

\bibitem[Provenzano et~al.(2012)Provenzano, Cuevas, Chang, Goel, Von~Hoff, and
  Hingorani]{provenzano_enzymatic_2012}
Paolo~P. Provenzano, Carlos Cuevas, Amy~E. Chang, Vikas~K. Goel, Daniel~D.
  Von~Hoff, and Sunil~R. Hingorani.
\newblock Enzymatic targeting of the stroma ablates physical barriers to
  treatment of pancreatic ductal adenocarcinoma.
\newblock \emph{Cancer cell}, 21\penalty0 (3):\penalty0 418--429, 2012.
\newblock Publisher: Elsevier.

\bibitem[Putelat et~al.(2018)Putelat, Recho, and
  Truskinovsky]{Putelat+Truskinovsky.2018.1}
T.~Putelat, P.~Recho, and L.~Truskinovsky.
\newblock Mechanical stress as a regulator of cell motility.
\newblock \emph{Phys. Rev. E}, 97, 2018.
\newblock \doi{10.1103/PhysRevE.97.012410}.

\bibitem[Rape et~al.(2011)Rape, Guo, and Wang]{rape_microtubule_2011}
Andrew Rape, Wei-hui Guo, and Yu-li Wang.
\newblock Microtubule depolymerization induces traction force increase through
  two distinct pathways.
\newblock \emph{Journal of Cell Science}, 124\penalty0 (24):\penalty0
  4233--4240, December 2011.
\newblock ISSN 0021-9533.
\newblock \doi{10.1242/jcs.090563}.
\newblock URL \url{https://doi.org/10.1242/jcs.090563}.

\bibitem[Recho and Truskinovsky(2013)]{Recho-Truskinovsky.2013.1}
P.~Recho and L.~Truskinovsky.
\newblock An asymmetry between pushing and pulling for crawling cells.
\newblock \emph{Phys. Rev. E}, 87:\penalty0 022720, 2013.

\bibitem[Recho et~al.(2013)Recho, Putelat, and Truskinovsky]{Recho+.2013.1}
P.~Recho, T.~Putelat, and L.~Truskinovsky.
\newblock Contraction-driven cell motility.
\newblock \emph{Phys. Rev. Lett.}, 111:\penalty0 108102, 2013.

\bibitem[Recho et~al.(2020)Recho, Fouchard, Wyatt, Khalilgharibi, Charras, and
  Kabla]{recho_tug--war_2020}
P.~Recho, J.~Fouchard, T.~Wyatt, N.~Khalilgharibi, G.~Charras, and A.~Kabla.
\newblock Tug-of-war between stretching and bending in living cell sheets.
\newblock \emph{Physical Review E}, 102\penalty0 (1):\penalty0 012401, July
  2020.
\newblock ISSN 2470-0045, 2470-0053.
\newblock \doi{10.1103/PhysRevE.102.012401}.
\newblock URL \url{https://link.aps.org/doi/10.1103/PhysRevE.102.012401}.

\bibitem[Reina and Conti(2014)]{reina2014kinematic}
Celia Reina and Sergio Conti.
\newblock Kinematic description of crystal plasticity in the finite kinematic
  framework: A micromechanical understanding of {F= FeFp}.
\newblock \emph{Journal of the Mechanics and Physics of Solids}, 67:\penalty0
  40--61, 2014.

\bibitem[Reina et~al.(2016)Reina, Schlömerkemper, and
  Conti]{reina2016derivation}
Celia Reina, Anja Schlömerkemper, and Sergio Conti.
\newblock Derivation of {F= FeFp} as the continuum limit of crystalline slip.
\newblock \emph{Journal of the Mechanics and Physics of Solids}, 89:\penalty0
  231--254, 2016.

\bibitem[Robert et~al.(2010)Robert, Nguyen, Gallet, and
  Wilhelm]{robert2010vivo}
Damien Robert, Thi-Hanh Nguyen, Fran{\c{c}}ois Gallet, and Claire Wilhelm.
\newblock In vivo determination of fluctuating forces during endosome
  trafficking using a combination of active and passive microrheology.
\newblock \emph{PloS one}, 5\penalty0 (4):\penalty0 e10046, 2010.

\bibitem[Rodriguez et~al.(1994)Rodriguez, Hoger, and
  McCulloch]{rodriguez1994stress}
Edward~K Rodriguez, Anne Hoger, and Andrew~D McCulloch.
\newblock Stress-dependent finite growth in soft elastic tissues.
\newblock \emph{Journal of biomechanics}, 27\penalty0 (4):\penalty0 455--467,
  1994.

\bibitem[Ronceray et~al.(2016)Ronceray, Broedersz, and
  Lenz]{ronceray_fiber_2016}
Pierre Ronceray, Chase~P. Broedersz, and Martin Lenz.
\newblock Fiber networks amplify active stress.
\newblock \emph{Proceedings of the national academy of sciences}, 113\penalty0
  (11):\penalty0 2827--2832, 2016.
\newblock Publisher: National Acad Sciences.

\bibitem[Saha et~al.(2016)Saha, Nishikawa, Behrndt, Heisenberg, J{\"u}licher,
  and Grill]{saha2016determining}
Arnab Saha, Masatoshi Nishikawa, Martin Behrndt, Carl-Philipp Heisenberg, Frank
  J{\"u}licher, and Stephan~W Grill.
\newblock Determining physical properties of the cell cortex.
\newblock \emph{Biophysical journal}, 110\penalty0 (6):\penalty0 1421--1429,
  2016.

\bibitem[Sahai et~al.(2020)Sahai, Astsaturov, Cukierman, DeNardo, Egeblad,
  Evans, Fearon, Greten, Hingorani, and Hunter]{sahai_framework_2020}
Erik Sahai, Igor Astsaturov, Edna Cukierman, David~G. DeNardo, Mikala Egeblad,
  Ronald~M. Evans, Douglas Fearon, Florian~R. Greten, Sunil~R. Hingorani, and
  Tony Hunter.
\newblock A framework for advancing our understanding of cancer-associated
  fibroblasts.
\newblock \emph{Nature Reviews Cancer}, 20\penalty0 (3):\penalty0 174--186,
  2020.
\newblock Publisher: Nature Publishing Group.

\bibitem[Salbreux et~al.(2012)Salbreux, Charras, and
  Paluch]{salbreux_actin_2012}
Guillaume Salbreux, Guillaume Charras, and Ewa Paluch.
\newblock Actin cortex mechanics and cellular morphogenesis.
\newblock \emph{Trends in cell biology}, 22\penalty0 (10):\penalty0 536--545,
  2012.
\newblock Publisher: Elsevier.

\bibitem[Salen{\c{c}}on(1994)]{salenccon1994mecanique}
Jean Salen{\c{c}}on.
\newblock \emph{M{\'e}canique des milieux continus. Tome II,
  Thermo{\'e}lasticit{\'e}}.
\newblock {\'E}cole polytechnique, 1994.

\bibitem[Savin et~al.(2011)Savin, Kurpios, Shyer, Florescu, Liang, Mahadevan,
  and Tabin]{Savin.2011.1}
Thierry Savin, Natasza~A Kurpios, Amy~E Shyer, Patricia Florescu, Haiyi Liang,
  L~Mahadevan, and Clifford~J Tabin.
\newblock On the growth and form of the gut.
\newblock \emph{Nature}, 476\penalty0 (7358):\penalty0 57--62, 2011.

\bibitem[Scarpa et~al.(2018)Scarpa, Finet, Blanchard, and
  Sanson]{Scarpa.2018.1}
Elena Scarpa, C{\'e}dric Finet, Guy~B Blanchard, and B{\'e}n{\'e}dicte Sanson.
\newblock Actomyosin-driven tension at compartmental boundaries orients cell
  division independently of cell geometry in vivo.
\newblock \emph{Developmental cell}, 47\penalty0 (6):\penalty0 727--740, 2018.

\bibitem[Schoetz et~al.(2013)Schoetz, Lanio, Talbot, and
  Manning]{schoetz2013glassy}
Eva-Maria Schoetz, Marcos Lanio, Jared~A Talbot, and M~Lisa Manning.
\newblock Glassy dynamics in three-dimensional embryonic tissues.
\newblock \emph{Journal of The Royal Society Interface}, 10\penalty0
  (89):\penalty0 20130726, 2013.

\bibitem[Shyer et~al.(2013)Shyer, Tallinen, Nerurkar, Wei, Gil, Kaplan, Tabin,
  and Mahadevan]{Shyer.2013.1}
Amy~E Shyer, Tuomas Tallinen, Nandan~L Nerurkar, Zhiyan Wei, Eun~Seok Gil,
  David~L Kaplan, Clifford~J Tabin, and L~Mahadevan.
\newblock Villification: how the gut gets its villi.
\newblock \emph{Science}, 342\penalty0 (6155):\penalty0 212--218, 2013.

\bibitem[Shyer et~al.(2017)Shyer, Rodrigues, Schroeder, Kassianidou, Kumar, and
  Harland]{shyer_emergent_2017}
Amy~E. Shyer, Alan~R. Rodrigues, Grant~G. Schroeder, Elena Kassianidou, Sanjay
  Kumar, and Richard~M. Harland.
\newblock Emergent cellular self-organization and mechanosensation initiate
  follicle pattern in the avian skin.
\newblock \emph{Science}, 357\penalty0 (6353):\penalty0 811--815, 2017.
\newblock Publisher: American Association for the Advancement of Science.

\bibitem[Sidhaye and Norden(2017)]{sidhaye_concerted_2017}
Jaydeep Sidhaye and Caren Norden.
\newblock Concerted action of neuroepithelial basal shrinkage and active
  epithelial migration ensures efficient optic cup morphogenesis.
\newblock \emph{elife}, 6:\penalty0 e22689, 2017.
\newblock Publisher: eLife Sciences Publications Limited.

\bibitem[Skalak et~al.(1997)Skalak, Farrow, and Hoger]{skalak1997kinematics}
R~Skalak, DA~Farrow, and A14782930883 Hoger.
\newblock Kinematics of surface growth.
\newblock \emph{Journal of mathematical biology}, 35\penalty0 (8):\penalty0
  869--907, 1997.

\bibitem[Stamenovi{\'c} et~al.(2002)Stamenovi{\'c}, Mijailovich,
  Toli{\'c}-N{\o}rrelykke, and Wang]{Stamenovic+Wang.2002.1}
D.~Stamenovi{\'c}, S.~M. Mijailovich, I.~M. Toli{\'c}-N{\o}rrelykke, and
  N.~Wang.
\newblock Cell prestress. {II}. {C}ontribution of microtubules.
\newblock \emph{Am. J. Physiol. Cell Physiol.}, 282:\penalty0 C617--C624, 2002.

\bibitem[Stylianopoulos et~al.(2012)Stylianopoulos, Martin, Chauhan, Jain,
  Diop-Frimpong, Bardeesy, Smith, Ferrone, Hornicek, Boucher,
  et~al.]{stylianopoulos2012causes}
Triantafyllos Stylianopoulos, John~D Martin, Vikash~P Chauhan, Saloni~R Jain,
  Benjamin Diop-Frimpong, Nabeel Bardeesy, Barbara~L Smith, Cristina~R Ferrone,
  Francis~J Hornicek, Yves Boucher, et~al.
\newblock Causes, consequences, and remedies for growth-induced solid stress in
  murine and human tumors.
\newblock \emph{Proceedings of the National Academy of Sciences}, 109\penalty0
  (38):\penalty0 15101--15108, 2012.

\bibitem[Szab{\'o} et~al.(2016)Szab{\'o}, Cobo, Omara, McLachlan, Keller, and
  Mayor]{Szabo.2016.1}
Andr{\'a}s Szab{\'o}, Isidoro Cobo, Sharif Omara, Sophie McLachlan, Ray Keller,
  and Roberto Mayor.
\newblock The molecular basis of radial intercalation during tissue spreading
  in early development.
\newblock \emph{Developmental cell}, 37\penalty0 (3):\penalty0 213--225, 2016.

\bibitem[Taber(1995)]{taber1995biomechanics}
Larry~A. Taber.
\newblock {Biomechanics of Growth, Remodeling, and Morphogenesis}.
\newblock \emph{Applied Mechanics Reviews}, 48\penalty0 (8):\penalty0 487--545,
  08 1995.
\newblock ISSN 0003-6900.
\newblock \doi{10.1115/1.3005109}.
\newblock URL \url{https://doi.org/10.1115/1.3005109}.

\bibitem[Taber(2009)]{taber2009towards}
Larry~A Taber.
\newblock Towards a unified theory for morphomechanics.
\newblock \emph{Philosophical Transactions of the Royal Society A:
  Mathematical, Physical and Engineering Sciences}, 367\penalty0
  (1902):\penalty0 3555--3583, 2009.

\bibitem[Tallinen et~al.(2016)Tallinen, Chung, Rousseau, Girard, Lef{\`e}vre,
  and Mahadevan]{Tallinen.2016.1}
Tuomas Tallinen, Jun~Young Chung, Fran{\c{c}}ois Rousseau, Nadine Girard,
  Julien Lef{\`e}vre, and Lakshminarayanan Mahadevan.
\newblock On the growth and form of cortical convolutions.
\newblock \emph{Nature Physics}, 12\penalty0 (6):\penalty0 588--593, 2016.

\bibitem[Tan et~al.(2003)Tan, Tien, Pirone, Gray, Bhadriraju, and
  Chen]{tan_cells_2003}
John~L. Tan, Joe Tien, Dana~M. Pirone, Darren~S. Gray, Kiran Bhadriraju, and
  Christopher~S. Chen.
\newblock Cells lying on a bed of microneedles: {An} approach to isolate
  mechanical force.
\newblock \emph{Proceedings of the National Academy of Sciences}, 100\penalty0
  (4):\penalty0 1484--1489, February 2003.
\newblock ISSN 0027-8424, 1091-6490.
\newblock \doi{10.1073/pnas.0235407100}.
\newblock URL \url{https://pnas.org/doi/full/10.1073/pnas.0235407100}.

\bibitem[Tetley et~al.(2016)Tetley, Blanchard, Fletcher, Adams, and
  Sanson]{Tetley-Blanchard+Sanson.2016.1}
R.~J. Tetley, G.~B. Blanchard, A.~G. Fletcher, R.~J. Adams, and B.~Sanson.
\newblock Unipolar distributions of junctional myosin ii identify cell stripe
  boundaries that drive cell intercalation throughout drosophila axis
  extension.
\newblock \emph{eLife}, 5:\penalty0 e12094, 2016.

\bibitem[Theriot et~al.(1992)Theriot, Mitchison, Tilney, and
  Portnoy]{theriot_rate_1992}
Julie~A. Theriot, Timothy~J. Mitchison, Lewis~G. Tilney, and Daniel~A. Portnoy.
\newblock The rate of actin-based motility of intracellular {Listeria}
  monocytogenes equals the rate of actin polymerization.
\newblock \emph{Nature}, 357\penalty0 (6375):\penalty0 257--260, 1992.
\newblock Publisher: Nature Publishing Group.

\bibitem[Tinevez et~al.(2009)Tinevez, Schulze, Salbreux, Roensch, Joanny, and
  Paluch]{tinevez_role_2009}
Jean-Yves Tinevez, Ulrike Schulze, Guillaume Salbreux, Julia Roensch,
  Jean-François Joanny, and Ewa Paluch.
\newblock Role of cortical tension in bleb growth.
\newblock \emph{Proceedings of the National Academy of Sciences}, 106\penalty0
  (44):\penalty0 18581--18586, 2009.
\newblock Publisher: National Acad Sciences.

\bibitem[Tlili et~al.(2020)Tlili, Durande, Gay, Ladoux, Graner, and
  Delanoë-Ayari]{tlili_migrating_2020}
S.~Tlili, M.~Durande, Cyprien Gay, B.~Ladoux, F.~Graner, and H.~Delanoë-Ayari.
\newblock Migrating epithelial monolayer flows like a {Maxwell} viscoelastic
  liquid.
\newblock \emph{Physical Review Letters}, 125\penalty0 (8):\penalty0 088102,
  2020.
\newblock Publisher: APS.

\bibitem[Tofangchi et~al.(2016)Tofangchi, Fan, and
  Saif]{tofangchi2016mechanism}
Alireza Tofangchi, Anthony Fan, and M~Taher~A Saif.
\newblock Mechanism of axonal contractility in embryonic drosophila motor
  neurons in vivo.
\newblock \emph{Biophysical journal}, 111\penalty0 (7):\penalty0 1519--1527,
  2016.

\bibitem[Tojkander et~al.(2012)Tojkander, Gateva, and
  Lappalainen]{tojkander_actin_2012}
Sari Tojkander, Gergana Gateva, and Pekka Lappalainen.
\newblock Actin stress fibers–assembly, dynamics and biological roles.
\newblock \emph{Journal of cell science}, 125\penalty0 (8):\penalty0
  1855--1864, 2012.
\newblock Publisher: Company of Biologists.

\bibitem[Tozluoǧlu and Mao(2020)]{tozluoglu_folding_2020}
Melda Tozluoǧlu and Yanlan Mao.
\newblock On folding morphogenesis, a mechanical problem.
\newblock \emph{Philosophical Transactions of the Royal Society B: Biological
  Sciences}, 375\penalty0 (1809):\penalty0 20190564, October 2020.
\newblock \doi{10.1098/rstb.2019.0564}.
\newblock URL
  \url{https://royalsocietypublishing.org/doi/full/10.1098/rstb.2019.0564}.
\newblock Publisher: Royal Society.

\bibitem[Tozluoǧlu et~al.(2019)Tozluoǧlu, Duda, Kirkland, Barrientos, Burden,
  Muñoz, and Mao]{tozluoglu_planar_2019}
Melda Tozluoǧlu, Maria Duda, Natalie~J. Kirkland, Ricardo Barrientos,
  Jemima~J. Burden, José~J. Muñoz, and Yanlan Mao.
\newblock Planar differential growth rates initiate precise fold positions in
  complex epithelia.
\newblock \emph{Developmental cell}, 51\penalty0 (3):\penalty0 299--312, 2019.
\newblock Publisher: Elsevier.

\bibitem[Truskinovsky and Zurlo(2019)]{truskinovsky2019nonlinear}
Lev Truskinovsky and Giuseppe Zurlo.
\newblock Nonlinear elasticity of incompatible surface growth.
\newblock \emph{Physical Review E}, 99\penalty0 (5):\penalty0 053001, 2019.

\bibitem[Vaishnav and Vossoughi(1987)]{vaishnav1987residual}
Ramesh~N Vaishnav and Jafar Vossoughi.
\newblock Residual stress and strain in aortic segments.
\newblock \emph{Journal of biomechanics}, 20\penalty0 (3):\penalty0 235--239,
  1987.

\bibitem[van~den Dries et~al.(2019)van~den Dries, Linder, Maridonneau-Parini,
  and Poincloux]{van_den_dries_probing_2019}
Koen van~den Dries, Stefan Linder, Isabelle Maridonneau-Parini, and Renaud
  Poincloux.
\newblock Probing the mechanical landscape–new insights into podosome
  architecture and mechanics.
\newblock \emph{Journal of Cell Science}, 132\penalty0 (24):\penalty0
  jcs236828, 2019.
\newblock Publisher: The Company of Biologists Ltd.

\bibitem[Van~Helvert et~al.(2018)Van~Helvert, Storm, and
  Friedl]{van2018mechanoreciprocity}
Sjoerd Van~Helvert, Cornelis Storm, and Peter Friedl.
\newblock Mechanoreciprocity in cell migration.
\newblock \emph{Nature cell biology}, 20\penalty0 (1):\penalty0 8--20, 2018.

\bibitem[Vignaud et~al.(2021)Vignaud, Copos, Leterrier, Toro-Nahuelpan, Tseng,
  Mahamid, Blanchoin, Mogilner, Théry, and Kurzawa]{vignaud_stress_2021}
Timothée Vignaud, Calina Copos, Christophe Leterrier, Mauricio Toro-Nahuelpan,
  Qingzong Tseng, Julia Mahamid, Laurent Blanchoin, Alex Mogilner, Manuel
  Théry, and Laetitia Kurzawa.
\newblock Stress fibres are embedded in a contractile cortical network.
\newblock \emph{Nature materials}, 20\penalty0 (3):\penalty0 410--420, 2021.
\newblock Publisher: Nature Publishing Group.

\bibitem[Wershof et~al.(2019)Wershof, Park, Jenkins, Barry, Sahai, and
  Bates]{wershof_matrix_2019}
Esther Wershof, Danielle Park, Robert~P. Jenkins, David~J. Barry, Erik Sahai,
  and Paul~A. Bates.
\newblock Matrix feedback enables diverse higher-order patterning of the
  extracellular matrix.
\newblock \emph{PLOS Computational Biology}, 15\penalty0 (10):\penalty0
  e1007251, October 2019.
\newblock ISSN 1553-7358.
\newblock \doi{10.1371/journal.pcbi.1007251}.
\newblock URL \url{https://dx.plos.org/10.1371/journal.pcbi.1007251}.

\bibitem[Wyatt et~al.(2015)Wyatt, Harris, Lam, Cheng, Bellis, Dimitracopoulos,
  Kabla, Charras, and Baum]{wyatt_emergence_2015}
Tom P.~J. Wyatt, Andrew~R. Harris, Maxine Lam, Qian Cheng, Julien Bellis,
  Andrea Dimitracopoulos, Alexandre~J. Kabla, Guillaume~T. Charras, and Buzz
  Baum.
\newblock Emergence of homeostatic epithelial packing and stress dissipation
  through divisions oriented along the long cell axis.
\newblock \emph{Proceedings of the National Academy of Sciences}, 112\penalty0
  (18):\penalty0 5726--5731, May 2015.
\newblock ISSN 0027-8424, 1091-6490.
\newblock \doi{10.1073/pnas.1420585112}.
\newblock URL \url{https://pnas.org/doi/full/10.1073/pnas.1420585112}.

\bibitem[Wyatt et~al.(2020)Wyatt, Fouchard, Lisica, Khalilgharibi, Baum, Recho,
  Kabla, and Charras]{wyatt_actomyosin_2020}
Tom~PJ Wyatt, Jonathan Fouchard, Ana Lisica, Nargess Khalilgharibi, Buzz Baum,
  Pierre Recho, Alexandre~J. Kabla, and Guillaume~T. Charras.
\newblock Actomyosin controls planarity and folding of epithelia in response to
  compression.
\newblock \emph{Nature materials}, 19\penalty0 (1):\penalty0 109--117, 2020.
\newblock Publisher: Nature Publishing Group.

\bibitem[Xie et~al.(1991)Xie, Liu, Yang, and Fung]{xie1991zero}
J.~P. Xie, S.~Q. Liu, R.~F. Yang, and Y.~C. Fung.
\newblock The zero-stress state of rat veins and vena cava.
\newblock \emph{Journal of Biomechanical Engineering}, 113:\penalty0 36--41,
  1991.
\newblock \doi{10.1115/1.2894083}.

\bibitem[Xie et~al.(2018)Xie, Yang, and Jiang]{xie2018controlling}
Kenan Xie, Yuehua Yang, and Hongyuan Jiang.
\newblock Controlling cellular volume via mechanical and physical properties of
  substrate.
\newblock \emph{Biophysical journal}, 114\penalty0 (3):\penalty0 675--687,
  2018.

\bibitem[Xiong et~al.(2014)Xiong, Ma, Hiscock, Mosaliganti, Tentner, Brakke,
  Rannou, Gelas, Souhait, Swinburne, et~al.]{Xiong.2014.1}
Fengzhu Xiong, Wenzhe Ma, Tom~W Hiscock, Kishore~R Mosaliganti, Andrea~R
  Tentner, Kenneth~A Brakke, Nicolas Rannou, Arnaud Gelas, Lydie Souhait, Ian~A
  Swinburne, et~al.
\newblock Interplay of cell shape and division orientation promotes robust
  morphogenesis of developing epithelia.
\newblock \emph{Cell}, 159\penalty0 (2):\penalty0 415--427, 2014.

\bibitem[Xue et~al.(2016)Xue, Li, Feng, and Gao]{xue2016biochemomechanical}
Shi-Lei Xue, Bo~Li, Xi-Qiao Feng, and Huajian Gao.
\newblock Biochemomechanical poroelastic theory of avascular tumor growth.
\newblock \emph{Journal of the Mechanics and Physics of Solids}, 94:\penalty0
  409--432, 2016.

\bibitem[Yamada et~al.(2011)Yamada, Tadano, and Fujisaki]{yamada2011residual}
Satoshi Yamada, Shigeru Tadano, and Kazuhiro Fujisaki.
\newblock Residual stress distribution in rabbit limb bones.
\newblock \emph{Journal of biomechanics}, 44\penalty0 (7):\penalty0 1285--1290,
  2011.

\bibitem[Yamamoto(1956)]{Yamamoto.1956.1}
M.~Yamamoto.
\newblock The visco-elastic properties of network structure: {I. G}eneral
  formalism.
\newblock \emph{J. Phys. Soc. Jpn}, 11:\penalty0 413--421, 1956.
\newblock \doi{10.1143/JPSJ.11.413}.

\bibitem[Yang et~al.(2021)Yang, Xue, Chan, Rempfler, Vischi, Maurer-Gutierrez,
  Hiiragi, Hannezo, and Liberali]{Yang.2021.1}
Qiutan Yang, Shi-Lei Xue, Chii~Jou Chan, Markus Rempfler, Dario Vischi,
  Francisca Maurer-Gutierrez, Takashi Hiiragi, Edouard Hannezo, and Prisca
  Liberali.
\newblock Cell fate coordinates mechano-osmotic forces in intestinal crypt
  formation.
\newblock \emph{Nature Cell Biology}, 23\penalty0 (7):\penalty0 733--744, July
  2021.
\newblock ISSN 1465-7392, 1476-4679.
\newblock \doi{10.1038/s41556-021-00700-2}.
\newblock URL \url{http://www.nature.com/articles/s41556-021-00700-2}.

\bibitem[Zallen and Wieschaus(2004)]{Zallen-Wieschaus.2004.1}
J.~A. Zallen and E.~Wieschaus.
\newblock Patterned gene expression directs bipolar planar polarity in
  {\textit{{d}rosophila}}.
\newblock \emph{Dev. Cell}, 6:\penalty0 343--355, 2004.

\end{thebibliography}
\end{document}